\documentclass[onecolumn,nofootinbib,showkeys,superscriptaddress]{revtex4}
\usepackage{amsmath,amssymb}     %
\usepackage{graphics,epsfig,psfrag}
\usepackage{color}
\usepackage{graphicx}
\usepackage{dcolumn}
\usepackage{bm}
\usepackage{hyperref}

\begin{document}

\title{Calculating black hole shadows: review of analytical studies}

\author{Volker Perlick}
\email{perlick@zarm.uni-bremen.de}
\affiliation{ZARM, University of Bremen, 28359 Bremen, Germany}

\author{Oleg Yu. Tsupko}
\email{tsupko@iki.rssi.ru; tsupkooleg@gmail.com}
\affiliation{Space Research Institute of Russian Academy of Sciences, Profsoyuznaya 84/32, Moscow 117997, Russia}

\begin{abstract}
In this article, we provide a review of the current state of the research of the black hole shadow, focusing on analytical (as opposed to numerical and observational) studies. We start with particular attention to the definition of the shadow and its relation to the often used concepts of escape cone, critical impact parameter and particle cross-section. For methodological purposes, we present the derivation of the angular size of the shadow for an arbitrary spherically symmetric and static space-time, which allows one to calculate the shadow for an observer at arbitrary distance from the center. Then we discuss the calculation of the shadow of a Kerr black hole, for an observer anywhere outside of the black hole. For observers at large distances we present and compare two methods used in the literature. Special attention is given to calculating the shadow in space-times which are not asymptotically flat. Shadows of wormholes and other black-hole impostors are reviewed. Then we discuss the calculation of the black hole shadow in an expanding universe as seen by a comoving observer. The influence of a plasma on the shadow of a black hole is also considered.
\end{abstract}


\maketitle


\vspace{-0.9cm}

\tableofcontents


\section{Introduction}

Observation of the light deflection during a solar eclipse in 1919 was the first experimental confirmation of a prediction from the general theory of relativity. Since then, significant progress has been made in the study of effects that are caused by the deflection of light in a gravitational field. These effects are now combined under the name of ``gravitational lensing" \cite{GL-1, Blandford-Narayan-1992, Petters-2001, Perlick-2004a, GL-2, Bartelmann-2010, Dodelson-2017, Congdon-Keeton-2018}. In most observable manifestations of the gravitational lens effect the gravitational field is weak and deflection angles are small. The theoretical investigation can then be based on a linearized formula for the deflection angle that was derived already in 1915 by Albert Einstein. More recently, however, observations in the regime of strong deflection became possible.

A major breakthrough was made when, exactly a century after the first observation of gravitational light deflection, in 2019 the Event Horizon Telescope (EHT) Collaboration \cite{EHT-1, EHT-2, EHT-3, EHT-4, EHT-5, EHT-6} published an image of a black hole (BH): If light passes close to a BH, the rays can be deflected very strongly and even travel on circular orbits. This strong deflection, together with the fact that no light comes out of a BH, has the effect that a BH is seen as a dark disk in the sky; this disk is known as the BH shadow. The idea that this shadow could actually be observed was brought forward in the year 2000 in a pioneering paper by Falcke et al. \cite{Falcke-2000}, also see Melia and Falcke \cite{Melia-Falcke-2001}. Based on numerical simulations they came to the conclusion that observations at wavelengths near $1 \,$mm with Very Long Baseline Interferometry could be successful. This article was focused on the supermassive BH at the center of our Galaxy which is associated with the radio source Sagittarius A$^*$. The predicted size of this shadow was about 30 $\mu$as, which was shown to be comparable with the resolution of a global network of radio interferometers. (Based on the best available data for the mass and the distance of the black hole at the center of our Galaxy, the angular diameter of its shadow is now estimated as about 54 $\mu$as.)
For subsequent detailed general-relativistic magnetohydrodynamic (GRMHD) models of this object see, e.g., \cite{Broderick-Loeb-2005, Moscibrodzka-2009, Dexter-2009, Broderick-Fish-2011, Broderick-Johannsen-2014}. After remarkable achievements, both on the technological side for making the observations possible and on the computational side for evaluating them, the EHT Collaboration was then able to produce a picture that shows the shadow, not of the black hole at the center of our Galaxy, but rather of the one at the center of the galaxy M87, see \cite{EHT-1, EHT-2, EHT-3, EHT-4, EHT-5, EHT-6} and also \cite{Gralla-2019, Narayan-2019, Johnson-2020, Bronzwaer-2021,Kocherlakota-2021, Broderick-2021, Bronzwaer-Falcke-2021}. Inspired by this achievement, great attention is now focused on the investigation of various aspects of BH shadows.

In addition to setting up the ground-based Event Horizon Telescope network, which then turned out to be a great success, there have also been discussions about using space-based radio interferometers for observing the shadows of black holes. However, until now we do not have any space-based instrument appropriate for such observations. As demonstrated by Falcke et al. \cite{Falcke-2000}, scattering of light would wash out the shadow at wavelengths bigger than a few mm. Therefore the satellite Radioastron, which was operating at wavelengths of more than 1 cm, could not be used for observations of the shadow, although the resolution of this instrument would have been good enough \cite{Zakharov-Paolis-2005-New-Astronomy}. By contrast, the prospective space observatory Millimetron is supposed to operate at wavelengths that are near 1 mm which is appropriate for shadow observations. The perspectives of imaging with Millimetron the shadow of a black hole were briefly mentioned in Ref. \cite{Zakharov-Paolis-2005-New-Astronomy}; for more details we refer, e.g., to \cite{Kardashev-Millimetron-2014, Andrianov-2021, Likhachev-2021, Novikov-Millimetron-2021}.

In this article, we attempt to provide an up-to-date review of the current state of the research of the shadow of BHs, focusing on analytical (as opposed to numerical and observational) results. For analytical calculations of the shadow one starts out from the (over-)idealized situation that we see a BH against a backdrop of light sources, with no light sources between us and the BH, and that light travels unperturbed by any medium along lightlike geodesics of the space-time metric. In this setting one can, indeed, analytically calculate the shape and the size of the shadow, for an observer anywhere outside the BH, for a large class of BH models that includes the Kerr space-time as the most important example. Of course, this approach cannot give an image of the shadow that is realistic in the sense that it fully describes what we actually expect to see in the sky: In reality, the above-mentioned idealized assumptions will be violated for two reasons: Firstly, there will be light sources between us and the BH; e.g., light coming from an accretion disk will partly cover the shadow in the sky. The first who actually calculated, numerically, the visual appearance of a Schwarzschild BH surrounded by a shining and rotating accretion disk was Luminet \cite{Luminet-1979}. Thereby he assumed that light travels on lightlike geodesics of the space-time metric. For a generalization of this work to the Kerr space-time we refer to Viergutz \cite{Viergutz-1993}. Secondly, the propagation of light will be partly influenced by a medium, i.e., the rays will deviate from lightlike geodesics of the space-time metric because they are influenced by refraction, and there may also be scattering or absorption. This happens, e.g., when the BH is surrounded by a plasma or a dust. Such effects cannot in general be taken into account if one wants to restrict to analytical calculations, but many ray tracing codes have been written for investigating such effects numerically. For such numerical studies, which are not the subject of this article, we refer e.g. to James et al. \cite{JamesEtAl2015} where an overview on such ray tracing methods is given before concentrating on the BH picture that was produced for the Hollywood movie ``Interstellar'', and also to the above-mentioned papers by the EHT Collaboration \cite{EHT-1, EHT-2, EHT-3, EHT-4, EHT-5, EHT-6} where the numerical models on which the evaluation of the observations was based are detailed.
The observational appearance of a black hole as obtained in numerical simulations strongly depends on the distribution of light sources and on the properties of the emitting matter in the vicinity of the black hole \cite{Luminet-1979,Falcke-2000, Melia-Falcke-2001, Broderick-Loeb-2005, Moscibrodzka-2009, Dexter-2009, Broderick-Fish-2011, Broderick-Johannsen-2014, Gralla-2019, Narayan-2019, Johnson-2020, Bronzwaer-Falcke-2021}. Resulting images can be very different, but the main feature -- the shadow -- will have the same size and the same shape, as determined by the propagation of light in the strong gravitational field of the black hole. Determining the visual appearance of the shining matter, most likely an accretion disk, is the subject of ongoing numerical studies. This important work is beyond the scope of this review.

Obviously, the great interest of the general public in the shadow of a black hole has its reason in the fact that it gives us a visual impression of how a black hole looks like.
In addition, there is also a high scientific relevance of shadow observations, in particular because they can be used for distinguishing different types of black holes from each other (thereby confronting standard general relativity with alternative theories of gravity, fundamental or effective), and also for distinguishing black holes from other ultracompact objects which are sometimes called \emph{black hole mimickers} or \emph{black hole impostors}. For some kinds of such black hole impostors an analytical calculation of the shadow is possible and we will discuss these cases in a separate chapter.

We believe that analytical investigations of the shadow are of great relevance although they are restricted to highly idealized situations. The reasons are that by way of an analytical calculation (i) one gets a good understanding of how certain effects come about, (ii) one sees in which way certain parameters of the model influence the result and (iii) one provides a test-bed for checking the validity of numerical codes with simple examples. In particular, we believe that for getting a solid understanding of how a BH shadow comes about one cannot do anything better than repeating Synge's simple analytical calculation of the Schwarzschild shadow which will be reviewed below. 

Having said this, it should be clear that the major part of this article is restricted to light propagation in vacuo. However, there is one particular type of medium whose refractive influence on the shadow can be analytically taken into account, namely a non-magnetized, pressureless electron-ion plasma. We will consider this case in the last section of this article. In all other parts of the review, the terms ``light ray'' and ``light orbit'' are synonymous with ``lightlike geodesic of the space-time metric''.\\

\section{Basics of black hole shadow: definition and related concepts}

\vspace{1mm}

A black hole captures all light falling onto it and it emits nothing. Therefore even a naive consideration suggests that an observer will see a dark spot in the sky where the black hole is supposed to be located. However, due to the strong bending of light rays by the BH gravity, both the size and the shape of this spot are different from what we naively expect on the basis of Euclidean geometry from looking at a non-gravitating black ball. In the case of a spherically symmetric black hole, the difference between the shadow and the imaginary Euclidean image of the black hole is only in angular size: the shadow is about two and a half times larger (see discussion below). For a rotating black hole, the shape of the shadow becomes different: it is deformed and flattened on one side. The size and the shape of the shadow depend not only on the parameters of the black hole itself, but also on the position of the observer.

There are many words that have historically been used to refer to the visual appearance of a black hole and related concepts\footnote{We try to give a complete list and to indicate, for each term, the authors who were the first to use it; however, we cannot exclude the possibility that we missed something.}:
\begin{itemize}
   
    \item escape cone, see Synge \cite{Synge-1966};
    
    \item cone of gravitational capture of radiation, see Zeldovich and Novikov \cite{Zeld-Novikov-1965};
   
    \item the apparent shape of black hole, see Bardeen \cite{Bardeen-1973}, Chandrasekhar \cite{Chandra-1983};

    \item cross section, see Young \cite{Young-1976};

    \item image, optical appearance, photograph, see Luminet \cite{Luminet-1979, Luminet-2018};

    \item cross section of photon gravitational capture, see Dymnikova \cite{Dymnikova-1986};

    \item black hole shadow, see Falcke, Melia and Agol \cite{Falcke-2000};
    
    \item silhouette, see Fukue \cite{Fukue-2003}, Broderick and Loeb \cite{Broderick-Loeb-2009};
    
    \item mirage around black hole, see Zakharov et al \cite{Zakharov-Paolis-2005-New-Astronomy};

    \item photon ring, see Johannsen and Psaltis \cite{Johannsen-2010}, Johnson et al \cite{Johnson-2020};

    \item apparent image of the photon capture sphere, see Bambi \cite{Bambi-2017};
    
    \item critical curve, see Gralla, Holz and Wald \cite{Gralla-2019}.
    
\end{itemize}

Despite different names and different physical formulation of the problem (including those related to observational issues), all these concepts are strongly intertwined. It seems important to us to point out this list here, so that the historical perspective is not lost over the use of different names. For example, in many cases later works are based at least on the mathematical apparatus developed in earlier works. We emphasize that this is a list of different names; it is not meant as a list of the most important works in the field.

By now the term `shadow' has become the most common one. The word `shadow' in different languages has several meanings. The most usual meaning is the dark area created on a surface (a screen or the ground) by an obstacle located between a source of light rays and this surface. For example, it may be the shadow of the human body or of a building on the ground at a sunny day. Another meaning of the word `shadow' is a dark silhouette of a body which occurs when we look at it against a bright background. In this case, we see not the details of this body, but only the shape. For example, this is what happens when we look from the outside at a bright window in the evening and see only dark silhouettes of the people in the room. In the case of the BH shadow, we use the word `shadow' rather in the second meaning. Thus, the shadow of a BH can be understood as \textit{a dark silhouette of the BH against a bright background} which, however, is strongly influenced by the gravitational bending of light.

\begin{figure*}[ht]
\begin{center}
\includegraphics[width=0.95\textwidth]{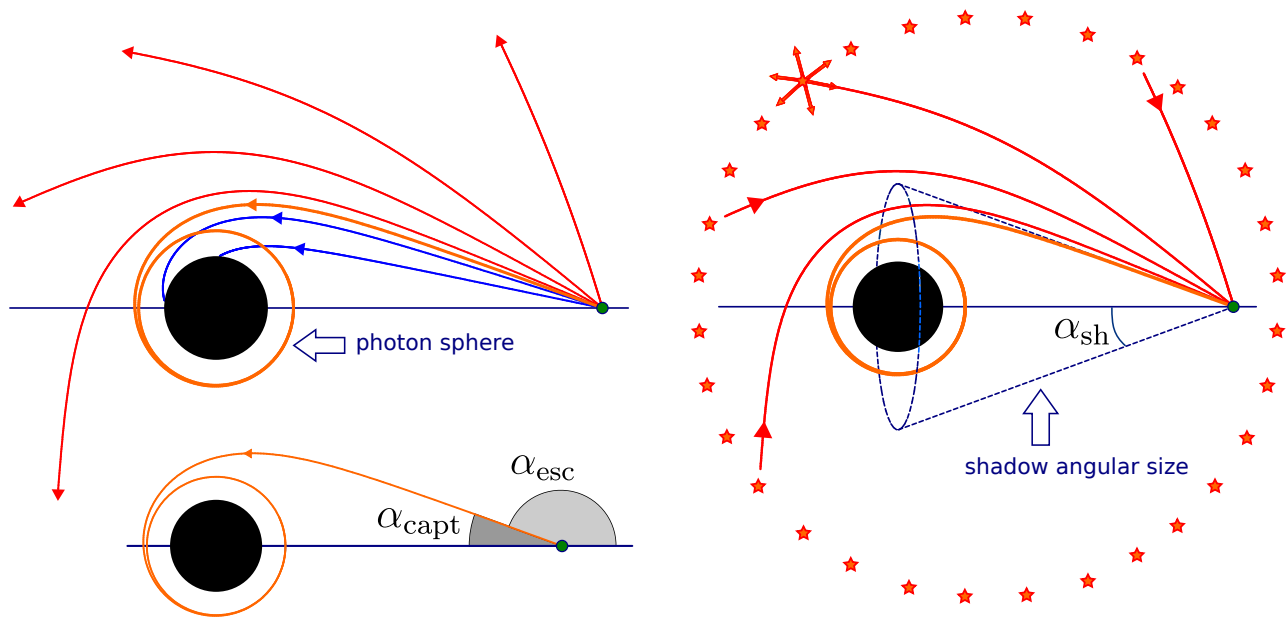}
\end{center}
\caption{(Color online) Formation of a black hole shadow in the case of the Schwarzschild BH. All paths shown are obtained numerically. LEFT: Imagine that the observer does not receive the rays, but emits them (as if time goes in backward direction). Part of the rays after being deflected by the BH will fly away to infinity (red ones in the figure), other rays will be captured by the BH (blue ones). The borderline case between these two classes is the rays that spiral asymptotically towards the photon sphere (orange curve). In the lower part of the image we show the angular size $\alpha_\textrm{capt}$ of the capture cone of light rays and the angular size $\alpha_\textrm{esc}$ of the escape cone. RIGHT: Now imagine that there are light sources all around the observer and the black hole, and each of these sources emits light in all directions. Some of the rays emitted, after being deflected by the BH, will reach the observer. Although very diverse trajectories of the rays arriving at the observer are possible, there will be no rays of the second type (blue) from the left picture among them. Therefore, the indicated cone will remain empty for the observer, this will be the BH shadow. Note that it is assumed that there are no light sources in the region through which rays of the second type travel. In the picture, we have extended the cone in the observer's sky up to the BH in order to demonstrate that the observed angular size of the shadow will be larger than both the Euclidean size of the BH and the Euclidean size of the photon sphere. See also Fig.3 in our paper \cite{BK-2019}. }
\label{fig:formation}
\end{figure*}

For understanding the theoretical construction of the shadow, one should consider an observer at some distance from the black hole. We can then divide all light rays that issue from this observer position into the past into two classes: Those which go to infinity after being deflected by the black hole and those which go to the horizon. We now consider the idealized situation that there are light sources densely distributed everywhere in the universe but not in the region between the observer and the black hole, i.e., we assume that past-oriented light rays of the first class will meet a light source somewhere but light rays of the second class will not. As each light ray corresponds to a point on the observer's sky, we would then assign brightness to a point on the observer's sky if the corresponding light ray goes to infinity and darkness otherwise. The dark part of the observer's sky is what we call the shadow. Its boundary corresponds to light rays that go neither to infinity nor to the horizon but are trapped within the space-time. In the Schwarzschild space-time, and in other similar space-times that are spherically symmetric, static and asymptotically flat, these light rays asymptotically approach an unstable photon sphere, i.e., a sphere that is filled with circular lightlike geodesics that are unstable with respect to radial perturbations. We sketch this situation for a Schwarzschild BH in Fig.~\ref{fig:formation}.    

Below we will also discuss the shadow of a non-static (rotating) black hole, which includes in particular the case of a Kerr black hole. We mention already now that then the situation is more complicated. There is no longer a photon sphere but rather a photon region, see Refs.~\cite{Perlick-2004a,Gren-Perlick-2014,Gren-Perlick-2015}.

\begin{figure*}[ht]
\begin{center}
\includegraphics[width=0.95\textwidth]{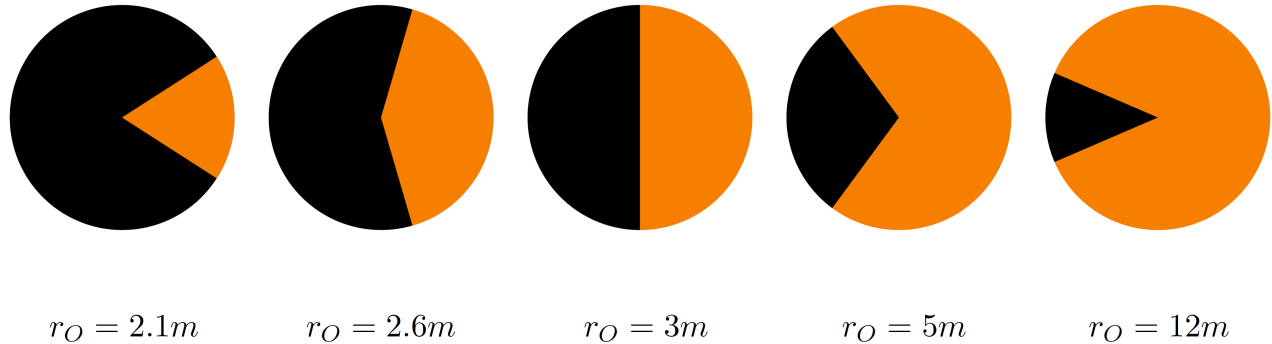}
\end{center}
\caption{(Color online) 
Angular region of the shadow (black) for static observers at different radius coordinates $r_{\mathrm{O}}$ in the Schwarzschild space-time. The black hole is to the left of the picture. For $r_{\mathrm{O}}=12m$ the shadow is a small circular area on the observer's sky. If the observer is closer to the black hole, the shadow is bigger. At $r_{\mathrm{O}}=3m$, when the observer is at the photon sphere, the shadow fills exactly one half of the observer's sky. For an observer between $2m$ and $3m$ the shadow is bigger than one half of the sky, and the entire sky becomes dark if the  observer position approaches the horizon, $\alpha_{\mathrm{sh}} \to \pi$ if $r_{\mathrm{O}} \to 2m$.   }
\label{fig:pacman}
\end{figure*}

The Schwarzschild metric
\begin{equation}  \label{schw-metric}
ds^2 = - \left(1 - \frac{2m}{r} \right) \, c^2 dt^2 + \frac{dr^2}{1-2m/r} + \, r^2
\left( d \theta^2 + \sin^2 \theta \, d \varphi^2 \right),   \;   m=\frac{GM}{c^2}   \,   ,
\end{equation}
features a horizon at radius $2m$ and an unstable photon sphere at radius $r_{\mathrm{ph}}=3m$. The angular radius $\alpha_{\mathrm{sh}}$ of the shadow (i.e., the opening angle of the shadow cone) was presented in equivalent forms in a paper by Synge \cite{Synge-1966} and in a review by Zeldovich and Novikov \cite{Zeld-Novikov-1965}. Neither of them actually used the word `shadow'. Synge calculated, for a static observer at radius $r_\mathrm{O}$, the `escape cone' of light which is the complement of the shadow cone, see Fig.\ref{fig:formation}:
\begin{equation} \label{synge-escape}
\sin^2 \alpha_{\mathrm{esc}} \, = \,
\frac{27m^2 (1-2m/r_{\mathrm{O}})}{r_{\mathrm{O}}^2} \,  .
\end{equation}
Zeldovich and Novikov calculated the cone of `gravitational capture of radiation', see Fig.\ref{fig:formation} in our paper and Fig.12 in their paper \cite{Zeld-Novikov-1965}:
\begin{equation} \label{zn-capture}
\left| \tan \alpha_{\mathrm{capt}} \right| \, = \,
\frac{ \left( 1-\frac{2m}{r_{\mathrm{O}}} \right)^{1/2}  }{ \left( \frac{2m}{r_{\mathrm{O}}} - 1 + \frac{r_\mathrm{O}^2}{27m^2} \right)^{1/2} } \, .
\end{equation}
Obviously, we have:
\begin{equation}
\alpha_{\mathrm{sh}} = \alpha_{\mathrm{capt}} = \pi - \alpha_{\mathrm{esc}} \, .
\end{equation}
Since $0< \alpha_\textrm{i} < \pi$
\begin{equation}
\sin \alpha_{\mathrm{sh}} = \sin \alpha_{\mathrm{capt}} = \sin \alpha_{\mathrm{esc}}
\end{equation}
and it does not make any difference if we write the formulas for the angular radius of the shadow $\alpha_{\mathrm{sh}}$ with the sine or with the square of the sine,
\begin{equation} \label{synge-shadow}
\sin^2 \alpha_{\mathrm{sh}} \, = \, \frac{27m^2 (1-2m/r_{\mathrm{O}})}{r_{\mathrm{O}}^2}  \, , \quad \mbox{or} \quad
\sin \alpha_{\mathrm{sh}} \, = \,
\frac{3\sqrt{3} m \sqrt{1-2m/r_{\mathrm{O}}} }{r_{\mathrm{O}}} \,  .
\end{equation}
However, we have to carefully choose the correct branch of the sine function: for $r_\textrm{O} < r_\textrm{ph}$ we must choose $\alpha_\textrm{sh} > \pi/2$. Light rays that leave the observer position under the angle $\alpha _{\mathrm{sh}}$ with respect to the inwards directed radial line will asymptotically spiral towards the photon sphere at $r_{\mathrm{ph}}=3m$, either from above ($r_\textrm{O}> 3m$, $\alpha _{\textrm{sh}}< \pi/2$) or from below ($r_\textrm{O} < 3m$, $\alpha _{\textrm{sh}}> \pi/2$). If we define the arcsin function such that it takes values in the interval between $-\pi/2$ and $\pi/2$, as usual, solving (\ref{synge-shadow}) for $\alpha_{\textrm{sh}}$ results in
\begin{equation} \label{alpha-schw-expl}
\alpha_\mathrm{sh}(r_\mathrm{O}) = \left\{\begin{array}{l}
 \pi - \mathrm{arcsin} \left( 3\sqrt{3}m \sqrt{1 - 2m/r_\mathrm{O}}/r_\mathrm{O} \right) \; 
 \mbox{for}  \;\;   2m  < r_\mathrm{O} \le 3m \, ,\\
 \mathrm{arcsin} \left( 3\sqrt{3}m \sqrt{1 - 2m/r_\mathrm{O}}/r_\mathrm{O} \right) \;  \mbox{for}  \;\;   r_\mathrm{O} \ge 3m    \, .
\end{array} \right.
\end{equation}
We must, of course, have $2m< r_{\mathrm{O}}$ because a static observer cannot exist beyond the horizon. In Fig. \ref{fig:pacman} we show the region of the observer's sky that is covered by the shadow for different values of $r_{\mathrm{O}}$.

\begin{figure}[ht]
\begin{center}
\includegraphics[width=0.65\textwidth]{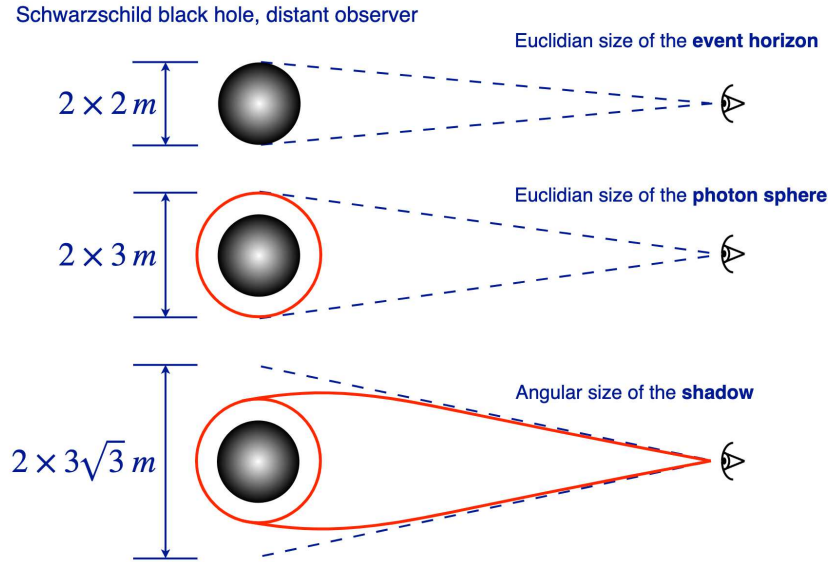}
\end{center}
\caption{(Color online) 
Comparison of Euclidean size of the black hole (event horizon), Euclidean size of the photon sphere and angular size of the shadow. See also figures in our papers \cite{BK-Tsupko-Universe-2017}, see Fig.10 there, and \cite{BK-2019}, see Fig.2 there.    }
\label{fig:misc}
\end{figure}

Here we want to comment on something we believe to be a common misconception. It is true that the defining property of a black hole is the existence of a horizon; so when talking about the 'shadow' of a black hole, which some people just call the 'image' of a black hole, it seems natural to associate this with the 'visual appearance of the horizon'. Indeed, statements of this kind can be found, e.g., in the media or in the popular literature on science. Moreover, based on such statements sometimes it is given the impression that one can calculate, or at least estimate, the apparent  angular radius of the shadow on the observer's sky by dividing the coordinate radius of the horizon by the coordinate distance to the observer. For an  observer at a large distance, this gives the angle \footnote{Note that here and everywhere below we refer to the angular radius and not to the angular diameter, see Fig.\ref{fig:misc}} under which the observer would see the coordinate radius of the horizon according to Euclidean geometry, i.e., according to the idea that light rays are straight lines in the chosen coordinate system. Actually, this consideration is incorrect for two reasons. Firstly, the boundary of the shadow corresponds to light rays that spiral towards the photon sphere (at $r = 3m$ in the Schwarzschild case) and not towards the horizon (at $r = 2m$ in the Schwarzschild case). Secondly, light rays that come close to the photon sphere are very much different from straight lines in Euclidean geometry. These two different errors have the consequence that the angular radius of the shadow is actually about two-and-a-half times bigger than the naive Euclidean estimate suggests, as can be seen by the following simple calculation. If light rays were straight lines in the chosen coordinate system, the horizon radius of a Schwarzschild black hole would be seen by an observer at radius $r_{\textrm{O}}$ under the angle $\alpha_\textrm{bh}$ given by
$\alpha_\textrm{bh} \approx 2m/r_\textrm{O}$
if $r_\mathrm{O} \gg m$.
Comparing with the angle $\alpha _{\textrm{sh}}$ from the correct formula (\ref{alpha-schw-expl}) shows that, for $r_{\textrm{O}} \gg m$, we have $\alpha_\textrm{sh} \approx (3 \sqrt{3}/2) \alpha_\textrm{bh}$, see also Fig.\ref{fig:misc}. In this context we quote from the press release \cite{EHT-press} of the Event Horizon Telescope Collaboration (April 10, 2019) 
about capturing the first image of a black hole: 'The shadow of a black hole is the closest we can come to an image of the black hole itself, a completely dark object from which light cannot escape. The black hole's boundary --- the event horizon from which the EHT takes its name --- is around 2.5 times smaller than the shadow it casts...'; see also \cite{EHT-1}.

Another important notion related to the BH shadow is the critical value $b_{\mathrm{cr}}$ of the impact parameter which separates the captured orbits from the flyby orbits of the incident light rays. As a light ray with $b_{\mathrm{cr}}$ asymptotically spirals towards an unstable circular orbit on the photon sphere with the radius $r_{\mathrm{ph}}$, it must have the same constants of motion as the latter. In the case of the Schwarzschild metric this value is equal to \cite{Hilbert-1917, MTW-1973}:
\begin{equation} 
b_{\mathrm{cr}} = 3\sqrt{3}m \, , \quad r_{\mathrm{ph}} = 3m \, .
\end{equation}

Circular lightlike geodesics are sometimes referred to as `light rings' \cite{Cunha-Herdeiro-2018}. By contrast, some authors use the word `photon ring' in a completely different sense, namely for a bright ring that surrounds the shadow in the observer's sky. More precisely, the concept of a photon ring is used in the literature in several closely related, but still different, meanings. In general, one has to distinguish primary, secondary and higher-order images, see e.g. \cite{Broderick-2021, Luminet-1979}: Primary images correspond to light rays that make less than a half turn around the center; for secondary images they make between a half turn and a full turn, and for higher-order images they make more than a full turn.  All higher-order images of a shining black hole environment together form an infinite sequence of very thin and faint rings in the observer's sky which cannot be expected to be isolated by observation, so they are often considered as a single ring. This is what some authors, e.g. \cite{Johannsen-2010}, call the 'photon ring'. As this ring practically coincides with the boundary of the shadow, i.e., with the critical curve on the observer's sky, one may use the formulas of the critical curve for calculating the shape and the size of this photon ring, as is done e.g. in \cite{Johannsen-2013}. Gralla et al \cite{Gralla-2019} also use the word ``photon ring'' for the higher-order images and they introduce in addition the name ``lensing ring'' for the secondary images. In the case that the illuminating matter is an accretion disk and that the observer is in an almost polar position, this lensing ring is also fairly close to the boundary of the shadow, although not quite as close as the photon ring, and much brighter than the latter. Numerical simulations show that then the observer would see a single, unresolved, bright ring in which the leading contribution comes from secondary images \cite{Johnson-2020}. Johnson et al \cite{Johnson-2020} call this bright ring the 'photon ring', i.e. they include the ring of secondary images ('lensing ring' of \cite{Gralla-2019}) and the higher-order rings into it. This terminology makes only a small difference with respect to the previous one \cite{Johannsen-2010, Johannsen-2013} as far as the position of the photon ring is concerned, but it makes a considerable difference as to its brightness. Finally, some authors use the term 'photon ring' for each individual ring rather than for the entire set \cite{Broderick-2021}.

The notion of 'photon ring' should also not be confused with the `Einstein rings' that occur in gravitational lensing of distant source in cases of perfect rotational symmetry about a central view line. In the case of a spherically symmetric black hole there is a primary Einstein ring and an infinite sequence of higher-order Einstein rings, also known as `relativistic Einstein rings' \cite{Virbhadra-2000, Bozza-2001, BK-Tsupko-2008}.

For large distances, the angular radius (\ref{alpha-schw-expl}) of the sha\-dow can be simplified to
\begin{equation} \label{eq: L1a}
\alpha_{\mathrm{sh}}
\, \approx \, 
\frac{b_{\mathrm{cr}}}{r_\textrm{O}}  \, , \quad r_\textrm{O} \gg m \, .
\end{equation}
Therefore, to calculate the shadow at large distances, it is sufficient to find the critical value of the impact parameter: the angular size of the shadow is then given simply by dividing by the radial coordinate of the observer.

With this approximation in mind, the shadow is often calculated in terms of critical impact parameters. For a spherically symmetric black hole, the critical value of the impact parameter is often referred to simply as \textit{the radius of the black hole shadow} $r_\mathrm{sh}$, see, e.g., \cite{Psaltis-2020}. Note, however, that the impact parameter has the dimension of a length, so one has to convert this into an angle to get the radius of the shadow in the observer's sky. Bardeen \cite{Bardeen-1973} has introduced this approach for the shadow of a Kerr black hole. As the latter is not circular, one needs \emph{two} impact parameters; the angular radii of the shadow are then approximately given by dividing these impact parameters by the (Boyer-Lindquist) radius coordinate $r_\textrm{O}$ of the observer. However, the principal point is that the described relationship between the angular size of the shadow and the critical impact parameter is valid only for metrics that are asymptotically flat at infinity. Of course, one can calculate the critical impact parameters also in space-times that are not asymptotically flat, but then these parameters will not give us directly the necessary information about the size and the shape of the shadow, see Sections \ref{sec:spher-symm}, \ref{sec:generalize}, \ref{sec:expanding}.

\begin{figure*}[t]
	\begin{center}
	\includegraphics[width=0.75\textwidth]{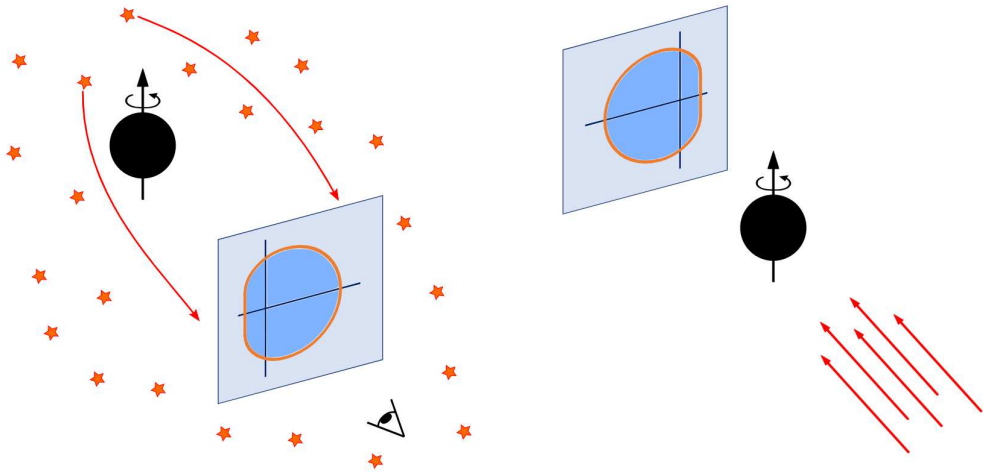}
	\end{center}
	\caption{(Color online) Mirror difference in shape of shadow (left) and capture cross-section (right).}
	\label{fig:cross-section}
\end{figure*}

Owing to its described property for an observer at a large distance, the shadow is sometimes referred to as the \textit{capture cross-section} of a black hole for photons incident from infinity. Identifying the two notions is, indeed, justified in asymptotically flat space-times if the observer is far away from the black hole, as is the case in all practical situations. It should be noted, however, that for the definition of the capture cross section one considers light rays that come in from infinity and fly towards the black hole whereas for the definition of the shadow one has to consider the time-reversed situation. In static space-times this makes no difference. In stationary but non-static space-times such as the Kerr space-time, however, it leads to the fact that the capture cross-section is flattened on the other side than the shadow, see Fig.\ref{fig:cross-section}, compare with Figs. in Young \cite{Young-1976}, Dymnikova \cite{Dymnikova-1986}.

In view of astrophysically relevant situations, the definition of the shadow introduced in this Section was based on a highly idealized description, so it has to be complemented with more detailed, largely numerical, investigations for explaining actual observations.
However, this idealized consideration is a necessary first step which, in particular, gives us the correct understanding of how the shadow basically comes about.

A noticeable example where this idealized description is not valid is the following. In reality, the condition that there are light sources conveniently placed behind the black hole to provide a bright backdrop but not between the observer and the black hole will not be satisfied. So in practice it is not possible to see the shadow as a perfectly black disk surrounded by a bright area. For actually seeing the shadow, some special conditions must be satisfied. E.g., it is sufficient to have more light sources behind the black hole than between the black hole and the observer.\footnote{OYuT is thankful to Sandra Boeschenstein for a discussion of this point.} Bardeen \cite{Bardeen-1973} considered 'a source of illumination behind the black hole whose angular size is large compared with the angular size of the black hole'. Luminet \cite{Luminet-1979} showed a picture of the shadow against the background of a luminous, geometrically thin and optically thick accretion disk, where the lower part of the shadow is 'covered' by the accretion disk. Falcke et al \cite{Falcke-2000} argued that under realistic conditions a black hole can be observed 'when the emission occurs in an optically thin region surrounding the black hole'. In all these situations, even in complicated cases where numerical simulations are required, the construction of the shadow as given above remains qualitatively and quantitatively correct, although the actual image will depend on the emission profile and will be `blurred' by various effects.

Let us consider this in more detail on the example of a famous picture (Fig.11) from the work of Luminet \cite{Luminet-1979}. Luminet calculated the primary and secondary images of a geometrically thin accretion disk. (In this context we also refer to earlier work \cite{Cunningham-1973} where the visible shapes of circular orbits around black hole were presented.) In the center of Luminet's picture, one can see a dark spot; this is the shadow of a black hole. The lower part of the shadow is obscured by the primary image of the accretion disk, while the upper part of the shadow is framed by an arc from secondary images that is located very close to the boundary of the shadow. The size of the shadow in this illustration is completely determined not by the distribution of the luminous matter (in this case, a geometrically thin accretion disk), but by the motion of the rays in the strong gravitational field of the black hole.

Another good example is provided in Fig.12 of Bronzwaer and Falcke \cite{Bronzwaer-Falcke-2021}. The left panel shows a Schwarzschild black hole illuminated by a rotating thin disk. A superficial observer might be tempted to identify the entire ``Central Brightness Depression'', as it is called in this paper, with the shadow. This, however, is false. If looking more carefully one sees the shadow as a smaller dark disk, surrounded by a thin bright ring.

To get an idea of the variety of different physical scenarios for observing a shadow and their comparison with each other, we refer the reader to the following interesting works that were published after observing a shadow: \cite{Gralla-2019, Narayan-2019, Dokuchaev-2020, Chael-2021, Bronzwaer-2021, Bronzwaer-Falcke-2021}. \\


\section{Derivation of the angular size of the shadow for the general case of a spherically symmetric and static metric}
\label{sec:spher-symm}

The general method of constructing the shadow in a spherically symmetric and static spacetime consists of two steps:

(i) First of all, we write down a general expression for the inclination angle of a light ray emitted from the observer into the past. This expression is general in the sense that it will contain some constant of motion the value of which will not be specified; therefore it will apply to all emitted light rays. Below, we use the radius coordinate of the point of closest approach as the relevant constant of motion. 

(ii) Secondly, we need to distinguish those light rays which asymptotically go to unstable circular light orbits and substitute the appropriate value of the constant of motion for these rays into the general formula for the angle, see eq.(\ref{eq:sinalpha}) below. If there is only one photon sphere, and if it consists of unstable circular light rays, this construction is unambiguous. If there are several photon spheres, one has to specify where precisely one assumes light sources to be situated.

Here we will not consider objects without a photon sphere. Such an object casts no shadow at all if it is transparent; if it is opaque, it casts a shadow which is determined by light rays grazing its surface, i.e., in  a way quite similar to shadows in every-day life, although of course the light bending has to be taken into account. We mention in passing that in the class of spherically symmetric and static metrics the ones with a photon sphere are exactly the ones where an observer can see (theoretically) infinitely many images of a light source, see Hasse and Perlick \cite{HassePerlick2002}.

Let us consider a spherically symmetric and static metric
\begin{equation}\label{eq:g}
g_{\mu \nu} dx^{\mu} dx^{\nu} 
= - A(r) \, c^2 dt^2 + B(r) \, dr^2 
+ \, D(r) \big( d \vartheta ^2 + \mathrm{sin} ^2 \vartheta \,
d \varphi ^2 \big) \, ,
\end{equation}
where $A(r)$, $B(r)$ and $D(r)$ are positive.

In this metric, the Lagrangian $\mathcal{L} (x , \dot{x} ) = (1/2) g_{\mu \nu} \dot{x}^{\mu} \dot{x}^{\nu}$ for the geodesics takes the form:
\begin{equation}\label{eq:Lagr}
\mathcal{L} (x , \dot{x} ) \, = \, 
\dfrac{1}{2} \, \left(- \, A(r) \,  c^2 \, \dot{t}{}^2
+ B(r) \, \dot{r}^2 + D(r) ( \dot{\vartheta}^2 + \sin^2 \vartheta \, \dot{\varphi}^2 )  \right) 
\,.
\end{equation}
Because of the symmetry, it suffices to consider geodesics in the equatorial plane: $\vartheta=\pi/2$, $\sin \vartheta = 1$. The $t$ and $\varphi$ components of the Euler-Lagrange equation
\begin{equation}
\frac{d}{d \lambda} \left( \frac{\partial \mathcal{L}}{\partial \dot{x}^\mu}  \right)  - \frac{\partial \mathcal{L}}{\partial x^\mu} = 0
\end{equation}
give us two constants of motion,
\begin{equation}\label{eq:com}
E \, = \, A(r) \, c^2 \, \dot{t} \, , \qquad
L \, = \, D(r) \,   \dot{\varphi} \, .
\end{equation}
Instead of using the $r$ component of the Euler-Lagrange equation, for our purposes it is easier to use a first integral of the geodesic equation, namely (for light) $g_{\mu \nu} \dot{x}^\mu \dot{x}^\nu  =  0$. Hence
\begin{equation}
- \, A(r) \,  c^2 \, \dot{t}{}^2
+ B(r) \, \dot{r}^2 + D(r) \, \dot{\varphi}^2 = 0
\,.
\end{equation}
After inserting (\ref{eq:com}) this equation can be solved for 
$\dot{r}{}^2/\dot{\varphi}{}^2 = (dr/d \varphi )^2$ which gives us the orbit equation for lightlike geodesics,
\begin{equation}\label{eq:ol}
\left( \dfrac{dr}{d\varphi} \right) ^2 \, = \, \frac{D(r)}{B(r)}   \left(  \frac{D(r)}{A(r)}
\dfrac{E^2 }{c^2 L^2} - 1 \right) \, .
\end{equation}
We see that the orbit equation for a given metric depends only on one constant of motion, for example on the impact parameter $cL/E = b$. Note that eq.(\ref{eq:ol}) is of the same form as an energy conservation law in one-dimensional classical mechanics, $(dr/d \varphi )^2 + V_{eff}(r)=0$, where the effective potential depends on the impact parameter $b =cL/E$, with $\varphi$ playing the role of the time variable. If the impact parameter has been fixed, we may thus visualize the radial motion by a motion in the classical potential $V_{eff}(r)$. The circular orbits can then be determined by solving the equations $V_{eff}=0$ and $dV_{eff}/dr = 0$ with respect to $r$ and $b$.
In situations where the light ray approaches the center and then goes out again after reaching a minimum radius $R$, it is convenient to rewrite the orbit equation (\ref{eq:ol}) using $R$ instead of $b$. As $R$ corresponds to the turning point of the trajectory, the condition $dr/d\varphi |_R =0$ has to hold. Using (\ref{eq:ol}), we obtain the relation between $R$ and the constant of motion $E/(cL)$:
\begin{equation} \label{eq:turn-point-vac}
\dfrac{E^2 }{c^2 L^2} = \frac{A(R)}{D(R)}     \, , \quad \mbox{or} \quad  \dfrac{1}{b^2} = \frac{A(R)}{D(R)}  \, .
\end{equation}
Let us also introduce the function
\begin{equation} \label{eq: def-h-spher}
h(r) ^2 =  \frac{D(r)}{A(r)}  \, .
\end{equation}
In the Schwarzschild case, the function $h(r)$ is equivalent to the 'effective potential' introduced in eq.(25.58) of Misner et al. \cite{MTW-1973} for photon motion in Schwarzschild gravity.

It is easy to see that the impact parameter and the function $h(r)$ are related by
\begin{equation} \label{eq:b-h-R}
b = h(R) \, ,
\end{equation}
compare, e.g., with eq.(4) of Bozza \cite{Bozza-2002}.
Using (\ref{eq:turn-point-vac}) and (\ref{eq: def-h-spher}) in (\ref{eq:ol}), we get
\begin{equation}\label{eq:ol1}
\left( \dfrac{dr}{d\varphi} \right) ^2 \, = \, \frac{D(r)}{B(r)}   \left(  \frac{h(r)^2}{h(R)^2} - 1 \right) \, .
\end{equation}

\begin{figure}[ht]
	\begin{center}
		\includegraphics[width=0.75\textwidth]{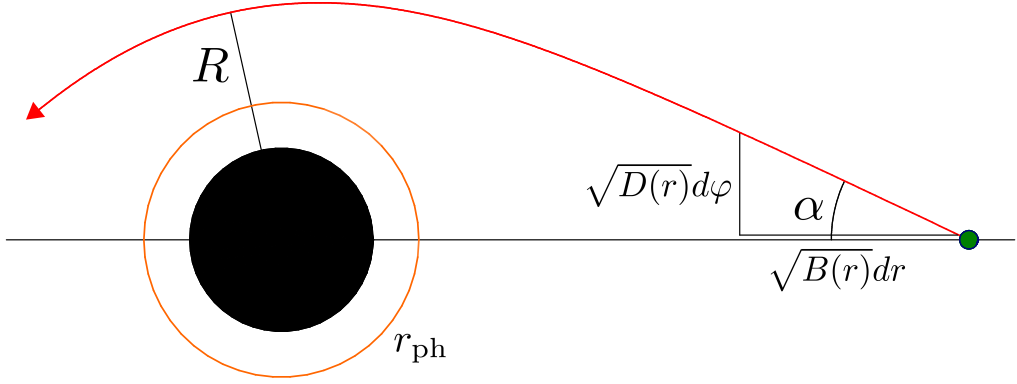}
	\end{center}
	\caption{Example of calculation of light ray emitted from the observer's position (small disk) into the past under an angle $\alpha$. The black hole horizon and the photon sphere are shown; $r_{\mathrm{ph}}$ is the photon sphere radius. The trajectory is calculated in the Schwarzschild space-time, $R$ denotes the radius coordinate at the point of closest approach.}
	\label{fig:initial}
\end{figure}

For constructing the shadow we assume that a static  observer at radius coordinate $r_{\mathrm{O}}$ sends light rays into the past. As can be seen from Fig.~\ref{fig:initial}, the angle $\alpha$ between such a light ray and the radial direction is given by
\begin{equation}\label{eq:alpha1}
\cot \, \alpha \, = 
\left. \frac{\sqrt{g_{rr}}}{\sqrt{g_{\varphi \varphi}}}  
\, \dfrac{dr}{d \varphi} \right|_{r=r_{\mathrm{O}}} = \left.
\dfrac{\sqrt{B(r)}}{\sqrt{D(r)}} \,
\dfrac{dr}{d \varphi} \right|_{r=r_{\mathrm{O}}} \, .
\end{equation}
With the help of (\ref{eq:ol1}), we obtain 
\begin{equation}
\cot^2 \alpha \, = \frac{h(r_{\mathrm{O}})^2}{h(R)^2 }  - 1 \, .
\end{equation}
By elementary trigonometry, we get
\begin{equation}\label{eq:sinalpha}
\sin^2 \alpha \, = \, \dfrac{h(R)^2}{h(r_{\mathrm{O}})^2}   \, .
\end{equation}
The boundary curve of the shadow corresponds to past-oriented light rays that asymptotically approach one of the unstable circular light orbits at radius $r_{\mathrm{ph}}$. Therefore we have to consider the limit $R \rightarrow r_{\mathrm{ph}}$ in (\ref{eq:sinalpha}) for getting the angular radius $\alpha_{\mathrm{sh}}$ of the shadow, 
\begin{equation} \label{eq:shadow}
\sin^2 \alpha_{\mathrm{sh}} \, = \, 
\dfrac{h(r_{\mathrm{ph}})^2}{h(r_{\mathrm{O}})^2}  \, .
\end{equation}
Here $h(r)$ is given by the formula (\ref{eq: def-h-spher}). Note that the critical value $b_\textrm{cr}$ of the impact parameter is connected with $r_\textrm{ph}$ by
\begin{equation} \label{eq:b-crit-h}
b_\textrm{cr} = h(r_\textrm{ph}) \,  .
\end{equation}
Therefore we can also write eq.(\ref{eq:shadow}) as
\begin{equation}\label{eq:shadow-2}
\sin^2 \alpha_{\mathrm{sh}} \, = \, 
\dfrac{ b_\textrm{cr}^2  }{h(r_{\mathrm{O}})^2}  \, ,
\quad \mbox{or} \quad
\sin^2 \alpha_{\mathrm{sh}} \, = \, 
\dfrac{ b_\textrm{cr}^2 A(r_\mathrm{O})  }{D(r_{\mathrm{O}})}   \, .
\end{equation}
In different notations, the formula for the angular size of the shadow in a spherically symmetric and static spacetime can be found in Pande and Durgapal \cite{Pande-1986}, see their eq.(6), and in Cveti\v{c}, Gibbons and Pope \cite{Cvetic-2016}, see their eq.(2.59).

The only thing we still have to find is the radius of the photon sphere $r_{\mathrm{ph}}$ for the given metric (\ref{eq:g}). Along a circular light orbit the two conditions  $dr/d \varphi = 0$ and $d^2r/d \varphi ^2 =0$ have to hold simultaneously. The condition of $d^2r/d \varphi ^2 =0$ can be obtained by differentiation of eq. (\ref{eq:ol}). Solving the two equations together, we find the equation for the radius of a circular light orbit in the form
\begin{equation}\label{eq:circ}
0 \, = \, \dfrac{d}{dr} h(r)^2 
\end{equation}
with the function $h(r)$ from (\ref{eq: def-h-spher}).
Eq.(\ref{eq:circ}) was first given, in a different notion, as eq.(33) in Atkinson \cite{Atkinson-1965}. Equivalent equations, again in different notation, can also be found, e.g. as eq.(9) of Virbhadra and Ellis \cite{Virbhadra-Ellis-2001}, as eq. (54) of Claudel, Virbhadra and Ellis \cite{Claudel-Virbhadra-2001} and as eq.(3) of Bozza \cite{Bozza-2002}.
This equation may be satisfied for several radius values $r$, so there may be several photon spheres. If this is the case, one has to distinguish unstable photon spheres, towards which light rays may spiral, from stable photon spheres about which light rays may oscillate. The construction of the shadow has to be discussed for each such case separately and the result strongly depends on where exactly the light sources are situated.

Let us now consider a spherically symmetric and static metric (\ref{eq:g}) that is asymptotically flat, i.e., 
$A(r) \to 1$, $B(r) \to 1$ and $D(r)/r^2 \to 1$ for $r \to \infty$. Then for an observer at a large distance the angular size (\ref{eq:shadow}) of the shadow can be approximated by
\begin{equation}\label{eq:large-dist}
\alpha_{\mathrm{sh}} \, = \, 
\dfrac{h(r_{\mathrm{ph}})}{r_{\mathrm{O}}}  \, , \quad \mbox{or} \quad  \alpha_{\mathrm{sh}} \, = \, 
\dfrac{b_\textrm{cr}}{r_{\mathrm{O}}}    \,  .
\end{equation}
Therefore, for calculating the shadow for large distances in an asymptotically flat spherically symmetric and static space-time, it is sufficient to determine the function $h(r)$ from (\ref{eq: def-h-spher}) with $r=r_{\mathrm{ph}}$ found from eq.(\ref{eq:circ}). If we additionally assume that $D(r)=r^2$ and introduce the function $w(r)=\sqrt{A(r)} = \sqrt{-g_{tt}(r)}$, then the equation (\ref{eq:circ}) can be simplified to
\begin{equation}
r w'(r) = w(r)    
\end{equation}
(cf. eq.(3) of Psaltis \cite{Psaltis-2020}). Also, we have already mentioned that several authors refer to our $b_\mathrm{cr}$ from eq.(\ref{eq:b-crit-h}) just as to ``the radius of the black hole shadow''. In the approximations stated above it can be further simplified to give (cf, e.q., eq.(4) of \cite{Psaltis-2020}):
\begin{equation}
r_\mathrm{sh} \equiv b_\mathrm{cr} = \frac{r_\mathrm{ph}}{w(r_\mathrm{ph})} \, .  
\end{equation}
We emphasize again that $r_{\mathrm{sh}}$ has the dimension of a length and should not be confused with the non-dimensional \textit{angular} radius of the shadow $\alpha_\mathrm{sh}$ that can actually be measured by an observer.

Above in this Section we have calculated the shadow radius using the Lagrange approach. The same results can be obtained using the Hamiltonian approach. Calculations for this approach can be found in our paper about the shadow in a plasma \cite{Perlick-Tsupko-BK-2015}. To use the formulas from this paper for the vacuum case, one should just set the plasma frequency equal to zero, and all final formulas presented here will be recovered. This way of calculation was used in a series of works by Konoplya et al. \cite{Konoplya-2019, Konoplya-et-al-2020, Konoplya-2020}.

All formulas in this section apply to the case that the observer is static, i.e., moving on a $t$-line in the metric (\ref{eq:g}). With these results at hand, the shadow as seen by a moving observer can be easily calculated: One just has to apply, at each event on the observer's worldline, the standard aberration formula on the tangent space \cite{Grenzebach-2015,Grenzebach-2016-book}. As the aberration formula maps circles in the sky onto circles in the sky, in a spherically symmetric and static spacetime \emph{any} observer sees a circular shadow, independently of his state of motion.

Below we present some important particular cases of sha\-dow calculations in spherically symmetric and static space-times.\\

\subsection*{Example 1: Shadow in the Schwarzschild space-time for a static observer}
\vspace{2mm}

For the Schwarzschild space-time (\ref{schw-metric}),
\begin{equation}\label{eq:Ss}
A(r) = B(r)^{-1} = 1 - \dfrac{2m}{r} \, , \quad
D(r) = r^2 \, ,
\end{equation}
the function $h(r)$ and the radius of the photon sphere $r_{\mathrm{ph}}$ specify to
\begin{equation}\label{eq:Ssh}
h(r) ^2= \frac{r^2}{1 - 2m/r} \, , \quad    r_{\mathrm{ph}} = 3m \, , \quad b_{\mathrm{cr}} = 3\sqrt{3}m \, ,\end{equation}
and, after substitution into (\ref{eq:shadow}), Synge's formula (\ref{synge-shadow}) is recovered.

For large distances we have:
\begin{equation} \label{alpha-3-3-m}
\alpha_{\mathrm{sh}} \, = \, 
\dfrac{3\sqrt{3} m}{r_{\mathrm{O}}}  \, , \quad r_{\textrm{O}} \gg m \, .
\end{equation}

\subsection*{Example 2: Shadow in the Reissner-Nordstr\"{o}m  space-time for a static observer}

\vspace{2mm}

In the Reissner-Nordstr\"{o}m space-time we have:
\begin{equation}\label{eq:rn-01}
A(r) = B(r)^{-1} = 1 - \dfrac{2m}{r} + \frac{q^2}{r^2} \, , \quad
D(r) = r^2 \, ,
\end{equation}
and the function $h(r)$ takes the form
\begin{equation}
h(r) ^2 = \frac{r^2}{1 - 2m/r + q^2/r^2} \, .
\end{equation}
Eq.(\ref{eq:circ}) gives:
\begin{equation}
r_{\mathrm{ph}}  = \frac{3m}{2} + \frac{1}{2} \sqrt{9m^2 - 8q^2} \, .
\end{equation}
For the critical value of the impact parameter we obtain:
\begin{equation}
b_{\mathrm{cr}}^2  = \frac{\left(3m + \sqrt{9m^2 - 8q^2}\right)^4 }{8 \left(3m^2 - 2q^2 + m \sqrt{9m^2 - 8q^2}\right)} \, .
\end{equation}
This expression can be found in eq.(36) of Eiroa et al \cite{Eiroa-2002} and in eq.(62) of Bozza \cite{Bozza-2002}. We can further transform this into the following form
\begin{equation}
b_{\mathrm{cr}}^2  = \frac{m \sqrt{(9m^2-8q^2)^3} + 8 q^4 - 36 q^2 m^2 + 27 m^4}{2(m^2 - q^2)}  \, ,
\end{equation}
given in eq.(25) of Zakharov \cite{Zakharov-2014}. Even earlier, this expression was found in the equivalent form in the paper of Zakharov \cite{Zakharov-1994}, see eq.(26) there. The critical impact parameter with its connection with shadows was discussed in \cite{Zakharov-2014, Zakharov-2005-AA}. For further discussion see \cite{Zakharov-2012-review, Zakharov-2021}. We also refer to Alexeyev et al \cite{Alexeyev-2019} who numerically calculated the shadow for a metric that generalizes the Reissner-Nordstroem metric by adding a cubic term in $1/r$ to the metric coefficient $g_{tt}=-g_{rr}^{-1}$.
With known $b_{\mathrm{cr}}$, the angular radius $\alpha_\textrm{sh}$ of the shadow can now be found for an arbitrary position of the observer using eq.(\ref{eq:shadow-2}). For observers at large distances, eq.(\ref{eq:large-dist}) can be used.\\

\subsection*{Example 3: Shadow in the Kottler space-time for a static observer}

\vspace{2mm}

For the Kottler space-time \cite{Kottler-1918},
\begin{equation}\label{eq:Ss-2}
A(r) = B(r)^{-1} = 1 - \dfrac{2m}{r} - \frac{\Lambda}{3} r^2 \, , \quad
D(r) = r^2 \, ,
\end{equation}
we have
\begin{equation}\label{eq:Ssh-2}
h(r) ^2 = \frac{r^2}{1 - 2m/r - \Lambda r^2/3} \, ,
\end{equation}
and find
\begin{equation}
r_{\mathrm{ph}} = 3m \, , \quad b_\textrm{cr} = \frac{3\sqrt{3}}{(1 - 9 \Lambda m^2)^{1/2}} \, .
\end{equation}
This value of $b_\textrm{cr}$ can be found, e.g., in Lake and Roeder \cite{Lake-Roeder-1977} and Stuchl{\' \i}k \cite{Stuchlik-1983}. The angular radius of the shadow equals (see Stuchl{\' \i}k and Hled{\' \i}k \cite{Stuchlik-1999}):
\begin{equation} \label{eq:kottler-static}
\sin^2 \alpha_\textrm{sh} = \frac{1 - \frac{2m}{r_\textrm{O}} - \frac{\Lambda}{3} r_\textrm{O}^2 }{ \left(\frac{1}{27m^2} - \frac{\Lambda}{3} \right) r_\textrm{O}^2 } \, .
\end{equation}
Let us rewrite (\ref{eq:kottler-static}) as
\begin{equation}
\sin^2 \alpha_\textrm{sh} = \left( 1 - \frac{2m}{r_\textrm{O}} - \frac{\Lambda}{3} r_\textrm{O}^2 \right)  \frac{b_\textrm{cr}^2}{r_\textrm{O}^2} \, .
\end{equation}
We see that with $r_\textrm{O} \gg m$:
\begin{equation}
\alpha_\textrm{sh} \not\approx  \frac{b_\textrm{cr}}{r_\textrm{O}} \, .
\end{equation}
Here we have to observe that in the case $0< \Lambda < (3m)^{-2}$ there are two horizons and that the vector field $\partial _t$ is timelike only between these two horizons. Therefore, as long as we discuss the position of an observer who is static, $r_{\mathrm{O}}$ is bounded above by the radius coordinate of the outer horizon.

This calculation of the shadow in the Kottler spacetime exemplifies the important fact that, in a space-time which is not asymptotically flat, the critical impact parameter does not provide us with information about the shadow size: Dividing the critical impact parameter by the radial coordinate of the observer does not give us the angular size of the shadow, not even for an observer at a large distance, cf. \cite{Perlick-Tsupko-BK-2018}. For example, if we want to compare the size of the shadow for the Schwarzschild and the Kott\-ler case, it is not sufficient to compare the expressions for $b_\textrm{cr}$ in these two cases: we have to use formulas with the radial coordinate of the observer specified. We also note that for studying the influence of the expansion of the universe on the observable size of the shadow, we do not get the correct result by considering a static observer in the Kott\-ler spacetime; we rather have to consider an observer comoving with the cosmic expansion, see \cite{Perlick-Tsupko-BK-2018} and further discussions in Subsection \ref{sec:kottler}. \\

\section{Shadow of a  Kerr black hole}

\vspace{2mm}

\subsection{Calculation of the shadow for arbitrary position of the observer}
\vspace{2mm}

As we learned from the previous subsection, in order to construct the shadow in a spherically symmetric and static space-time it is crucial that there are unstable circular light orbits that can serve as limit curves for light rays that approach them asymptotically in a spiral motion. Because of the spherical symmetry such circular orbits necessarily form a sphere (or several spheres). The crucial question we have to answer is: What happens to these `photon spheres' if the black hole is rotating? 

To that end we now consider a Kerr black hole in Boyer-Lindquist coordinates, i.e. the metric 
\[
g_{\mu \nu} dx^{\mu} dx^{\nu}  = - \left( 1- \dfrac{2m}{r^2+a^2 \mathrm{cos}^2 \vartheta} \right)  dt^2
+ \, \dfrac{r^2+a^2 \mathrm{cos}^2 \vartheta}{r^2-2mr+a^2} \, dr ^2
+  \, d \vartheta ^2
\]
\begin{equation} \label{eq:Kerr}
+ \, \mathrm{sin}^2 \vartheta \, 
\left( r^2+a^2+\dfrac{2mra^2 \mathrm{sin}^2 \vartheta}{r^2+a^2 \mathrm{cos}^2 \vartheta} \right)
d \varphi ^2
- \, \dfrac{4mra \, \mathrm{sin}^2 \vartheta}{r^2+a^2 \mathrm{cos}^2 \vartheta} \, dt \, d \varphi
\end{equation}
with $a^2 \le m^2$. In this case the construction of the shadow is more complicated. However, it can be done fully ana\-ly\-tic\-ally because the geodesic equation is completely integrable: There are four constants of motion which, for lightlike geodesics, are $g_{\mu \nu} \dot{x}{}^{\mu} \dot{x}{}^{\nu} = 0$, the energy $E$, the $z$ component of the angular momentum $L$ and the Carter constant $K$. Whereas $E$ and $L$ are associated with the Killing vector fields $\partial _t$ and $\partial _{\varphi}$, respectively, $K$ results from the separability of the Hamilton-Jacobi equation for geodesics, see Carter \cite{Carter1968}. We will make use of these constants of motion in the following.

\begin{figure}[ht]
	\begin{center}
		\includegraphics[width=0.75\textwidth]{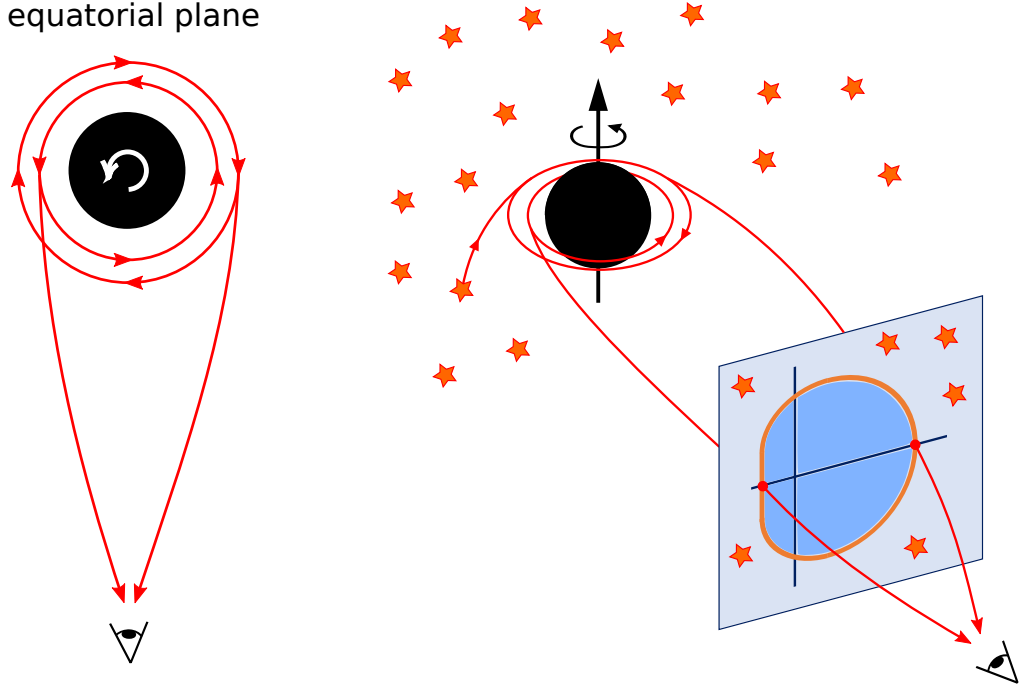}
	\end{center}
	\caption{Two unstable circular light orbits in the equatorial plane of a Kerr black hole. The one at the smaller radius is corotating, the one at the bigger radius is counterrotating. Light rays asymptotically approaching them give two points of the boundary curve of the shadow for an observer in the equatorial plane. The spherical light orbits outside the equatorial plane are not planar, see the text for more details.}
	\label{fig:kerr-orbits}
\end{figure}

To begin with, let us consider light rays in the equatorial plane (Fig. \ref{fig:kerr-orbits}). In the domain of outer communication (i.e., between the outer horizon at $r = m + \sqrt{m^2-a^2}$ and infinity) there are always exactly two circular light orbits; one is corotating with the black hole (prograde), the other one counterrotating (retrograde). Both of them are unstable with respect to radial perturbations, i.e., they can serve as limit curves for lightlike geodesics that spiral towards them. Whereas for the Schwarzschild black hole the radius coordinate of all the unstable circular light orbits equals $3m$, for a Kerr BH the value of the radius decreases from $3m$ to $m$ for the prograde orbit and it increases from $3m$ to $4m$ for the retrograde one, if the BH spin parameter $|a|$ varies from 0 to $m$. Since these two unstable circular orbits are in one plane only, light rays asymptotically approaching them give us only two points of the boundary curve of the shadow and only for an observer in the equatorial plane (Fig. \ref{fig:kerr-orbits}).

Now let us consider non-equatorial light rays. In the Kerr metric there is \textit{no photon sphere} but rather a \textit{photon region} \cite{Perlick-2004a}, \cite{Gren-Perlick-2014}, also sometimes referred to as \textit{photon shell} \cite{Johnson-2020}. Whereas a photon sphere is filled with circular light rays, the Kerr photon region is filled with \textit{spherical}, in general non-planar, light rays (see, e.g., Fig.3 in Teo \cite{Teo-2021}). Here the word ``spherical'' means that these light rays stay on a sphere $r$ = constant. For a detailed discussion of such spherical light orbits in the Kerr space-time we refer to Teo \cite{Teo-2003}, cf. \cite{Hod-2013, Igata-2019, Cunha-Thesis-2015, Teo-2021}. An online tool for their visualization has been developed by Stein \cite{stein-orbits}. Pictures of the photon region in the Kerr spacetime can be found, e.g., in \cite{HassePerlick2006, Perlick-2004a, Gren-Perlick-2014}.

We emphasize that, in view of the shadow in spacetimes that are not spherically symmetric and static, the notion of a photon region is the relevant generalization of the notion of a photon sphere. We mention here only briefly that there is another such generalization which is called a ``photon surface''. This notion was introduced and discussed by Claudel, Virbhadra and Ellis \cite{Claudel-Virbhadra-2001}. By definition, a photon surface is a nowhere spacelike hypersurface with the property that every lightlike geodesic is completely contained in this hypersurface if it is tangential to the hypersurface at one point. A timelike hypersurface is a photon surface if and only if it is \emph{totally umbilic} \cite{Perlick2005}, i.e., if and only if the second fundamental form is everywhere a multiple of the first fundamental form. In the Schwarzschild spacetime, the horizon at $2m$ is a lightlike photon surface and the photon sphere at $3m$ is a timelike photon surface. Intuitively one might think that, if the Schwarzschild spacetime is perturbed, the photon sphere at $3m$ turns into a (non-spherical) photon surface which then plays the same role for the construction of the shadow as the photon sphere in the Schwarzschild case. This, however, is not the case.  Cederbaum \cite{Cederbaum2015} proved the following uniqueness theorem: Assume that a static solution to Einstein's vacuum field equation is asymptotically flat, that it is foliated into hypersurfaces $g_{tt} = \mathrm{const.}$ and that the innermost of these hypersurfaces is a timelike photon surface; then it is the Schwarzschild solution. A similar uniqueness result was proven for solutions to the Einstein equation coupled to a Maxwell field and/or a scalar field \cite{YazadjievLazov2015,CederbaumGalloway2016,Yazadjiev2015,YazadjievLazov2016,Rogatko2016}. Yoshino \cite{Yoshino2017} proved a theorem to a similar effect: He showed that an asymptotically flat and static perturbation of the Schwarzschild metric that satisfies the vacuum field equation cannot contain a non-spherical photon surface. These results can be summarized as saying that, at least for vacuum metrics and for metrics with a Maxwell field and/or a scalar field as the source, photon surfaces are of interest only in the case of spherical symmetry.

\begin{figure}[ht]
	\begin{center}
		\includegraphics[width=0.65\textwidth]{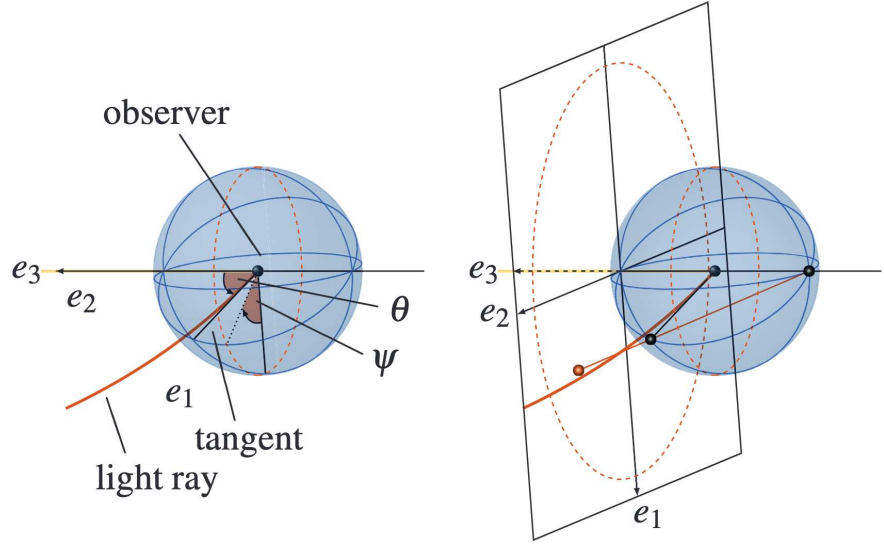}
	\end{center}
	\caption{Definition of the celestial coordinates $\psi$ and $\theta$ on the observer's sky. This picture was already used in Grenzebach, Perlick and L{\"a}mmerzahl \protect\cite{Gren-Perlick-2014}, see Fig.8 there.
	}
	\label{fig:sky}
\end{figure}

After this digression we now return to the Kerr metric.
As all spherical lightlike geodesics in the domain of outer communication turn out to be unstable, similarly to the circular lightlike geodesics in the Schwarzschild spacetime, they can serve as limit curves to which past-oriented light rays from an observer position spiral asymptotically. In this way the photon region determines the boundary curve of the shadow. For all light rays issuing from the observer position into the past the initial direction is determined by two angles in the observer's sky, a colatitude angle $\theta$ and an azimuthal angle $\psi$ which are defined with respect to the orthonormal tetrad
\[
e_0 = \dfrac{(r^2+a^2) \, \partial _t + a \, \partial _{\varphi}}{\sqrt{r^2+a^2 \mathrm{cos}^2 \vartheta}
\sqrt{r^2-2mr+a^2}} \, , \quad
e_1 = \dfrac{\partial_{\vartheta}}{\sqrt{r^2+a^2 \mathrm{cos}^2 \vartheta}} \, , 
\]
\begin{equation} \label{eq:tetrad}
e_2 = \dfrac{- \partial _{\varphi} - a \, \mathrm{sin}^2 \vartheta \, \partial _t
}{\sqrt{r^2+a^2 \mathrm{cos}^2 \vartheta} \sqrt{r^2-2mr+a^2}} \, , \quad
e_3 = \dfrac{- \sqrt{r^2-2mr+a^2} \, \partial _r}{\sqrt{r^2+a^2 \mathrm{cos}^2 \vartheta} }
\end{equation}
at the position $(r=r_{\mathrm{O}}, \vartheta = \vartheta _{\mathrm{O}})$
of the observer, see Fig. \ref{fig:sky}. This tetrad is chosen such that $e_3-e_0$ is tangent to the past-oriented ingoing principal null ray whereas $-e_3-e_0$ is tangent to the 
past-oriented outgoing principal null ray. 

If the observer position has been fixed, for every value of $\psi$ there exists exactly one spherical light orbit, at some radius coordinate $r_p$, towards which a light ray with azimuthal angle $\psi$ may spiral. This light ray has a unique colatitude angle $\theta$. The pair ($\theta$, $\psi$) gives us one point of the shadow boundary curve. If the azimuthal angle $\psi$ varies from $- \pi/2$ to $\pi /2$, the corresponding radius coordinate $r_p$ varies from its maximal value to its minimal value. This gives the lower half of the boundary curve of the shadow to which the upper half is symmetric. Note that $\theta =0$ corresponds to the ingoing principal null ray (i.e., the direction towards the black hole) and $\theta = \pi$ corresponds to the outgoing principal null ray (i.e., the direction away from the black hole.

For deriving the boundary curve of the shadow in the observer's sky, we have to determine, for each of the light rays issuing from the observer position into the past, the corresponding constants of motion. The spherical lightlike geodesic that serves as the limit curve for this light ray must have the same constants of motion which are determined by its radius coordinate $r_p$. Letting $r_p$ run over all possible values, from its minimal value to the maximal value and back, gives us an analytical expression for the boundary of the shadow as a curve parametrized by $r_p$ in the form $(\psi(r_p), \theta(r_p))$.

When working this out for an observer at the position $(r_{\textrm{O}}, \vartheta_{\textrm{O}})$, one finds the following equations which correspond to Eqs. (24)--(26) of Grenzebach, Perlick and L{\"a}mmerzahl \cite{Gren-Perlick-2014} specified to the Kerr metric:
\begin{equation} 
\sin \psi(r_p) = \frac{L_E (r_p)  - a \sin^2 \vartheta_\textrm{O}}{\sqrt{K_E(r_p)} \sin \vartheta_\textrm{O}} \, , \quad
\sin \theta(r_p) = \frac{\sqrt{r_\textrm{O}^2 + a^2 - 2mr_\textrm{O}} \sqrt{K_E(r_p)}}{r_\textrm{O}^2 - aL_E (r_p) +a^2} \, ,
\label{shadow-two-angles}
\end{equation}
where
\begin{equation} \label{shadow-two-constants}
K_E(r_p) = \frac{4r_p^2(r_p^2+a^2-2mr_p)}{(r_p-m)^2} \, , \quad
aL_E(r_p) = \frac{-r_p^2 (r_p-3m) -r_p a^2 - a^2 m}{r_p -m} 
\end{equation} 
are the constants of motion, $K_E = K/E^2$ and $L_E = L/E$, of the spherical light ray at $r_p$. Here we have used the fact that two geodesics must have the same constants of motion if one of them approaches the other asymptotically. In explicit analytical form via the parameter $r_p$ \cite{Tsupko-2017}:
\begin{equation} \label{shadow-theta}
\sin \theta(r_p) = \frac{2r_p \sqrt{r_p^2 +a^2 - 2mr_p} \sqrt{r_{\textrm{O}}^2 +a^2 -2m r_{\textrm{O}}}}{ r_{\textrm{O}}^2 r_p - r_{\textrm{O}}^2 m + r_p^3 - 3 r_p^2 m + 2 r_p a^2} \, ,
\end{equation} 
\begin{equation} \label{shadow-psi}
\sin \psi(r_p) =  - \frac{r_p^3 - 3 r_p^2 m + r_p a^2 + a^2 m + a^2 \sin^2 \vartheta_{\textrm{O}} (r_p - m)}{2 a r_p \sin \vartheta_{\textrm{O}} \sqrt{r_p^2 + a^2 - 2mr_p}} \, ,
\end{equation}
where $\theta$ and $\psi$ are the celestial coordinates of the observer, see Fig. \ref{fig:sky}. The minimal and maximal values, $r_{p, \mathrm{min}}$ and $r_{p, \mathrm{max}}$, of the parameter $r_p$ have to be found from the equations
\begin{equation} \label{min-max}
\sin \psi(r_p) = 1 \;\; \mbox{for} \;\; r_{p, \mathrm{min}} \, , \quad \mbox{and} \quad
\sin \psi(r_p) = - 1 \;\; \mbox{for} \;\; r_{p, \mathrm{max}} \, .
\end{equation}
In the Schwarzschild case ($a=0$, $r_p=3m$) eq. (\ref{shadow-theta}) gives a constant angle $\theta =\alpha _{\mathrm{sh}} $ where $\alpha _{\mathrm{sh}}$ is given by Synge's formula (\ref{synge-shadow}). Eqs. (\ref{shadow-two-angles}), which are reproduced here from Grenzebach et al. \cite{Gren-Perlick-2014}, give us the celestial coordinates of the boundary curve of the shadow explicitly in dependence of the parameter $r_p$. An implicit representation of this boundary curve was given already earlier by Perlick \cite{Perlick-2004a}. However, he restricted to observation events in the equatorial plane and he assumed a so-called zero angular momentum observer, rather than an observer adapted to the tetrad (\ref{eq:tetrad}). Even earlier Semer{\'a}k \cite{Semerak1996} had calculated the relation between constants of motion and celestial coordinates for an observer who rotates around the black hole and then determined the 'escape cone' of light (i.e., what we now call the shadow) numerically.

For plotting the shadow, we use a stereographic projection which maps the celestial sphere of the observer (except the pole at $\theta = \pi$) onto a plane that is tangent to this sphere at the pole $\theta =0$. In this plane we introduce (dimensionless) Cartesian coordinates,
\begin{equation}\label{eq:stereo}
x(r_p) = -2 \tan \left( \frac{\theta(r_p)}{2} \right)
		\sin \big( \psi(r_p) \big) \, , \quad
y(r_p) = -2 \tan \left( \frac{\theta(r_p)}{2} \right)
		\cos \big( \psi(r_p) \big) \, .
\end{equation}
In the plots the cross-hairs mark the origin of this coordinate system, i.e., the pole $\theta = 0$.

We now summarize the construction of the shadow as a step-by-step procedure.

1. Choose the mass of black hole $M$. This will give us the value of the mass parameter $m=GM/c^2$. It is convenient to use $m$ as a length unit and to express all other lengths in units of $m$. Choose the spin parameter $a$ such that $0 \le a \le m$. (Restriction to non-negative values of $a$ is no loss of generality because we are free to make a coordinate change $\varphi \to - \varphi$.)

2. Choose the position of an observer, namely the radial and angular coordinates $r_\mathrm{O}$ and $\vartheta_\mathrm{O}$. See Fig. 7 in \cite{Gren-Perlick-2014} for an illustration. 

3. In order to find $r_{p, \mathrm{min}}$ and $r_{p, \mathrm{max}}$, solve the equations (\ref{min-max}). If $\vartheta _{\mathrm{O}} = \pi /2$, for small $|a|$ it will be $r_{p, \mathrm{min}} \lesssim 3m$ and $r_{p, \mathrm{max}} \gtrsim 3m$, while for an almost extremal Kerr black hole ($|a| \lesssim m$) it will be $r_{p, \mathrm{min}} \gtrsim m$ and $r_{p, \mathrm{max}} \lesssim 4m$.

4. Determine $\sin \theta(r_p)$ and $\sin \psi(r_p)$ where the parameter $r_p$ ranges over $\,[ \, r_{p, \mathrm{min}} , r_{p, \mathrm{max}} \, ] \,$. The set of points ($\sin \theta(r_p)$, $\sin \psi(r_p)$) will give us the boundary curve of the shadow. The horizontal diameter of the shadow is equal to $\theta(r_{p, \mathrm{min}}) + \theta(r_{p, \mathrm{max}})$.

5. For plotting the boundary curve of the shadow on a flat sheet of paper, convert the angular celestial coordinates $(\theta , \psi )$ to dimensionless Cartesian coordinates $x$ and $y$ by stereographic projection, see eqs. (\ref{eq:stereo}). The upper half of the boundary curve ($\pi /2 \le \psi (r_p) \le 3 \pi /2$) is the mirror image of the lower half ($-\pi /2 \le \psi (r_p) \le \pi /2$). When going around the boundary curve in a clockwise sense, the parameter $r_p$ runs from $r_{p, \mathrm{min}}$ to $r_{p, \mathrm{max}}$ on the upper half and then back from $r_{p, \mathrm{max}}$ to $r_{p, \mathrm{min}}$ on the lower half.

\begin{figure*}[t]
\begin{center}
\includegraphics[width=0.95\textwidth]{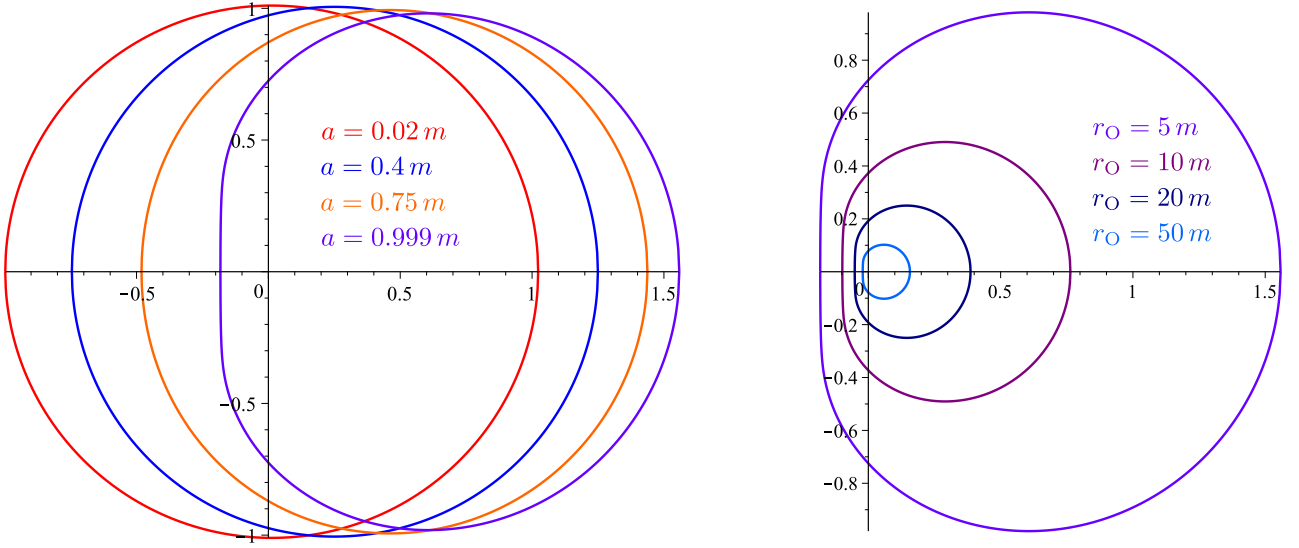}
\end{center}
\caption{(Color online) Black hole shadow as seen by an equatorial observer ($\vartheta_\textrm{O}=\pi/2$). The
curves are plotted with the help of (\ref{eq:stereo}) with (\ref{shadow-theta}) and (\ref{shadow-psi}). LEFT: Shadow as seen by an observer at $r_\textrm{O}=5m$ for different values of BH spin (from the leftmost to the rightmost): $a=0.02m$, $0.4m$, $0.75m$, $0.999m$. See also Fig.1 in Ref.~\cite{Perlick-Tsupko-2017}. RIGHT: Shadow of BH with fixed spin $a=0.999m$ as seen by observer at different positions (from the innermost to the outermost): $r_\textrm{O}=5m$, $10m$, $20m$, $50m$.}
\label{fig:shadow-close}
\end{figure*}

Examples of Kerr shadows are presented in Fig.\ref{fig:shadow-close}. \\

\subsection{Calculation of the shadow for an observer at large distance}
\vspace{2mm}

\begin{figure*}[t]
\begin{center}
\includegraphics[width=0.95\textwidth]{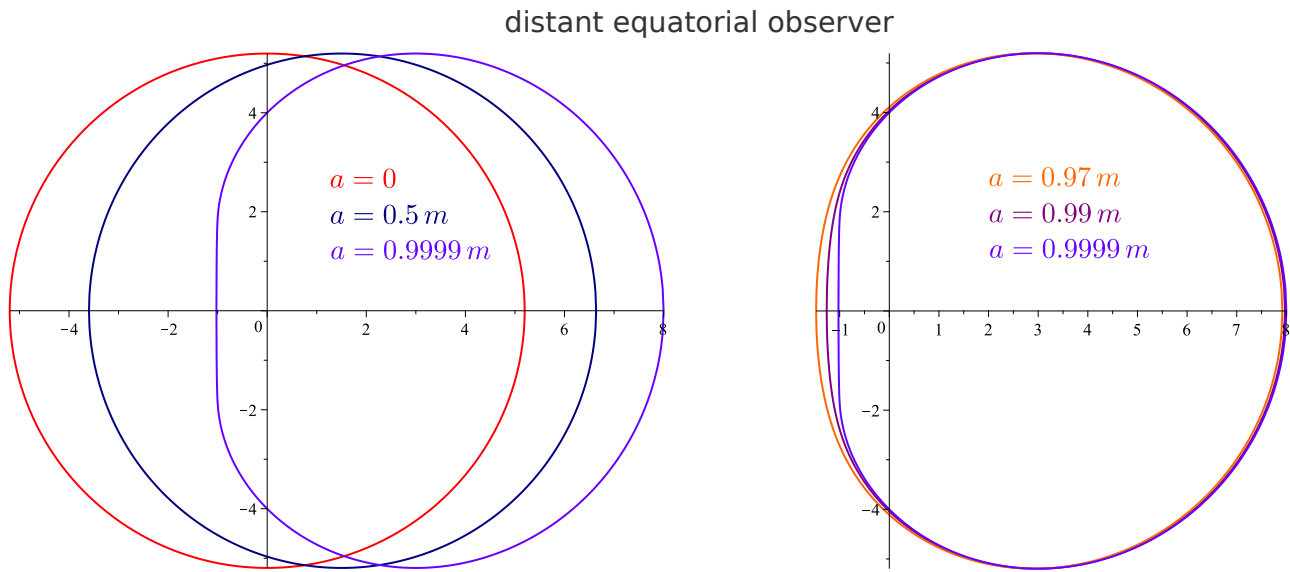}
\end{center}
\caption{(Color online) Kerr shadow as seen by a distant equatorial observer ($r_{\textrm{O}} \gg m$, $\vartheta_\textrm{O}=\pi/2$). Curves are plotted with the help of Eqs (\ref{Cartesian}) with (\ref{shadow-two-constants}). The axes are in units of $m/r_\textrm{O}$. LEFT: Kerr shadow for different values of spin (from the leftmost to the rightmost): $a=0$, $0.5m$, $0.9999m$. RIGHT: Shadow of nearly extreme Kerr BH for different values of spin (from the leftmost to the rightmost): $a=0.97m$, $0.99m$, $0.9999m$.}
\label{fig:shadow-large}
\end{figure*}

For a distant observer, $r_{\textrm{O}} \gg m$, the formulas for the shadow can be simplified. Firstly, we can solve the second equation in (\ref{shadow-two-angles}) for $\theta ( r_p)$ and then linearize with respect to $m/r_{\mathrm{O}}$,
\begin{equation} \label{theta-simple}
\theta(r_p) = \dfrac{\sqrt{K_E (r_p )}}{r_{\mathrm{O}}} \, .
\end{equation} 
This equation demonstrates that linearization with respect to $m/r_{\mathrm{O}}$ is tantamount to linearization with respect to $\theta (r_p)$, which is geometrically comprehensible. Secondly, to within the same approximation Eqs. (\ref{eq:stereo}) simplify to
\begin{equation} \label{Cartesian}
x(r_p) = 
\dfrac{a \, \mathrm{sin} ^2 \vartheta _{\mathrm{O}} -L_E (r_p )
}{r_{\mathrm{O}} \, \mathrm{sin} \, \vartheta _{\mathrm{O}}} \, , \quad
y(r_p) = \pm
\dfrac{1}{r_{\mathrm{O}}} \sqrt{ K_E (r_p)  - \dfrac{ 
\big( L_E (r_p) -a \, \mathrm{sin} ^2 \vartheta _{\mathrm{O}} \big) ^2}{\mathrm{sin} ^2 \vartheta _{\mathrm{O}}}} \, .
\end{equation}
Examples of shadow calculations based on Eqs. (\ref{Cartesian}), with (\ref{shadow-two-constants}), are presented in Fig.\ref{fig:shadow-large}.

\begin{figure*}[h]
\begin{center}
\includegraphics[width=0.95\textwidth]{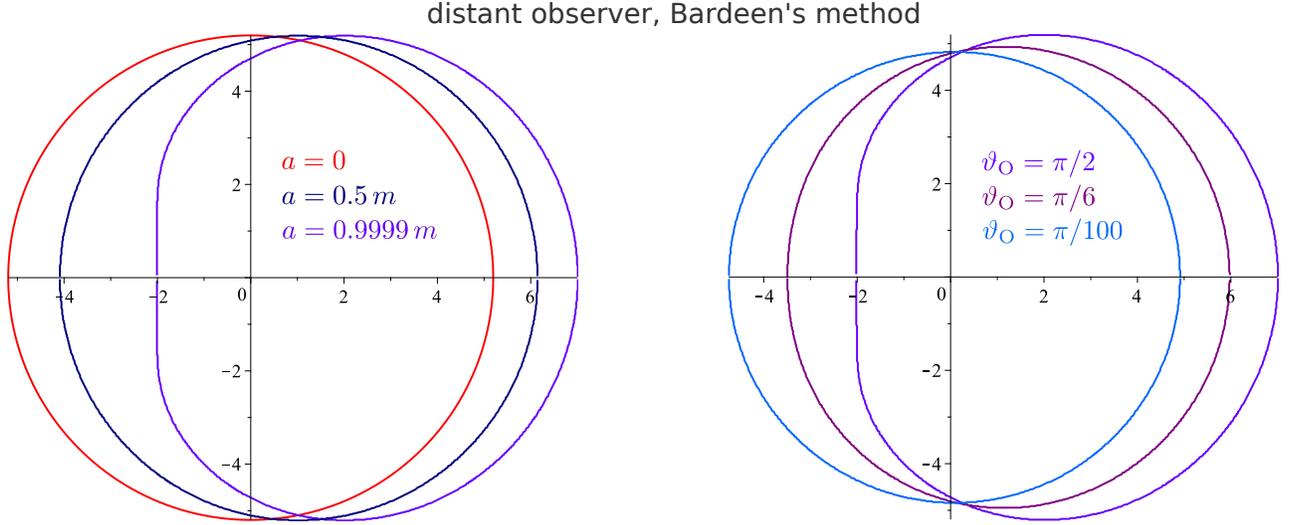}
\end{center}
\caption{(Color online) Kerr shadow as seen by a distant observer, plotted using Bardeen's approach, see Eqs (\ref{bardeen-alpha-beta}) together with (\ref{bardeen-lambda-eta}). The axes are in units of $m$. LEFT: Shadow of a Kerr black hole, as seen by an equatorial observer for different values of the spin (from the leftmost to the rightmost): $a=0$, $0.5m$, $0.9999m$. Compare with Fig.~\ref{fig:shadow-large} (left). RIGHT: Shadow of a nearly extreme Kerr BH with spin $a=0.9999m$ as seen by an observer with different inclinations (from the leftmost to the rightmost): $\vartheta_\textrm{O}= \pi/2$, $\pi/6$, $\pi/100$.}
\label{fig:bardeen}
\end{figure*}

Another approach for plotting the shadow, which is often used in the literature, is based on two impact parameters. This method, which is appropriate only for observers at large distances, was introduced by Bardeen \cite{Bardeen-1973}, see also the book by Chandrasekhar \cite{Chandra-1983}.

Bardeen represented every point on the celestial sphere by two impact parameters, $\alpha$ and $\beta$, see  \cite{Bardeen-1973, Cunningham-1972, Cunningham-1973}. They are defined as the apparent displacement of the image perpendicular to the projected axis of symmetry of the black hole and the apparent displacement parallel to the axis of symmetry in the sense of the angular momentum of the black hole \cite{Cunningham-1973}. The same variables are used in Chandrasekhar's book \cite{Chandra-1983}. These impact parameters have the dimension of a length. Most authors use units with $G = c =1$ and give all lengths in units of the black-hole mass.

As $\alpha$ and $\beta$ have the dimension of a length, they cannot be directly identified with angles in the observer's sky. Actually, when  calculating the shadow for an observer who is far away from a Kerr black hole Bardeen begins his analysis with considering angles in the sky and then introduces the impact parameters $\alpha$ and $\beta$ by multiplying these angles with the radius coordinate $r_{\mathrm{O}}$ of the observer, see the paragraph above eqs. (42) in Bardeen \cite{Bardeen-1973} and also p.277 of Frolov and Zelnikov \cite{Frolov-Zelnikov-2011}. So when using the Bardeen approach one always has to keep in mind that the impact parameters have to be divided by $r_{\mathrm{O}}$ to get the measured angles in the sky. As a good example for elucidating the methodology, we mention the articles by Bozza et al \cite{Bozza-2005, Bozza-2006, Bozza-2008, Bozza-2010} where really angular values are used, together with the distance to the BH.

We now link our shadow formula, in the linearized form (\ref{Cartesian}) for distant observers, to Bardeen's approach. We first observe that, because linearization with respect to $m/r_{\mathrm{O}}$ is tantamount to linearization with respect to $\theta (r_p)$, by (\ref{eq:stereo}) in this approximation the dimensionless Cartesian coordinates $x(r_p)$ and $y(r_p)$ are identical with the angles measured in the sky.
On the other hand, as we have already said, in Bardeen's approach the impact parameters $(\alpha , \beta)$ give angles in the sky if we divide them by $r_{\mathrm{O}}$. Moreover, as Bardeen chooses the origin of the coordinate system as given by light rays with $L=0$, rather than by the principal null rays, we have to add a shift of the origin. This gives the following transformation from the \textit{non-dimensional} variables $(x,y)$ used above to Bardeen's \textit{dimensional} variables $(\alpha,\beta)$:
\begin{equation}\label{alphabeta}
\alpha = r_{\mathrm{O}} \, x - a \, \mathrm{sin} \, \vartheta _{\mathrm{O}} \, , \quad
\beta = r_{\mathrm{O}} \, y \, .
\end{equation}
The shift of the origin is, of course, a matter of convention. Moreover, we note that Bardeen uses, instead of the Carter constant $K$, the modified Carter constant $Q = K - (L - a E)^2$. Correspondingly, he uses instead of our $L_E$ and $K_E$ the constants of motion
\begin{equation} \label{bardeen-00}
\lambda = \frac{L}{E} \, , \quad \eta = \frac{Q}{E^2}  \, ,
\end{equation}
see Eqs (41a) and (41b) in Bardeen\cite{Bardeen-1973}. Note that $\lambda$ and $\eta$ have different dimensionalities. In our notation, we have:
\begin{equation} \label{bardeen-02}
\lambda = L_E \, , \quad
\eta = K_E - \Big( L_E  - a \Big) ^2 \, .
\end{equation}
Then we find from Eqs. (\ref{Cartesian}) and (\ref{alphabeta}) 
\begin{equation} \label{bardeen-alpha-beta}
\alpha (r_p) = - \frac{\lambda (r_p) }{\sin \vartheta_{\textrm{O}}} \, , \quad
\beta (r_p ) = \pm \sqrt{\eta (r_p) + a^2 \cos^2 \vartheta_{\textrm{O}}  - 
\lambda (r_p) ^2 \cot^2 \vartheta_{\textrm{O}} } \, ,
\end{equation}
where $\lambda (r_p)$ and $\eta (r_p)$ are given, in agreement with (\ref{shadow-two-constants}), by
\begin{equation} \label{bardeen-lambda-eta}
\lambda (r_p) = \frac{-r_p^2 (r_p-3m) -a^2 (r_p+m) }{a(r_p -m)} \, , \quad
\eta(r_p) = \frac{r_p^3\Big(4a^2m-r_p(r_p-3m)^2\Big)}{a^2(r_p-m)^2} \, .
\end{equation} 
Eqs (\ref{bardeen-alpha-beta}) together with (\ref{bardeen-lambda-eta}) give us Bardeen's formulas for the shadow curve, see Eqs. (42a) and (42b) together with (48) and (49) in Bardeen \cite{Bardeen-1973}. Note, however, that the Eq.(42b) is misprinted there.

Bardeen's impact parameters $\alpha$ and $\beta$ together with the conserved quantities $\lambda$ and $\eta$ have been used by many authors. However, few authors follow the notations exactly. There is such an amazing variety of notations for the Bardeen method in the literature that we thought it necessary to present a table comparing these notations, see Table~\ref{ta-bardeen}.

\begin{table*}[t]
\caption{Comparison of notations used in literature for shadow plotting by Bardeen's method.}
\begin{ruledtabular}
{\begin{tabular}{@{}m{4.5cm}m{2.5cm}m{3.2cm}m{2.0cm}m{3.2cm}@{}}
Refs & Impact parameters \cite{Bardeen-1973}, or 'celestial coordinates' \cite{Chandra-1983} & Equations equivalent to (42a) and (42b) of Bardeen; note that some authors use (42a) with another sign & Related pair of constants of motion, see Eqs.(\ref{bardeen-00}) & Equations equivalent to (48) and (49) of Bardeen (if present) \\
\colrule
\\
Cunningham and Bardeen \cite{Cunningham-1973} & $\alpha$ and $\beta$ & (28a) and (28b)  & $\lambda$ and $q^2$ & --- \\\\

Bardeen \cite{Bardeen-1973} & $\alpha$ and $\beta$ & (42a) and (42b)  &  $\lambda$ and $\eta$ & (48) and (49) \\\\

Chandrasekhar \cite{Chandra-1983} & $\alpha$ and $\beta$ & (192), p.347 &  $\xi$ and $\eta$ & (224) and (225), p.351  \\\\

Dymnikova \cite{Dymnikova-1986} & $\rho_\perp$ and $\rho_\parallel$  & (4.1) &  $L_z/E$, $Q/E^2$ & --- \\\\

Rauch and Blandford \cite{Rauch-Blandford-1994} & $\alpha$ and $\beta$ & Appendix A & $l$ and $q^2$ & --- \\\\
 
Takahashi \cite{Takahashi-2004} & $x$ and $y$ & (8) and (9) & $p$ and $q^2$ & ---  \\\\

Bozza et al \cite{Bozza-2006} & $D_{OL} \theta_1$, $D_{OL} \theta_2$ & (13) and (14) & $J$ and $Q$ & (8) and (9) \\\\

Frolov and Zelnikov \cite{Frolov-Zelnikov-2011} & $x$ and $y$ & (8.6.13), p.277 & $\ell$ and $\mathbb{Q}$ & (8.6.20), p.279 \\\\

Johannsen \cite{Johannsen-2013} &  $x'$ and $y'$ & (26) and (27) & $\xi$ and $\eta$ & (A3) and (A4)   \\\\

Bambi \cite{Bambi-2017} & $X$ and $Y$ & (10.9), p.197 & $\lambda$ and $q^2$ & (10.13), p.198 \\\\

Cunha and Herdeiro \cite{Cunha-Herdeiro-2018} & $x$ and $y$ & see Eqs (7) & $\eta$ and $\xi$ & (4) and (5) \\\\

Gralla and Lupsasca \cite{Gralla-Lups-2020c} & $\alpha$ and $\beta$ & (44a) and (44b)  & $\lambda$ and $\eta$ & (45a) and (45b) \\\\

\end{tabular} \label{ta-bardeen}}
\end{ruledtabular}
\end{table*}

In our next Table \ref{ta-comparison}, we compare the two methods of calculating and plotting the shadow in the Kerr metric described in this Section. In Table \ref{ta3} we list the horizontal and vertical diameters of the Kerr shadow in terms of impact parameters.\\

\subsection{Analytical properties of the shadow of a Kerr black hole}
\label{sec:properties}
\vspace{2mm}

Generally speaking, the size and the shape of the shadow depend on
\begin{itemize}
    \item the parameters of the black hole; for example, the shadow becomes deformed if the black hole is rotating;
    
    \item the position of the observer; for example, the shadow is different for equatorial and polar observers;
    
    \item the properties of the regions through which light travels on its way to the observer; for example, the shadow is influenced by the expansion of the universe (see Sec.\ref{sec:expanding}) or by the presence of a plasma (see Sec.\ref{sec:plasma}).
\end{itemize}
In this Section, we will look at some of the properties of the black hole shadow curve for the Kerr metric in vacuum. A discussion of analytical properties of the Kerr black hole shadow can be found, e.g., in Bardeen \cite{Bardeen-1973}, Chandrasekhar \cite{Chandra-1983}, Dymnikova \cite{Dymnikova-1986}, Perlick \cite{Perlick-2004a}, Zakharov et al \cite{Zakharov-Paolis-2005-New-Astronomy}, Frolov and Zelnikov \cite{Frolov-Zelnikov-2011}, Grenzebach, Perlick and L{\"a}mmerzahl \cite{Gren-Perlick-2014, Gren-Perlick-2015}, Tsupko \cite{Tsupko-2017}, Cunha and Herdeiro \cite{Cunha-Herdeiro-2018}, Gralla et al. \cite{Gralla-Lups-2018}, Gralla and Lupsasca \cite{Gralla-Lups-2020c}.

First of all, the shadow of a Kerr black hole is smaller if the observer is farther from the black hole. This is illustrated in the right panel of Fig. \ref{fig:shadow-close}. Note that this is not the case in an expanding universe for a comoving observer at a large cosmological distance, see Section \ref{sec:expanding}.

\begin{table*}[t]
\caption{Comparison of two methods of calculating and plotting the Kerr shadow}
\begin{ruledtabular}
{\begin{tabular}{@{}m{8.5cm}m{8.5cm}@{}}
Bardeen \cite{Bardeen-1973} and Chandrasekhar \cite{Chandra-1983} & Grenzebach, Perlick and L{\"a}mmerzahl \cite{Gren-Perlick-2014, Gren-Perlick-2015} \\
\colrule \\
The method can only be applied to observers at a large distance from the black hole. 
& Any position of the observer outside of the black hole can be considered. \\\\
The shape of the shadow is described in terms of critical impact parameters which have the dimension of a length. If these parameters are divided by the radial coordinate of the observer, we get the angular dimensions of the shadow. & The shape of the shadow is given in terms of angular coordinates on the observer's sky which are directly measurable. For plotting they are converted into dimensionless Cartesian coordinates by way of stereographic projection. \\\\
The origin of the coordinate system is located at the point where both impact parameters are equal to zero; this corresponds to a null ray with $L = p_{\varphi} = 0$. For example, the shadow of an extreme Kerr BH as seen by an observer in the equatorial plane lies between $-2m$ and $7m$, see the left panel of Fig.\ref{fig:shadow-large}. & The origin of the coordinate system corresponds to a principal null ray which has $L=p_{\varphi} = - a \sin \vartheta_{\textrm{O}}$. For example, the shadow of an extreme Kerr BH as seen by an observer in the equatorial plane lies between $-m/r_\textrm{O}$ and $8 \, m/r_\textrm{O}$, see the left panel of Fig.\ref{fig:bardeen}.\\
\end{tabular} \label{ta-comparison}}
\end{ruledtabular}
\end{table*}

The shadow of the Kerr black hole is always symmetrical about the horizontal axis, regardless of the observer's viewing angle and the distance to the black hole. This remarkable fact, which could not be anticipated on the basis of the space-time symmetries alone, follows for an observer at large distances from Bardeen's article \cite{Bardeen-1973}, whereas Grenzebach et al. \cite{Gren-Perlick-2014} showed it for an arbitrary position of the observer. Horizontal and vertical angular diameters of the Kerr shadow for an observer at arbitrary distance in the equatorial plane of a Kerr black hole were calculated in the paper of Grenzebach et al. \cite{Gren-Perlick-2015}, see Sec.5 there. In particular, they showed that in this situation to within linear approximation with respect to $m/r_{\mathrm{O}}$ the vertical diameter of the shadow equals $2 \alpha _{\mathrm{sh}}$, with $\alpha _{\mathrm{sh}}$ given by Synge's formula (\ref{synge-shadow}).

In the following we will further discuss the properties of the shadow of a Kerr black hole for observers at large distances. For this case two methods of calculating and plotting the shadow have been discussed in the previous subsection and pictures have been presented in Figs. \ref{fig:shadow-large} and \ref{fig:bardeen}. In Bardeen's approach, the shadow is calculated as a parametric curve $(\alpha(r_p), \beta(r_p))$. For an extreme Kerr BH, it becomes possible to write the shadow curve as an explicit function $\beta(\alpha)$, which is discussed in pp.357-358 of Chandrasekhar \cite{Chandra-1983} and pp. 282-283 of Frolov and Zelnikov \cite{Frolov-Zelnikov-2011}. Taking $a=m$ in Eqs.(\ref{bardeen-lambda-eta}), one finds:
\begin{equation}
\lambda(r_p) = \frac{m^2 + 2mr_p - r_p^2}{m} \, , \quad \eta(r_p) = \frac{r_p^3 (4m - r_p)}{m^2} \, .
\end{equation}
Solving the first equation for $r_p$ in terms of $\lambda$, we obtain:
\begin{equation}
r_p = m + \sqrt{m(2m - \lambda)} \, .
\end{equation}
Substituting $r_p$ into the expression for $\eta$, we find $\eta$ as a function of $\lambda$:
\begin{equation}
\eta(\lambda) = \frac{1}{m^2} \left( m + \sqrt{m(2m - \lambda)} \right)^3 \left( 3m - \sqrt{m(2m - \lambda)} \right)   \, .
\end{equation}
Using (\ref{bardeen-alpha-beta}), we can rewrite this curve in the form of a function $\beta(\alpha)$ :
\begin{equation}
\beta^2 = \frac{1}{m^2} \left( m + \sqrt{m(2m + \alpha \sin \vartheta_{\textrm{O}})} \right)^3 \left( 3m - \sqrt{m(2m + \alpha \sin \vartheta_{\textrm{O}})} \right) + (m^2 - \alpha^2) \cos^2 \vartheta_{\textrm{O}}  \, . 
\end{equation}
This can be simplified to the following form:
\begin{equation}
\beta^2 = m^2 \cos^2 \vartheta_{\textrm{O}} + 2 \alpha m \sin \vartheta_{\textrm{O}} + 11m^2 + \, 8m \sqrt{m(2m + \alpha \sin \vartheta_{\textrm{O}})} - \alpha^2 \, .
\end{equation}

For equatorial observers ($\vartheta_{\textrm{O}} =\pi/2$) we obtain \cite{Chandra-1983, Frolov-Zelnikov-2011}
\begin{equation}
\beta^2 = \frac{1}{m^2} \left( m + \sqrt{m(2m + \alpha)} \right)^3 \left( 3m - \sqrt{m(2m + \alpha) } \right) \, ,
\end{equation}
or
\begin{equation}
\beta^2 =  2 \alpha m  + 11m^2 + 8m \sqrt{m(2m + \alpha) } - \alpha^2 \, .
\end{equation}

Interestingly, Cunha and Herdeiro \cite{Cunha-Herdeiro-2018} have recently presented a way of calculating the shadow curve of a Kerr black hole in the form of a function $\beta(\alpha)$ by solving a cubic equation. Their method works for any value of the BH spin. We will now describe it in our notation. For given $\alpha$, one can calculate $\lambda$ using the first equation in (\ref{bardeen-alpha-beta}). Then the following two quantities are introduced:
\begin{equation}
\mathcal{A} \equiv m^2 - \frac{1}{3} a (\lambda + a) \, , \quad \mathcal{B} \equiv m(m^2- a^2) |\mathcal{A}|^{-3/2} \, .
\end{equation}
They can be calculated if $\lambda$ is known. Then the expression for $r_p$ is given by \cite{Cunha-Herdeiro-2018}:
\begin{equation}
\mathcal{A} > 0, \quad \mathcal{B} \le 1 : \quad r_p = m + 2 \sqrt{\mathcal{A}} \cos \left( \frac{1}{3} \arccos \mathcal{B} \right) ,
\end{equation}
\begin{equation}
\mathcal{A} \ge 0, \; \mathcal{B} > 1 : \quad r_p = m + 2 \sqrt{\mathcal{A}} \cosh \left( \frac{1}{3} \log \left[ \sqrt{\mathcal{B}^2 - 1} + \mathcal{B} \right] \right) ,
\end{equation}
\begin{equation}
\mathcal{A} < 0 : \quad r_p = m - 2 \sqrt{|\mathcal{A}|} \sinh \left( \frac{1}{3} \log \left[ \sqrt{1 + \mathcal{B}^2} - \mathcal{B} \right] \right) .
\end{equation}
With known $r_p$, one can calculate $\eta$ with the help of the second equation in (\ref{bardeen-lambda-eta}), and then find $\beta$ using the second equation in (\ref{bardeen-alpha-beta}).

The shape of the shadow depends on the black hole spin. For a Schwarzschild black hole the shadow is circular. If the black hole is rotating, the shadow becomes deformed and flattened on one side. The larger the spin of the black hole, the greater the deformation of its shadow (Figs.\ref{fig:shadow-close}, \ref{fig:shadow-large}, \ref{fig:bardeen}). For the same viewing angle, the shadow will be maximally deformed for an extreme Kerr black hole ($a=m$). At the same time, the deformation of the shadow depends on the viewing angle of the observer. With a fixed black hole spin, the shadow will be maximally deformed for an observer in the equatorial plane. For a polar observer, the shadow always remains circular, regardless of the spin value (Fig.\ref{fig:bardeen}). For a collection of Kerr shadow curves with different spins and viewing angles we also refer, e.g., to Fig.8.15 of Frolov and Zelnikov \cite{Frolov-Zelnikov-2011}, Fig.5 of Chan et al \cite{Chan-Psaltis-2013} and Fig.1 of Takahashi \cite{Takahashi-2004}. For the more general case of a Kerr-Newman-NUT space-time see Figs. 9, 10 in \cite{Gren-Perlick-2014} and Figs. 4.5, 4.6, 4.7 in \cite{Grenzebach-2016-book}.

In the case of small spin, $a \ll m$, the deformation of the shadow occurs only in the second order of $a$. It was shown by Bozza et al \cite{Bozza-2006} and Gralla and Lupsasca \cite{Gralla-Lups-2020c} that the shadow curve remains a circle to within the first order of $a$ and becomes an ellipse in second order of $a$. The semi-axes of this ellipse are \cite{Gralla-Lups-2020c}:
\begin{equation}
\alpha_1 = 3\sqrt{3} m \left( 1 - \frac{a^2}{18m^2} \right)  \, , \quad
\beta_1 = 3\sqrt{3} m \left( 1 - \frac{a^2}{18m^2}  \cos^2 \vartheta_{\textrm{O}} \right) \, ,
\end{equation}
see also Eq.(7) of Psaltis \cite{Psaltis-2019-review}.

A nearly extreme black hole, $a=(1-\delta)m$ with $\delta \ll 1$, is discussed in Tsupko \cite{Tsupko-2017}. Here the following property occurs: in comparison with the extreme case ($a=m$), the left boundary of the shadow is shifted proportionally to $\sqrt{\delta}$, while the right boundary is shifted proportionally to $\delta$. Such a different behaviour of left and right boundaries of the shadow curve can be seen from the right panel of Fig.\ref{fig:shadow-large} where the shadow is plotted for three different spin values close to the extreme one.

Another interesting property of the shadow curve was found in a paper by Zakharov et al. \cite{Zakharov-Paolis-2005-New-Astronomy}. They found that $\Delta \beta= 6 \sqrt{3} m$ for any spin parameter $0 \le a \le 1$ and the maximum of $\beta(\alpha)$ is located in the point $\alpha=2a$. This claim was based on an earlier paper \cite{Zakharov-1986} where the author analysed the critical curve $\eta (\xi)$ which separates scattering from capturing of photons in the Kerr metric.

A peculiar feature of the shadow of an extreme Kerr black hole is the vertical line in the left part of the shadow \cite{Bardeen-1973, Chandra-1983, Dymnikova-1986}. Now it is sometimes referred to as the 'NHEK line', where NHEK stands for Near Horizon Extreme Kerr, after the work of Gralla et al \cite{Gralla-Lups-2018} where the observational signatures of a high black-hole spin are discussed in detail.

Since the deformation of the shadow depends on the spin of the black hole, measuring the parameters of this deformation may allow one to determine the spin of the black hole. This idea is discussed with different analytical and numerical techniques, e.g., in \cite{Takahashi-2004, Zakharov-Paolis-2005-New-Astronomy, Bozza-2006, Hioki-Maeda-2009, Johannsen-2013, Li-Bambi-2014, Tsukamoto-Li-Bambi-2014, Abdu-Rezzolla-Ahmedov-2015, Gren-Perlick-2015, Yang-Li-2016, Tsupko-2017, Wei-2019-Rapid, Dokuchaev-2020, Kumar-Ghosh-2020}.

Different approximating curves which can describe the Kerr black hole shadow curve have been proposed, see the papers of Cunha and Herdeiro \cite{Cunha-Herdeiro-2018}, Farah et al \cite{Farah-2020} and Gralla and Lupsasca \cite{Gralla-Lups-2020c}.

\begin{table*}[t]
\caption{Diameters of Kerr black hole shadow for observer at large distances}
\begin{ruledtabular}
{\begin{tabular}{@{}m{2.0cm}m{2.5cm}m{3.0cm}m{3.5cm}m{4.0cm}@{}}
Refs & spin $a$  & viewing angle $\vartheta_{\textrm{O}}$  & diameter $\Delta \alpha$ & diameter $\Delta \beta$ \\
\colrule \\
Ref. \cite{Hilbert-1917} & $0$ & any & $6\sqrt{3} \, m$ & $6\sqrt{3} \, m$  \\

Refs. \cite{Bardeen-1973, Chandra-1983} & $m$ & $\pi/2$ (equat. view) & $9 \, m$ & ---  \\

Ref. \cite{Dymnikova-1986} & $m$ & $\pi/2$ (equat. view) & $9 \, m$ & $6\sqrt{3} \, m$  \\

Ref. \cite{Zakharov-Paolis-2005-New-Astronomy} & any & $\pi/2$ (equat. view) & --- & $6\sqrt{3} \, m$  \\

Refs.\cite{Dymnikova-1986, Frolov-Zelnikov-2011} & $m$ & $0$ (polar view)& $4(1+\sqrt{2}) \, m$ & $4(1+\sqrt{2}) \, m$ \\\\

Ref.\cite{Gralla-Lups-2020c} & $a \ll m$ & any & $6\sqrt{3} \left( 1 - \frac{a^2}{18m^2} \right) m $  & $ 6\sqrt{3} \left( 1 - \frac{a^2}{18m^2}  \cos^2 \vartheta_{\textrm{O}} \right) m $   \\\\

Ref.\cite{Tsupko-2017} & $(1-\delta) \, m$, $\delta \ll 1$ & $i = \pi/2 - \vartheta_{\textrm{O}} \ll 1$ & $\left(9 + \sqrt{6} \sqrt{\delta} + o(i^2) \right) m $ & $ \left(6\sqrt{3} -\sqrt{3} i^2/3 + o(\sqrt{\delta}) \right) m$  \\

\end{tabular} \label{ta3}}
\end{ruledtabular}
\end{table*}

Note that the Kerr shadow is close to circular. The maximal deviation from a circular shape occurs in the extreme case $a^2=m^2$ for an observer in the equatorial plane. However, even in this case the deformation (vertical shadow diameter -- horizontal shadow diameter)/(vertical shadow diameter + horizontal shadow diameter) is only $(6 \sqrt{3}-9)/(6 \sqrt{3}+9) \approx 0.07$, 
if we neglect terms of order $m/r_O$, see Table 3. This number is a relevant guide to estimate the accuracy that one should demand for analytical studies using spherically symmetric geometries.

We end this section on the shadow in the Kerr space-time with a short remark on naked singularities. What we have said until now referred to a Kerr black hole, i.e., to a Kerr metric with $a^2 \le m^2$. In the case $a^2>m^2$ there is no longer a horizon and the ring singularity at $(r=0, \vartheta = \pi /2)$ is naked. In this case, light rays that have already entered the central region or even the region of negative $r$ values may come back to the region of big positive $r$ values. The shadow is then completely different from a black-hole shadow; in particular, it is no longer a (two-dimensional) black disk on the observer's sky but rather a (one-dimensional) black arc, see de Vries \cite{Vries-2000}. The existence of naked singularities in Nature is, of course, highly speculative. \\

\section{Generalization of the results for the Kerr space-time to other rotating black holes}
\label{sec:generalize}

For analytically calculating the boundary curve of the shadow in the Kerr metric it was crucial that the equation for lightlike geodesics was completely integrable, i.e., that it admitted, in addition to the constants of motion $\mathcal{L}$, $E$ and $L$ the Carter constant. Therefore, by introducing a tetrad in analogy to (\ref{eq:tetrad}), we can determine the boundary curve of the shadow for any observer in the domain of outer communication of a black hole provided that the space-time is stationary and axi\-symmetric and that the equation for lightlike geodesics is separable so that there exists a (generalized) Carter constant. This was carried through by Grenzebach et al. \cite{Gren-Perlick-2014, Gren-Perlick-2015} for all space-times of the Pleba{\'n}ski-Demia{\'n}ski class. This class of space-times describes black holes with a mass $m$, a spin $a$, electric and magnetic charges $q_e$ and $q_m$, a NUT parameter $\ell$, a cosmological constant $\Lambda$ and an acceleration parameter $\alpha$. A Carter constant also exists for some other black-hole metrics: Konoplya, Rezzolla and Zhidenko \cite{KonoplyaRezzollaZhidenko2016} had introduced a certain class of parametrized metrics for which the shadow was discussed by Younsi et al. \cite{YounsiEtAl2016}. Some metrics in this class satisfy the separability condition, see Konoplya et al. \cite{KonoplyaStuchlikZhidenko2018}, so that an analytical treatment of the shadow is possible, see Konoplya and Zhidenko \cite{KonoplyaZhidenko2021}. Tsukamoto \cite{Tsukamoto-2018} considered the class of metrics that results if one replaces in the Kerr metric the mass parameter $m$ by a function $m(r)$; as the separability condition is still satisfied, he could calculate the shadow for an observer at a large distance with Bardeen's method. Glampedakis and Pappas \cite{GlampedakisPappas2019} found an interesting connection between the existence of spherical lightlike geodesics and the separability condition: If an axisymmetric and static metric admits no non-equatorial spherical lightlike geodesics, in no coordinate system, then the separability condition cannot be satisfied.
As to the more special question of
whether \emph{circular} lightlike geodesics exist, Cunha
and Herdeiro \cite{Cunha-Herdeiro-PRL-2020} have
shown that this is always the case in the domain of outer
communication of a black hole that is axisymmetric,
stationary and asymptotically flat with a horizon that
is topologically spherical and non-degenerate.

As particular cases, the Pleba{\'n}ski-Demia{\'n}ski class covers Schwarzschild, Reissner-Nordstr\"{o}m, Kerr, Kerr-Newman, Taub-NUT, Kerr-NUT, Kerr-Newman-NUT, Kottler (Schwarzschild-(anti-)de Sitter) and other metrics, see, e.g., Table 2.1 in Grenzebach \cite{Grenzebach-2016-book}. Therefore, for calculating the shadow in these space-times one just has to set some of the parameters in the formulas of Refs. \cite{Gren-Perlick-2014, Gren-Perlick-2015} equal to zero. We emphasize that the original formulas are based on the assumption that the observer is in a particular state of motion, given by the vector $e_0$ of the chosen tetrad. For observers in a different state of motion the aberration formula has to be applied, see Grenzebach
\cite{Grenzebach-2015,Grenzebach-2016-book}. Moreover, we also emphasize that metrics of the 
Pleba{\'n}ski-Demia{\'n}ski class with $\Lambda$, $\ell$ or $\alpha$ different from zero are \emph{not} asymptotically flat. 

Having an analytical formula for the shadow in Ple\-ba{\'n}ski-Demia{\'n}ski metrics, depending on the various parameters of these metrics and on the position of the observer, it is an interesting question to investigate if these parameters and the position of the observer can be reproduced from the boundary curve of the shadow. Important partial results in this direction have been found by Mars et al. \cite{Mars-2018}.   

Also for Bardeen's method, which we reviewed in the preceding section, it was crucial that the Kerr metric admits the Carter constant. In addition, this method also made use of the fact that the Kerr metric is asymptotically flat. We will now discuss the generalization of Bardeen's method to asymptotically non-flat metrics. We start with a reminder that in this approach, applied for Kerr, it was sufficient to divide the impact parameters by the radial coordinate of the observer to obtain the angles in the observer's sky. Such a simple relation between impact parameters and angles holds only if the space-time is asymptotically flat and if the observer is far away from the center, as was assumed by Bardeen. The crucial fact is that, if the space-time is not asymptotically flat, the impact parameters are not related in such a simple way to angles measured by an observer in the sky, not even for observers far away from the black hole. Moreover, Bardeen's \cite{Bardeen-1973} eqs. (42), which relate the impact parameters $(\alpha , \beta )$ to the constants of motion $(\lambda,\eta)$, do not carry over to space-times that are not asymptotically flat. Therefore, these formulas should not be mechanically applied in space-times that are not asymptotically flat.

We emphasize that in space-times that \emph{are} asymptotically flat the methods of Grenzebach et al. \cite{Gren-Perlick-2014, Gren-Perlick-2015} and of Bardeen \cite{Bardeen-1973} are completely equivalent for observers that are far away from the center; this is of course the situation which is most relevant in view of applications to astronomical observations. The asymptotically flat metrics in the Pleba{\'n}ski-Demia{\'n}ski class are the Kerr-Newman metrics (i.e., the metrics with mass $m$, spin $a$ and charges $q_e$ and $q_m$). Then the two methods give the same shape and, if impact parameters are properly converted into angles, also the same size of the shadow. The only difference is that the origins of the coordinate systems are different, so the shadow plotted by one method is shifted in the horizontal direction in comparison to the other method, see Eq.(\ref{alphabeta}). Also compare the left panel of Fig.\ref{fig:shadow-large} and the left panel of Fig.\ref{fig:bardeen}: We see that the shapes of the curves are the same but the origin is horizontally shifted; this is especially noticeable for the extreme Kerr case.

There are a few other black-hole models, in addition to the Pleba{\'n}ski-Demia{\'n}ski ones, that are described by stationary and axisymmetric metrics with a Carter constant. This includes, e.g., the Kerr-Sen black hole which was derived from heterotic string theory by Sen \cite{Sen-1992} in 1992. In this metric the shadow can be calculated with the same methods as in the Kerr metric; for a comparison of the two cases we refer to Xavier et al. \cite{XavierEtAl-2020}. Other rotating black-hole  metrics for which the Hamilton-Jacobi equation for lightlike geodesics separate and for which the shadow has been calculated are a black hole surrounded by dark matter, see Haroon et al.  \cite{Haroon-2019}, a class of regular black holes, see Neves \cite{Neves-2020a}, and several braneworld black holes, see Amarilla and Eiroa, \cite{AmarillaEiroa2012}, Eiroa and Sendra \cite{Eiroa-2018} and Neves \cite{Neves-2020b}.

We also mention that many authors have produced rotating black-hole metrics by applying the standard Newman-Janis formalism \cite{NewmanJanis-1965} to a spherically symmetric and static black hole. This mechanism produces the Kerr metric from the Schwarzschild metric. In general, however, the rotating metric that is produced in this way does \emph{not} admit a Carter constant, so one cannot analytically calculate its shadow. Therefore, Azreg-A{\"i}nou \cite{Azreg-2014} suggested a modified Newman-Janis formalism which always produces a metric that admits a Carter constant. The relevance for the shadow of this approach was detailed by Lima Junior et al. \cite{LimaEtAl-2020}. Note, however, that neither the original nor the modified Newman-Janis method necessarily preserves the form of the stress-energy tensor when inserted into Einstein's field equation. So if one starts, e.g., with a perfect-fluid solution, one has to check by hand if the resulting rotating metric is again a perfect-fluid solution.

Special care is necessary if in the domain of outer communication there are spherical photon orbits some of which are \emph{stable} with respect to radial perturbations. Such stable photon orbits cannot serve as limit curves for light rays, but light rays may oscillate about them. In such cases one has to specify very carefully where the light sources are situated in order to determine if a certain part of the observer's sky is in the (dark) interior or in the (bright) exterior of the shadow. E.g., if a light ray oscillates about a stable photon orbit, then its initial direction would have to be associated with ``darkness'' if there are no light sources along its path and with ``brightness'' otherwise.

A fairly large number of black holes have been investigated where, at least in part of the space-time around the black hole, the geometry is influenced by self-gravitating matter. For all these cases the shadow can be calculated only numerically, so we mention them here only in passing. There are the so-called ``distorted'' or ``bumpy'' black holes, which are static (i.e., non-rotating) and axisymmetric but surrounded by (unspecified) gravitating matter. One can characterize the vacuum parts of such space-times in terms of their Weyl multipole moments. For the case that, in addition to the monopole term, only the quadrupole is different from zero, the shadow was investigated by Abdolrahimi et al. \cite{AbdolrahimiEtAl2015}, assuming that the observer is in the vacuum region between the horizon and the surrounding matter. The shadows of such distorted black holes have also been investigated by Grover et al. \cite{Grover-2018}. Another class of  black holes distorted by matter are the so-called black holes with ``scalar hair''. These are solutions to the Einstein equation coupled to a Klein-Gordon field. In such space-times, the shadow can show features that are qualitatively quite different from the Kerr shadow, as was demonstrated by Cunha et al. \cite{CunhaEtAl2015} and Vincent et al. \cite{VincentEtAl2016a}.

Finally we mention the possibility of considering the shadow of binary and multiple black holes. The physically most interesting situation refers to the case that two black holes orbit around each other in a spiral motion and then merge. This dynamical problem, however, can be investigated only numerically, see Bohn et al. \cite{BohnEtAl2015}. In this paper the authors visualized what an observer sees by introducing a (big) sphere with four parts of different colour and a coordinate grid, see Fig.3 in \cite{BohnEtAl2015}. An observer inside this sphere and at a certain distance from the center sees the shadow surrounded by  distorted multiple images of this sphere; in the case of two merging black holes, unsurprisingly, the observer would see a time-dependent double-shadow, see Fig.6 in \cite{BohnEtAl2015}. The same method of visualization can be used for any shadow calculations, numerical or analytical.  Partial analytical results for multiple shadows are possible for the (idealized) cases of head-on colliding black holes, see Nitta et al. \cite{NittaEtAl2011}, and for multiple black holes that are held in equilibrium by electrostatic repulsion that balances the gravitational attraction, see \cite{YumotoEtAl2012}. Moreover, the inspiral of two Schwarzschild black holes can be considered in an ``adiabatic approximation'', see Cunha et al. \cite{CunhaEtAl2018}. Interestingly, Shipley and Dolan \cite{Shipley-Dolan-2016} observed that one-dimensional slices through the shadow of binary black holes could be regular, self-similar like the Cantor set or chaotic. Self-similarities had also been found by Bohn et al. \cite{BohnEtAl2015}. When considering the shadow of black holes one usually thinks of supermassive black holes. This is true also for black holes in a binary system. However, Gott et al. \cite{Yunes-2019} suggested that also in the case of a \emph{stellar} black hole with a shining companion the resulting change of the magnitude could become observable in the foreseeable future. \\

\section{Shadows of wormholes and other compact objects that are not black holes}
\label{sec:wormholes}

We have seen that in spherically symmetric and static space-times it is the photon sphere, not the horizon, that determines the shadow. Therefore, also some objects that are not black holes may cast a shadow that is very similar to, or even identical with, that of a black hole. As such objects could be easily mistaken for black holes, we call them \emph{black-hole impostors}. Some other authors prefer the term \emph{black-hole mimickers}. For general aspects of how to distinguish black hole impostors from black holes we refer to Abramowicz et al. \cite{AbramowiczEtAl2002}.

If we consider the spherically symmetric and static case, then the best black hole impostor is an \emph{ultracompact star}, i.e., a star with a radius between $2m$ and $3m$. If such a star has a dark surface, it will cast a shadow that is literally identical to the shadow of a Schwarzschild black hole with the same mass; this is obvious from the way in which the shadow is mathematically constructed. However, it should be mentioned that the existence of ultracompact stars is questionable. It was shown by Cardoso et al. \cite{CardosoEtAl2014} that they have a strong tendency to be unstable. So if we see a shadow, it is much more likely that it comes from a black hole rather than from an ultracompact star.

\begin{figure}[t]
	\begin{center}
		\includegraphics[width=0.75\textwidth]{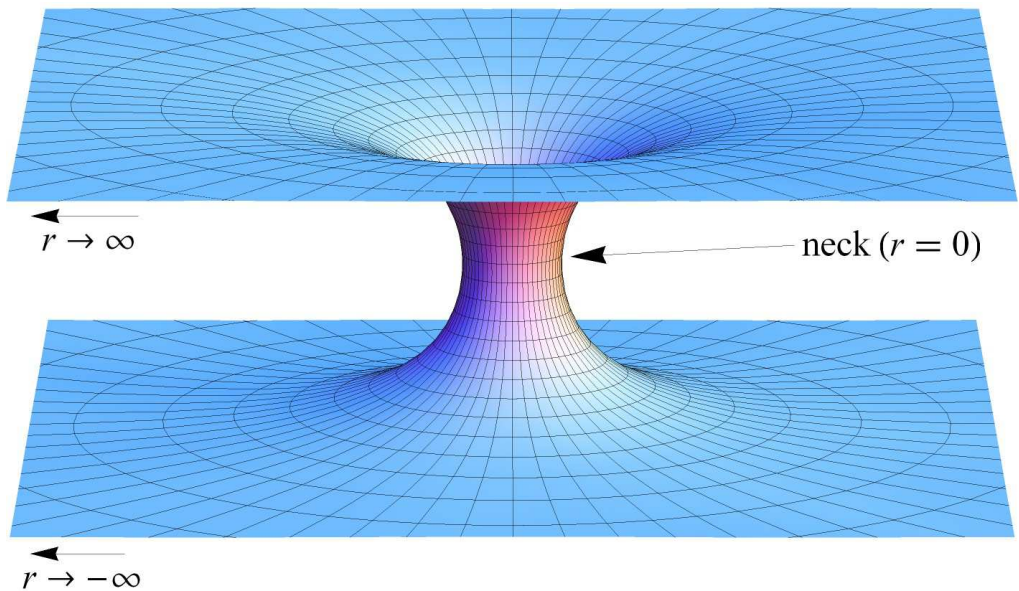}
	\end{center}
	\caption{Spatial section of the equatorial plane ($t = \mathrm{constant}$, $\vartheta = \pi /2$) of an Ellis wormhole}
	\label{fig:wormhole}
\end{figure}

Other good black hole impostors are wormholes, i.e., 
space-time models where two asymptotically flat ends are connected by a ``throat'' or ``neck''; if the two ends are glued together one gets a space-time with a shortcut between two regions which would be very far apart from each other without this connection. Historically, the first wormhole space-time that was found was the \emph{Einstein-Rosen bridge} \cite{EinsteinRosen1935} which connects two exterior Schwarzschild regions. The Einstein-Rosen bridge is a ``non-traversable wormhole'', i.e., it is impossible to travel from one asymptotic end to the other through the throat at subluminal speed. The interesting wormholes are the ``traversable'' ones because they allow for a certain kind of time travel. The best known traversable wormhole is the Ellis wormhole with the metric
\begin{equation}\label{eq:Ellisg}
g_{\mu \nu} dx^{\mu} dx^{\nu} = -c^2 dt^2 +dr^2 + 
(r^2 +a^2) \big( 
d \vartheta ^2 + \mathrm{sin} ^2 \vartheta \, d \varphi ^2 
\big)
\end{equation}
where $a$ is a constant. For $a =0$ this is just the 
Minkowski metric in spherical polars, with a regular center
at $r=0$. For $a \neq 0$, however, $r=0$ is a sphere of
finite radius; the $r$ coordinate runs from one asymptotic end at $- \infty$ to another asymptotic end at $+ \infty$, passing through a neck at $r =0$, see Figure \ref{fig:wormhole}.

The Ellis wormhole was discovered by H. Ellis \cite{Ellis1973} as the simplest member of a larger class of wormholes (or ``drainholes'', as Ellis called them). If viewed as a solution of Einstein's field equation, the Ellis wormhole is associated with a matter source of negative energy density. Later Morris and Thorne \cite{MorrisThorne1988} discussed the entire class of spherically symmetric and static wormholes that are traversable and they showed that, on the basis of Einstein's field equation,  all of them must have a negative energy density near the throat, i.e., that they require the existence of exotic matter which is usually considered as questionable.  However, this is no longer true if one allows for gravity theories more general than Einstein's. Then it is possible to construct traversable wormholes without the need of exotic matter, see e. g. Kanti et al \cite{Kanti-2011}, Bronnikov et al \cite{Bronnikov-2017} and Antoniou et al \cite{Antoniou-2020}. In any case, although somewhat speculative, wormholes have attracted great interest because of their intriguing perspectives of time travel.

The lensing features of wormholes and, in particular, their shadows are very similar to that of black holes. We exemplify this with the Ellis wormhole whose lensing features are similar to those of a a Schwarzschild black hole: Both metrics have an asymptotic end (at $r = \infty$), both have an unstable photon sphere (at $r=0$ in the wormhole case and at $r=3m$ in the black-hole case) and another end that can be approached only asymptotically by past-oriented light rays (at $r = - \infty$ in the wormhole case and at $r=2m$ in the black-hole case). Correspondingly, in both cases there is a circular shadow whose boundary is determined by light rays that asymptotically spiral towards the unstable photon sphere. In the case of the Ellis wormhole the angular radius of this shadow is given by inserting the metric coefficients of the Ellis space-time into (\ref{eq:sinalpha}), which results in
\begin{equation}\label{eq:Ellisalpha}
\sin^2 \alpha_\textrm{sh} \, = \, \dfrac{a^2}{r_\textrm{O}^2+a^2}
\end{equation}
where $r_\mathrm{O}$ is the radius coordinate of the observer. Here it is assumed that $r_\mathrm{O} > 0$ and that there are no light sources between the observer and the asymptotic end at $r= - \infty$. The lensing features of the Ellis wormhole have been discussed by Chetouani and Cl{\'e}ment \cite{ChetouaniClement1984}, Perlick \cite{Perlick-2004a,Perlick-2004b}, Nandi et al. \cite{NandiZhangZakharov2006}, M{\"u}ller \cite{Mueller2008} and Nakajima and Asada \cite{NakajimaAsada2012}. The latter paper corrects some errors that have occurred in other references. An Ellis wormhole surrounded by (not self-gravitating) matter, taking the radiative transfer in this matter into account, was discussed by Ohgami and Sakai \cite{Ohgami-2015}.

In the Ellis wormhole there is only one photon sphere, which is unstable and situated at the throat. More generally, one can construct spherically symmetric and static wormholes with several photon spheres, unstable ones and stable ones, and they need not be located symmetrically with respect to the wormhole throat. The shadow of such wormholes was discussed by Wielgus et al. \cite{WielgusEtAl2020}, also see Tsukamoto \cite{Tsukamoto-2021}. Note that, in principle, also black-hole spacetimes may have arbitrarily many photon spheres, as long as we do not care about energy conditions or other properties of the matter content. Therefore, our statement that wormhole lensing is qualitatively similar to black-hole lensing remains true also for wormholes with more than one photon sphere.

The spherically symmetric and static traversable wormholes of Morris and Thorne have been generalized to axisymmetric and stationary (i.e., rotating) traversable wormholes by Teo \cite{Teo1998}. For a subclass of these wormholes, the geodesic equation is completely integrable, see Nedkova et al. \cite{NedkovaEtAl2013}, so the shadow can be calculated analytically. As pointed out by Shaikh \cite{Shaikh2018}, for the construction of the shadow it is important to take \emph{all} existing spherical photon orbits into account. This question was further discussed by Gyulchev et al. \cite{GyulchevEtAl2018}. 

An important difference between the shadow of a black hole and that of a black-hole impostor is in the fact that, for a non-rotating isolated object, the shadow is circular in the first case whereas it may be deformed in the latter. In particular, a non-rotating and isolated black-hole impostor may have a non-zero quadrupole moment which would lead to a non-circular shadow. Of the many static and asymptotically flat vacuum solutions of Einstein's field equation with a quadrupole moment, the gamma metric, also known as the Zipoy-Vorhees metric or the q-metric, is the simplest one. If maximally extended it features a naked singularity, but if the inner part is replaced with some kind of fluid solution this metric makes a very good black-hole impostor. Of course, one has to assume that the fluid star is non-transparent and dark. The photon region in axisymmetric and static vacuum space-times, and in particular in  the gamma metric, was discussed by Gal'tsov and Kobialko \cite{Galtsov-2019}.

It was argued by Vincent et al \cite{VincentEtAl2016b} that also boson stars are quite good black-hole impostors, or black-hole mimickers. To that end they considered a boson star that is illuminated by an accretion disk. Strictly speaking, in this case there is no shadow because the boson star is assumed to be transparent. However, since the luminosity is very low in the central part, the visual appearance is actually quite similar to that of a black hole, with an (almost completely) dark disk in the middle. As boson stars are solutions to the Einstein-Klein-Gordon equations, which even in the case of spherical symmetry can be found only numerically, this ``pseudo-shadow'' cannot be calculated analytically. Equally good black-hole impostors are Proca stars, where the Klein-Gordon (scalar) equation is replaced with the Proca (vector) equation, see Herdeiro et al. \cite{HerdeiroEtAl2021}. Similarly, a Fermion star can be considered as an alternative to a black hole, see e.g. Gomez et al. \cite{GomezEtAl2016}. Also in these cases one has to resort to numerical studies. It should be emphasized that from the observational side there is no evidence that boson stars, Proca stars or Fermion stars actually exist in Nature. On the other hand, it is also true that at present their existence cannot be ruled out. Therefore, we believe that the continuation of theoretical studies of these objects is of great importance. \\

\section{The shadow of a collapsing star}
\label{sec:collapsing}

Up to now we have considered only ``eternal black holes'', i.e., black holes that exist for all time in a space-time that is stationary. In such a space-time any stationary observer sees a time-independent shadow. However, we believe that the (stellar or supermassive) black holes we actually observe in Nature have come into existence by gravitational collapse. Clearly, in such a situation the shadow would not exist forever; it would rather gradually form in the course of time, even for an observer who is at a constant distance from the center of the black hole. 

There are numerous papers, beginning with the pioneering work of Ames and Thorne \cite{AmesThorne1968}, where the visual appearance of a collapsing \emph{non-transparent} star is discussed, see e.g., Jaffe \cite{Jaffe1969}, Lake and Roeder \cite{LakeRoeder1979} and Frolov et al. \cite{FrolovKimLee2007}. In all these works the emphasis is on the frequency shift of light coming from the surface of the collapsing star. More recent papers by Kong et al.
\cite{KongMalafarinaBambi2014,KongMalafarinaBambi2015} and
by Ortiz et al. \cite{OrtizSarbachZannias2015a,OrtizSarbachZannias2015b} considered the frequency shift of light passing through a collapsing \emph{transparent} star, thereby contrasting the collapse to a black hole with the collapse to a naked singularity. However, none of these papers actually discussed how the shadow was formed in the course of time and how its angular radius, as seen by an observer at some distance from the collapsing star, would change in the course of time. The latter question was investigated by Schneider and Perlick \cite{SchneiderPerlick2018}. They considered the simplest possible model, namely a non-transparent spherically symmetric collapsing ball of dust whose surface doesn't emit any light. In this case the space-time outside the ball is just given by the Schwarzschild metric, and the temporal behaviour of the star's surface was calculated in a classical paper by Oppenheimer and Snyder \cite{OppenheimerSnyder1939}. Based on these facts, Schneider and Perlick found that the shadow was formed in a finite time, i.e., that after a finite time an observer at any fixed position farther away from the center than the Schwarzschild radius would see a circular shadow with an angular radius given by Synge's formula. This result couldn't have been easily anticipated without a calculation; actually, based on intuition one might have guessed that the Synge formula would be approached only asymptotically. \\

\section{Shadow in an expanding universe}
\label{sec:expanding}

Since we live in an expanding universe, cosmological expansion is expected to affect the observed size of the black hole shadow. Such effects become significant at large cosmological distances. In particular, it is known that at high redshifts
the observed angular size of any object can become larger due to cosmological expansion \cite{Mattig-1958, Zeldovich-1964, Dashevsk-Zeldovich-1965, Zeldovich-Novikov-book-2, Hobson, Mukhanov-book}. We can expect a similar behavior for the size of the shadow of a black hole.

Actually, calculating the angular size of the shadow of a black hole in an expanding universe is rather nontrivial. This is primarily due to the fact that, when calculating the trajectories of light rays, we must simultaneously take into account both the cosmological expansion and the influence of the black hole at all points of the trajectory. Quantitatively, the latter is relevant in the regime of strong gravity, since near the black hole the influence of its gravitational field is very significant. This complicates analytical calculations, and an exact analytical solution (valid for arbitrary observer positions, arbitrary black hole parameters and arbitrary cosmological models) has not yet been found. However, some progress has been achieved very recently in this field.

For calculating the BH shadow in an expanding universe we have to consider light rays that are influenced simultaneously by the gravitational field of the black hole and by the cosmic expansion. Because we want to do this calculation exactly, we restrict to the case that the black hole is spherically symmetric (i.e., non-rotating) and the ambient cosmological space-time is given by a Friedmann-Lema{\^\i}tre-Robertson-Walker (FLRW) metric. The best known models of this kind are the one by Einstein and Straus \cite{Einstein-Straus-1945, Einstein-Straus-1946} which was further discussed in \cite{Schucking-1954, Stuchlik-1984, Balbinot-1988, Schucker-2009, Schucker-2010} and the one by McVittie \cite{McVittie1933} which was further discussed in \cite{Nolan-1, Nolan-2, Nolan-3, Kaloper-2010, Carrera-Giulini-2010a, Carrera-Giulini-2010-review, Anderson-2011, Lake-Abdelqader-2011, Hobson-2012a, Hobson-2012b, Silva-2013, Nolan-2014, Piattella-PRD-2016, Piattella-Universe-2016, Nolan-2017, Faraoni-2017, Aghili-2017}.
Following the general strategy of this article, it is our aim to calculate the shadow analytically which requires to calculate the trajectories of light rays analytically. The crucial point for the analytical construction of the shadow is the existence of sufficiently many constants of motion that allow to reduce the differential equations for lightlike geodesics, which are originally of second order, to a system of \emph{first-order} equations. The Einstein-Straus model consists of a Schwarzschild space-time embedded as a `vacuole' into an FLRW space-time. Geodesics in this model have to be calculated piecewise, matching the Schwarzschild part to the FLRW part at the so-called Sch\"{u}cking radius \cite{Schucking-1954, Schucker-2009, Schucker-2010}. So here one has sufficiently many constants of motion for integrating the lightlike geodesic equation, both in the Schwarzschild and in the FLRW part, but the matching procedure can be rather awkward. By contrast, in the McVittie space-times one doesn't match an exact Schwarzschild space-time to an exact FLRW space-time at a particular radius; the McVittie space-times rather describe models that interpolate between Schwarzschild and FLRW in a continuous manner: For sufficiently small radius coordinates the metric is close to the Schwarzschild metric, for sufficiently large radius coordinates it is close to an FLRW model, and in between it changes gradually from one to the other. A major difficulty that arises with the McVittie metrics is in the fact that the lightlike geodesic equation is not completely integrable, i.e., there are not sufficiently many constants of motion that would allow to reduce the equation for lightlike geodesics to a system of first-order differential equations, see e.g. \cite{Aghili-2017, Carrera-Giulini-2010-review}. As this reduction was crucial for analytically calculating the shadow in Kerr (and related) space-times, we now have to face the problem that the techniques used for Kerr do not carry over to McVittie. For gravitational lensing in the McVittie metric with some approximations see Refs.~ \cite{Aghili-2017, Lake-Abdelqader-2011, Piattella-PRD-2016, Piattella-Universe-2016, Faraoni-2017}. \\

\subsection{Shadow in the Kottler space-time for comoving observers}
\label{sec:kottler}

\vspace{2mm}

Fortunately, there is a special McVittie metric where the situation is considerably simpler. We have said that the McVittie metric approaches \emph{some} FLRW model for sufficiently big radius coordinates. The metric becomes particularly simple if this is an FLRW model where the cosmic expansion is driven only by a cosmological constant, i.e., where it is the de Sitter metric. Then the resulting McVittie space-time is the Kottler (also known as the Schwarzschild-de Sitter) space-time. It is the unique solution to Einstein's vacuum field equation with a cosmological constant $\Lambda$, here assumed to be positive, that is spherically symmetric. Moreover, and this is crucial for the following discussion, the Kottler metric admits a hypersurface-orthogonal Killing vector field that is timelike on part of the space-time, i.e., it is static. Note that a general McVittie metric is spherically symmetric, so it admits the z-component and the modulus of the angular momentum as two constants of motion that are in involution. Together with the Lagrangian of the geodesic equation itself this gives three constants of motion in involution which is one too less for complete integrability. In the special case of the Kottler metric, however, we have a fourth constant of motion, namely the energy associated with the Killing vector field that is timelike on part of the space-time. This allows complete integrability of the geodesic equation, i.e., we may analytically calculate the paths of all (lightlike) geodesics and thereby, in complete analogy to the work of Synge, the angular size of the shadow for static observers, see (\ref{eq:kottler-static}). This was done in a paper of Stuchl\'{i}k and Hled\'{i}k \cite{Stuchlik-1999}. In subsequent papers different moving observers were considered for the Schwarzschild-de-Sitter case  \cite{Stuchlik-2006, Stuchlik-2007} and for the Kerr-de-Sitter case \cite{Stuchlik-2018}. Based on these observations, it was then possible for Perlick, Tsupko and Bisnovatyi-Kogan \cite{Perlick-Tsupko-BK-2018} to calculate the time-dependent shadow as seen by an observer who is comoving with the cosmic expansion. To that end they started out from the shadow formula for a static observer, see (\ref{eq:kottler-static}), and applied the aberration formula to get the shadow for the comoving observer. The result was \cite{Perlick-Tsupko-BK-2018, Tsupko-BK-2020b}
\begin{equation}
\alpha_\textrm{sh} = \alpha_\textrm{Schw} + \arcsin \left(\frac{3\sqrt{3}m H_0}{c} \right) \, , \quad \mbox{or} \quad    
\alpha_\textrm{sh} = \alpha_\textrm{Schw} + \arcsin \left( 3m\sqrt{\Lambda}\right) \, ,    
\end{equation}
where $\alpha_\textrm{Schw}$ is defined by formula (\ref{synge-shadow}) with $r_{\mathrm{O}}$ now time-dependent. $H_0= c \sqrt{\Lambda /3}$ is the Hubble constant of the de Sitter universe. Whereas the static observers exist only between the two horizons, this shadow formula for the comoving observer makes sense up to $r_{\mathrm{O}}$ to infinity. Nothing particular happens if the outer horizon is crossed. Here it is of particular importance that the angular radius of the shadow approaches a finite, non-zero, value if the comoving observer goes to infinity. The calculation of the shadow for a comoving observer in the Kottler space-time in \cite{Perlick-Tsupko-BK-2018} has been recently generalized to an arbitrary spherically symmetric black hole in an asymptotic de Sitter space-time by Roy and Chakrabarti \cite{Roy-2020}.

Here one should add a remark on the notion of a 'comoving observer'. This notion is unambiguous in an FLRW universe, but not so in a McVittie space-time. Actually, in a McVittie space-time all families of observers that approach the standard ('comoving') observers in the asymptotic FLRW space-time could be called 'comoving with the cosmic expansion'. In the above-mentioned paper by Perlick, Tsupko and Bisnovatyi-Kogan \cite{Perlick-Tsupko-BK-2018} the socalled McVittie observers were chosen for this family.
It should be noted that these observers are freely falling (i.e., geodesic) only asymptotically. However, the result that the shadow shrinks to a finite value is true for \emph{all} observers that asymptotically approach the (unambiguously defined) comoving observers in the FLRW universe. Therefore, this special choice of observers should be viewed as a matter of convention only. For example, Chang and Zhu \cite{Chang-Zhu-2020} have considered the shadow as viewed by \emph{freely falling} observers that asymptotically approach the standard observers and have obtained agreement with our results \cite{Perlick-Tsupko-BK-2018}, for observers at large distances.

For a discussion of light propagation in the Kottler space-time see, e.g., Refs. \cite{Lake-Roeder-1977, Stuchlik-1983, Islam-1983, Rindler-Ishak-2007, Hackmann-2008a, Hackmann-2008b, Lebedev-2013, Lebedev-2016}. We emphasize, in particular, the work of Rindler and Ishak \cite{Rindler-Ishak-2007} which clarified an important issue: It had been known since quite some time that, in the static coordinates, the coordinate representation of light rays in the Kottler space-time is independent of $\Lambda$. However, in spite of this fact and in contrast to earlier beliefs, Rindler and Ishak have shown that light deflection in the Kottler space-time \emph{is} influenced by $\Lambda$, because the measurement of angles between coordinate lines depends on $\Lambda$. \\

\subsection{Shadow in a general expanding universe}

\vspace{2mm}

Naively one might have expected that, in an expanding universe, the shadow for an observer who is comoving with the cosmic expansion shrinks to zero in the course of time because the distance of the observer to the black hole increases to infinity. This argument, however, is actually erroneous. In particular, it disregards the fact that the notion of 'distance' in general relativity and in particular in cosmology is ambiguous. What is relevant here is the 'angular-diameter distance' or 'area distance'. In an FLRW universe that starts with a big bang, the area distance between comoving observers is \emph{not} in general monotonically increasing in the course
of time \cite{Mattig-1958, Zeldovich-1964, Dashevsk-Zeldovich-1965, Zeldovich-Novikov-book-2, Hobson, Mukhanov-book, Jones-book}. 
This is reflected by the relation between angular size and redshift, 
\begin{equation} \label{eq:angular-size}
\alpha = \frac{L}{D_A(z)}  \, ,
\end{equation}
see e.g. \cite{Hobson, Mukhanov-book}.
Here $\alpha$ is the apparent angular size of an object of known linear physical size $L$ in an expanding universe, and $D_A(z)$ is the angular-diameter distance. The angle $\alpha$ decreases to a minimum at some value of $z$ and then starts to increase with increasing $z$. (In the de Sitter universe $\alpha$ goes to a constant.) Keeping the non-monotonicity of this relation in mind, the result that in an expanding universe the BH shadow does \emph{not} necessarily go to zero should, actually, not come as a surprise.

We emphasize that this argument was meant only as an intuitive explanation; it could \emph{not} replace a rigorous derivation as it was given in the above-mentioned paper \cite{Perlick-Tsupko-BK-2018} for comoving observers in the Kottler space-time. The relation between area distance and redshift is based on the assumption that light rays propagate in an FLRW universe without any additional sources of gravity. Moreover, it is assumed that these light rays were emitted or reflected by a real object with a given physical size. By contrast, in the case of the BH shadow, the light rays propagate through a space-time with the black hole, and the gravity of the black hole influences their propagation in addition to the cosmic expansion. Recall that near a black hole light rays are strongly bent. Additionally, one has to keep in mind that the black hole shadow is a region on the observer's sky with an angular size which should not be confused with the physical size of some object. Therefore the relation between angular size and redshift in an FLRW universe does not provide an exact solution for the shadow size.

However, relation (\ref{eq:angular-size}) can help to obtain an approximate expression for the shadow size if a number of conditions hold. First of all, in practice, the observer is always very far from the black hole. At distances much larger than the horizon, the gravity of the black hole can be neglected. Secondly, the expansion of the universe manifests itself only on large cosmological scales; in the vicinity of the black hole it is negligibly small. Therefore, we have a region of strong gravitation in the vicinity of the black hole (where the cosmological expansion can be neglected) and a region of significant cosmological expansion at large distances from the black hole (where the black hole gravity can be neglected). Also, we have an intermediate region of almost flat space-time in between. In this intermediate region, the black hole gravity is already negligible and the cosmological expansion is still not significant.

If these conditions are met, then the influence of the black hole and of the cosmological expansion occur at different parts of the trajectory, and the calculations can be done separately. This makes it possible to find a simple approximate formula for the angular size of the shadow as seen by an observer at cosmological distances from a black hole \cite{BK-Tsupko-2018}:
\begin{equation} \label{eq:shadow-BK-2018}
\alpha_{\mathrm{sh}}(z) \simeq \frac{3\sqrt{3}m}{D_A(z)} \, .
\end{equation}
In the described approximations, the influence of the black hole gravity on the whole path of a ray travelling from the vicinity of the black hole to an observer at a cosmological distance is reduced to using the effective linear size $3\sqrt{3}m$ of the black hole in formula (\ref{eq:angular-size}) instead of Schwarzschild radius $2m$.

It has been shown \cite{BK-Tsupko-2018} that the expression (\ref{eq:shadow-BK-2018}) gives us a sufficiently good approximate formula for the shadow size in a general expanding universe. The formula has been verified by comparing it with the exact solution for Kottler, as well as with the numerical calculation of rays in the McVittie metric, see also \cite{BK-2019}.

The idea of obtaining the solution (\ref{eq:shadow-BK-2018}) can be explained by matching two asymptotic solutions which has been presented in a subsequent work \cite{Tsupko-BK-2020b}. It has been shown that the physical conditions of the problem (see eq.(22) in \cite{Tsupko-BK-2020b}) allows us to use the method of matched asymptotic expansions known for solving differential equations. The Synge formula (\ref{synge-shadow}) and the cosmological formula for the angle (\ref{eq:angular-size}) are matched in an intermediate region of almost flat space-time to give a value of constant $L$ in (\ref{eq:angular-size}). Moreover, this method allows us to write down a continuous solution that is valid for any distance from the black hole to the observer, and can be applied to a Schwarzschild black hole embedded into any FLRW model where an approximately flat intermediate region exists \cite{Tsupko-BK-2020b}.

Recently, the results of \cite{BK-Tsupko-2018} have been generalized from the Schwarzschild to the Kerr case in \cite{Li-Guo-2020}. Obviously, for a comoving observer far away from the black hole the shadow increases as a whole with the distance, i.e. the size increases but keeping the shape (to a very good approximation) unchanged. \\

\subsection{Shadow as standard ruler in cosmological studies}

\vspace{2mm}

A special role in cosmology is played by objects with certain known, or as usually said, `standard' properties. Their standard properties allow us to determine the distance to them, which helps us to study and compare different cosmological models. The most famous example of such objects is the so-called standard candles, namely, objects with a known luminosity. Observations of supernovae of Ia type, which are considered as standard candles in cosmology, made it possible to discover the accelerated expansion of the Universe \cite{SN-1998, SN-1999}. Shortly after the recent discovery of gravitational waves, measurements of the Hubble constant were performed using sources of gravitational waves as standard sirens \cite{sirens-1986, sirens-2005, sirens-2017}.

Tsupko, Fan and Bisnovatyi-Kogan \cite{Tsupko-Fan-BK-2020} have suggested to use the shadow of black holes at cosmological distances as a standard ruler, i.e. as objects with known physical size. While the standard-ruler technique based on BAO measurements requires a statistical analysis, in the case of the black-hole shadow it is possible to use an individual object. If the mass of a BH is known independently, its shadow has a known \textit{effective} linear size $3\sqrt{3}m$. This physical size can be used as standard ruler. If we measure the angular size of the shadow $\alpha_\textrm{sh}$ by formula (\ref{eq:shadow-BK-2018}) we can obtain the angular-diameter distance $D_A$. If it is possible to obtain $z$ and $D_A$ independently, such observations provide the possibility to extract the cosmological parameters.

\begin{figure}[t]
	\begin{center}
		\includegraphics[width=0.75\textwidth]{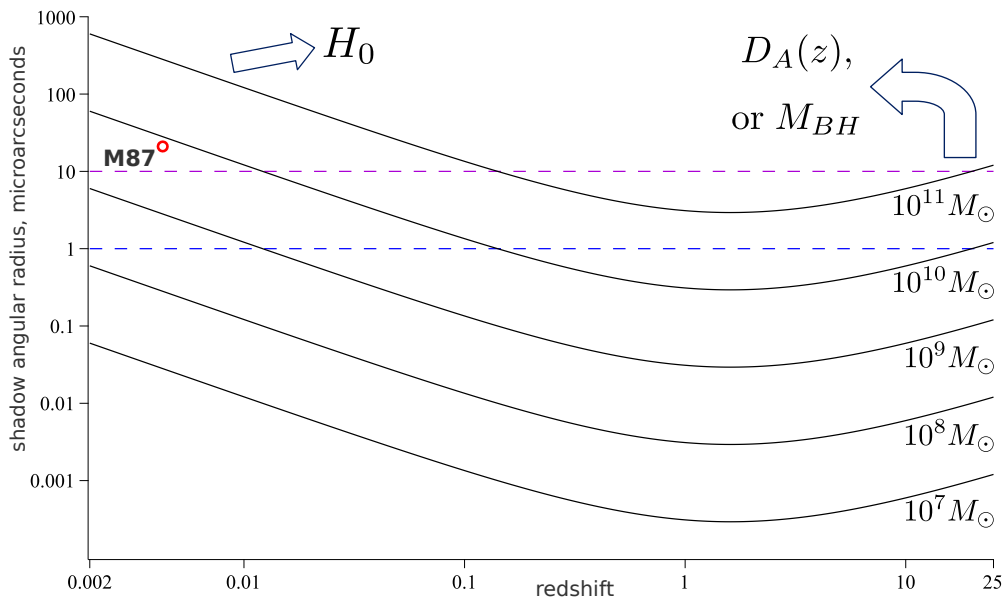}
	\end{center}
	\caption{Expected angular radius of the shadow of black holes with masses $10^7$, $10^8$, $10^9$, $10^{10}$ and $10^{11}$ $M_\odot$ as a function of redshift (black solid curves). The curves are calculated for the following present values of cosmological parameters: $\Omega_{\Lambda0}=0.7$ and $\Omega_{m0}=0.3$. Two horizontal dashed lines are located at 10 $\mu$as and 1 $\mu$as. With small redshift observation we can get the Hubble value, for high redshifts determination of $D_A(z)$ or BH mass estimation can be possible. See also Fig.1 in \cite{Tsupko-Fan-BK-2020}.}
	\label{fig:ruler}
\end{figure}

In Fig.\ref{fig:ruler} we have shown the expected angular size of the shadow for a given mass, as a function of the redshift, calculated for current values of cosmological parameters. Based on this Figure, two regimes of observations can be suggested \cite{Tsupko-Fan-BK-2020}.

For small redshifts, the shadow can provide an independent method of measuring the Hubble constant. New methods of determining the Hubble constant are extremely important for cosmological studies due to a current discrepancy in the values of the Hubble constant measured by different methods, known as the Hubble tension. For $z \ll 1$, we write $D_A(z) \simeq c z / H_0$, and obtain:
\begin{equation}
H_0 = \frac{cz \, \alpha_\textrm{sh}}{3\sqrt{3}m} \, .    
\end{equation}
We emphasize that independent black-hole mass estimations are crucial for the method, see discussion in \cite{Tsupko-Fan-BK-2020}.

The second regime of observations refers to black holes at high redshifts. At high redshifts the angular size of the shadow starts to grow due to the cosmic expansion and might become observable in the future. If we know the mass and measure the angular size of the shadow, we find the angular diameter distance:
\begin{equation}
D_A = \frac{3\sqrt{3}m}{\alpha_\textrm{sh}}  \, . 
\end{equation}
Together with the observation of the redshift, we can find the dependence of the angular-diameter distance on the redshift, $D_A(z)$, which allows us to extract the cosmological parameters. Since it can be difficult to obtain a very accurate estimate of the mass for black holes at high redshifts, it may be advantageous to use this method in the opposite way: With accurate knowledge of cosmological parameters, i.e. with known $D_A(z)$, it becomes possible to estimate the BH mass by measuring the shadow angle, which could be more accurate than other methods for determination of BH masses at cosmological distances.

For subsequent discussions of the method, including observational difficulties, see Qi and Zhang \cite{Qi-Zhang-2019}, Vagnozzi, Bambi and Visinelli \cite{Vagnozzi-2020}, Eubanks \cite{Eubanks-2020}. \\


\section{Influence of a plasma on the shadow}
\label{sec:plasma}

We have already mentioned in the Introduction that one usually has to resort to numerical investigations (i.e., to ray tracing) if one wants to take the effect of a medium onto light propagation into account. Therefore this field is largely outside the scope of this review. However, there is one particular kind of medium whose effect on the shadow can be treated analytically, and actually has been treated analytically for several space-times, namely a non-magnetized pressure-less electron-ion plasma. In this section we review the results on calculating the influence of such a medium on the shadow.

We work within the framework of geometric optics so that light propagation is modeled in terms of rays. In this approach the plasma makes itself felt only by influencing the space-time trajectories of the rays. Thereby we obviously ignore all plasma effects that require a description in terms of wave optics and we also neglect e.g. absorption. In a plasma the rays are no longer lightlike geodesics of the space-time metric; they are rather influenced by the \emph{plasma frequency}
\begin{equation}
\omega_p(x)^2 = \frac{4\pi e^2}{m_e} N(x) \, ,
\end{equation}
where $e$ is the electron charge, $m_e$ is the electron mass, and $N(x)$ is the number density of the electrons. It is a very special feature of a non-magnetized pressure-less electron-ion plasma that the rays are influenced by the plasma \emph{only} in terms of the plasma frequency, i.e., of the electron density. The same is no longer true if a magnetic field or a pressure is taken into account or if one has an electron-positron plasma where the positron density also plays a role.

We use the Hamiltonian formalism which allows us to assign a momentum co\-vector (or wave co\-vector) to each ray whose temporal component with respect to an observer field can be interpreted as a frequency. On an arbitrary space-time, light rays are then the solutions to Hamilton's equations
\begin{equation}
\dot{x}{}^{\mu} = \dfrac{\partial \mathcal{H}(x,p)}{\partial p_{\mu}} \, , \quad
\dot{p}{}_{\mu} = - \dfrac{\partial \mathcal{H}(x,p)}{\partial x^{\mu}} \, , \quad
\mathcal{H} (x,p) = 0
\end{equation}
with the Hamiltonian
\begin{equation}\label{eq:H}
\mathcal{H} (x,p) \, = \, \dfrac{1}{2} \left( g^{\mu \nu} (x) p_{\mu} p_{\nu}
+ \omega _p (x)^2 \right) \, . 
\end{equation}
The first rigorous derivation of this result from Maxwell's equations on a general-relativistic space-time with the plasma modeled in terms of two charged fluids was given by Breuer and Ehlers \cite{Breuer-Ehlers-1980}. (Actually, Breuer and Ehlers even considered the more general case of a magnetized plasma. Also note that their derivation made use of the fact that the ions are much more inert than the electrons, so it does not apply to an electron-positron plasma.) From the form of the Hamiltonian  one reads that the light rays in the plasma are timelike geodesics of a conformally rescaled metric. With the help of the four-velocity $U^{\mu} (x)$ of an observer field one can associate to each light ray a frequency $\omega (x) = -U^{\mu} (x) p_{\mu}$ and characterize the propagation of light by an index of refraction
\begin{equation}
n\big( x , \omega (x) \big) ^2  = 
1 - \frac{\omega_p(x)^2}{\omega(x)^2} \, .
\label{eq:ior}
\end{equation}
The characteristic feature of a plasma is in the fact that it is a dispersive medium, i.e., that the index of refraction depends on the frequency. Light propagation in a medium with a frequency-dependent index of refraction was discussed already by Synge \cite{Synge-1960}. A plasma belongs to this class of media as a special case which, however, Synge did not specifically consider. See also the works of Bi\v{c}\'{a}k and Hadrava \cite{Bicak-1975} and of Kulsrud and Loeb \cite{Kulsrud-Loeb-1992}.

In a spherically symmetric and static metric, i.e., a metric of the form of eq. (\ref{eq:g}), the light rays in a plasma with given plasma frequency $\omega _p (r)$ can be analytically determined in terms of integrals. For the Schwarzschild metric, the light deflection in such a medium was calculated in the weak-field approximation by Muhleman and Johnston \cite{MuhlemanJohnston1966} (see also
\cite{Bliokh-Minakov-1989}) and without this approximation by Perlick \cite{Perlick-2000b} who also considered the generalization to the equatorial plane of the Kerr space-time. The gravitational light deflection in the special case of a homogeneous plasma has been considered in detail in a series of works by Bisnovatyi-Kogan and Tsupko \cite{BK-Tsupko-2009, BK-Tsupko-2010, Tsupko-BK-2013}. 

The influence of the plasma on the shadow was analytically determined, for the case that both the space-time and the plasma are spherically symmetric and static, by Perlick, Tsupko and Bisnovatyi-Kogan \cite{Perlick-Tsupko-BK-2015}. It was assumed that the space-time is asymptotically flat and that the observer is situated between infinity and an unstable photon sphere. Then the angular radius $\alpha_{\mathrm{sh}}$ of the shadow was found as \cite{Perlick-Tsupko-BK-2015}
\begin{equation}\label{eq:shadow-pl}
\sin^2 \alpha_{\mathrm{sh}} \, = \, 
\dfrac{h(r_{\mathrm{ph}})^2}{h(r_{\mathrm{O}})^2}  
\end{equation}
where the function 
\begin{equation} \label{eq:h-definition}
h(r)^2 = \dfrac{D(r)}{A(r)} \left( 1 - A(r)  
\dfrac{\omega _p (r)^2}{\omega _0^2} \right)  
\end{equation}
generalizes in a natural way the corresponding function from
(\ref{eq: def-h-spher}) to the plasma case. Here $\omega _0$ is the frequency of the light at infinity with respect to a static observer, $r_{\mathrm{O}}$ is the radius coordinate of the observer and $r_{\mathrm{ph}}$ is the radius coordinate of the photon sphere. We emphasize that the constant of motion $\omega_0$ should not be confused with the photon frequency $\omega(x)$ measured by a static observer at position $x$. These values are connected by the gravitational redshift relation 
\begin{equation} \label{eq:gr-redshift}
\omega(x) = \frac{c}{\sqrt{-g_{tt}(x)}} \,  \omega_0   
\end{equation}
where we have used that in an asymptotically flat space-time
$g_{tt} \to -c^2$ at infinity. This is crucial when one uses Synge's approach \cite{Synge-1960} with the refractive index (\ref{eq:ior}). With (\ref{eq:gr-redshift}) taken into account, for a plasma frequency $\omega _p (r)$ on a spherically symmetric and static space-time (\ref{eq:g}), the refractive index can be written as a function of $r$ and $\omega _0$,
\begin{equation}
n(r,\omega _0) ^2  = 
1 - A(r) \dfrac{\omega_p (r)^2}{\omega_0^2} \, .
\end{equation}

Moreover, the influence of a plasma on the shadow was calculated also in the Kerr space-time by Perlick and Tsupko \cite{Perlick-Tsupko-2017}. 
For a plasma density $\omega _p (r , \vartheta)$, the index of refraction (\ref{eq:ior}) is equal to
\begin{equation}
n(r , \vartheta , \omega _0)^2   =  
1 - \left( 1 - \frac{2mr}{r^2 + a^2 \cos^2 \vartheta} \right) 
\dfrac{\omega _p (r, \vartheta)^2}{\omega_0^2} \, .
\end{equation}
Here the problem arises that the Hamilton-Jacobi equation is not in general separable, i.e., there is no Carter constant which is used in the vacuum case for analytically determining the light rays. A Carter constant exists only if the plasma frequency is of the special form \cite{Perlick-Tsupko-2017}
\begin{equation}\label{eq:sepcon}
\omega _p (r , \vartheta ) ^2 = \dfrac{f_r(r)+f_{\vartheta} ( \vartheta )}{ 
r^2 + a^2 \mathrm{cos} ^2 \vartheta}
\end{equation}
with some functions $f_r$ and $f_{\vartheta}$. Note that for a spinning black hole ($a \neq 0$) this excludes the case of a non-constant $\omega _p$ that depends only on $r$. If the plasma frequency is of this special form (\ref{eq:sepcon}), the boundary curve of the shadow can be analytically determined by the same method as for the vacuum case. Several examples are worked out in the paper by Perlick and Tsupko \cite{Perlick-Tsupko-2017}, and several others, again for cases where the separability condition is satisfied, in a follow-up paper by Yan \cite{Yan-2018}. For some examples where the separability condition is not satisfied, the shadow was numerically calculated by Huang, Dong and Liu \cite{Huang-2018}. If the plasma frequency is not small in comparison to the frequency of the light, the shape of the shadow can be significantly different from the vacuum case. Note that the influence of the plasma is palpable only in the radio regime. Even then, for the known supermassive black-hole candidates, the influence is at the borderline of what could be observable in the near future.

For other investigations of gravitational lensing in the presence of a cold and non-magnetized plasma we refer to \cite{Bliokh-Minakov-1989, BK-Tsupko-2009, BK-Tsupko-2010, Tsupko-BK-2013, Morozova-2013, Er-Mao-2014, BK-Tsupko-2015, Rogers-2015, Rogers-2017a, Rogers-2017b, Er-Rogers-2018, Perlick-2017, Crisnejo-Gallo-2018, Crisnejo-Gallo-Rogers-2019, Crisnejo-Gallo-Villanueva-2019, Crisnejo-Gallo-Jusufi-2019, Kimpson-2019a, Kimpson-2019b, Sareny-2019, Matsuno-2020, Jin-2020, Tsupko-BK-2020a, Wagner-Er-2020, Chowdhuri-2020, Li-Wang-2020}; for a recent review see Bisnovatyi-Kogan and Tsupko \cite{BK-Tsupko-Universe-2017}.

We end this Section with two additional remarks. Firstly, we comment on more general plasma models. An important step in the direction of determining the shadow size in a \emph{magnetized} plasma was taken in a paper of Broderick and Blandford \cite{Broderick-Blandford-2003}, see also \cite{Broderick-Blandford-2004}. In particular, by numerical integration of the ray equations they calculated the photon capture cross-sections for a number of different plasma densities and magnetic field strengths. Recently, the analytical results for the shadow of a black hole surrounded by a plasma have been generalized to the case of an arbitrary transparent dispersive medium with a given refractive index \cite{Tsupko-2021}. More precisely, the shadow has been analytically calculated in a general spherically symmetric metric in the presence of a medium with a spherically symmetric refractive index. These results can serve as a basis for studies of various plasma models beyond the cold plasma case.

Secondly, we mention that in a recent paper \cite{EHT-2021} the EHT Collaboration reports that the observations of M87 are consistent with GRMHD models of magnetically arrested accretion disks. This model of an accretion flow in a large-scale magnetic field was suggested by Bisnovatyi-Kogan and Ruzmaikin \cite{BK-1974, BK-1976} and further investigated by Igumenshchev et al. \cite{Igumenshchev-2003} and Narayan et al. \cite{Narayan-2003}; see \cite{BK-2019-review} for a recent review. \\


\section{Concluding remarks}

(i) By now, the analytical calculation of the shadow has been achieved for a rather wide class of metrics. The angular size of the shadow in any spherically symmetric and static metric, for any position $r_\textrm{O}$ of a static observer, can be calculated by the formula (\ref{eq:shadow}). The radius of the photon sphere in this metric is found by the formula (\ref{eq:circ}). The linear size of the shadow at large distances in an asymptotically flat metric is given by the critical impact parameter, which can be found by the formula (\ref{eq:b-crit-h}).

(ii) The calculation of the shadow for rotating black holes is presented in the works of Grenzebach et al \cite{Gren-Perlick-2014, Gren-Perlick-2015} for a wide class of space-times, including Kerr, Kerr-Newman, Kerr-Newman-de Sitter and others. In particular, the results of these works make it possible to calculate the shadow of a Kerr black hole for an arbitrary position of the observer, which generalizes the work of Bardeen \cite{Bardeen-1973}, where an observer at a large distance from the black hole was considered. For both methods it is crucial that the space-time is axisymmetric and stationary and that the equation for lightlike geodesics admit a Carter constant. For Bardeen's methodology to be applicable it is also necessary that the spacetime is asymptotically flat. If these conditions are satisfied, the two methods are completely equivalent for observers at large distances; this is, of course, the case most relevant in view of observations.

(iii) It is common to represent the boundary curve of the shadow in terms of impact parameters, but one must never forget that actual observations refer to measurements of angles in the observer's sky. For observers far away from the center in an asymptotically flat space-time, we are on the safe side because dividing the impact parameters by the radial coordinate of the observer gives us the corresponding angles. However, if the space-time is not asymptotically flat (for example, in the presence of a cosmological constant), the calculation of the critical impact parameters is insufficient to obtain the size and the shape of the shadow.

(iv) We have emphasized throughout that the construction of the shadow is related to the existence of a photon region, not of a horizon. Therefore, the analytical methods of determining the shadow apply not only to black holes but also to other objects that are compact enough to feature a photon region. Such objects are often called ``ultracompact''; e.g. wormholes are among them. As long as no light is emitted from the interior region, or passing through the interior region, it is very difficult to distinguish black holes from other ultracompact objects by way of their shadows. However, whenever we see a shadow a black hole is the most likely explanation:  Whereas we are quite sure by now that black holes actually exist in Nature, there is no observational evidence for the existence of other ultracompact objects.
Having said this, we repeat that at present the existence of exotic objects such as wormholes, boson stars, Proca stars or fermion stars \emph{cannot} be ruled out. Therefore, investigating the observational consequences of their existence is certainly an important line of research.

(v) The influence of a cold and non-magnetized plasma on the shadow can be calculated in the spherically symmetric and static case by formula (\ref{eq:shadow-pl}) if there is a unique unstable photon sphere at radius $r_\textrm{ph}$. In axisymmetric and static space-times, the plasma frequency (and, thus, the electron density of the plasma) must be of a particular form to admit a Carter constant which is necessary for an analytical calculation of the shadow. For a Kerr BH surrounded by a plasma this is the case if and only if the plasma frequency is of the form (\ref{eq:sepcon}). For more general plasma models some preliminary results have been found.

(vi) To discuss shadow observations in an expanding universe, one must consider comoving observers. A number of analytical solutions for various cases in different approximations have recently been presented. Based on these results, it has been proposed to use the shadow as a standard ruler in cosmological studies \cite{Tsupko-Fan-BK-2020}.

(vii) We strongly advocate the idea of further advancing analytical methods of calculating the shadow, thereby complementing the ongoing numerical studies that are of course of high relevance for comparing with observations. Among such relevant future projects we mention a more detailed study of the shadow, and in particular of the photon region, in rotating wormhole space-times and of the influence on the shadow of a magnetized plasma. \\

\section*{Acknowledgments}

It is our pleasure to thank Gennady S. Bisnovatyi-Kogan for helpful discussions and valuable recommendations. The work on this review article was partially supported by the Russian Foundation for Basic Research (OYuT) and the Deutsche Forschungsgemeinschaft (VP) according to the research project No. 20-52-12053. Moreover, VP gratefully acknowledges support from the Deutsche Forschungsgemeinschaft within the Research Training Group 1620 'Models of Gravity'. \\



\begin{thebibliography}{300}


\bibitem{GL-1}
P. Schneider, J. Ehlers, E. E. Falco, Gravitational Lenses, Springer-Verlag, Berlin (1992)


\bibitem{Blandford-Narayan-1992}
R. D. Blandford, R. Narayan,
Cosmological applications of gravitational lensing,
Ann. Rev. Astron. Astrophys. 30 (1992) 311




\bibitem{Petters-2001}
A. O. Petters, H. Levine, J. Wambsganss, Singularity Theory and Gravitational Lensing, 
Birkh\"{a}user, Boston, MA  (2001)

\bibitem{Perlick-2004a}
V. Perlick,
Gravitational lensing from a space-time Perspective,
Liv. Rev. Relativity 7 (2004) 9

\bibitem{GL-2}
P. Schneider, C. S. Kochanek, J. Wambsganss, Gravitational Lensing: Strong, Weak and Micro, 
Springer, Berlin (2006) 


\bibitem{Bartelmann-2010}
M. Bartelmann, 
Gravitational lensing,
Class. Quantum Grav. 27 (2010) 233001

\bibitem{Dodelson-2017}
S. Dodelson, 
Gravitational Lensing, 
Cambridge Univ. Press, Cambridge, UK (2017) 

\bibitem{Congdon-Keeton-2018}
A. B. Congdon, C. Keeton, 
Principles of Gravitational Lensing: Light Deflection as a 
Probe of Astrophysics and Cosmology, 
Springer, Cham, Switzerland (2018) 



\bibitem{EHT-1}
The Event Horizon Telescope Collaboration, 
First M87 Event Horizon Telescope results. I.
The shadow of the supermassive black hole,
Astrophys. J. Lett. 875 (2019) L1

\bibitem{EHT-2}
The Event Horizon Telescope Collaboration, 
First M87 Event Horizon Telescope results. II.
Array and instrumentation,
Astrophys. J. Lett. 875 (2019) L2

\bibitem{EHT-3}
The Event Horizon Telescope Collaboration, 
First M87 Event Horizon Telescope results. III.
Data processing and calibration,
Astrophys. J. Lett. 875 (2019) L3

\bibitem{EHT-4}
The Event Horizon Telescope Collaboration, 
First M87 Event Horizon Telescope results. IV.
Imaging the central supermassive black hole,
Astrophys. J. Lett. 875 (2019) L4

\bibitem{EHT-5}
The Event Horizon Telescope Collaboration, 
First M87 Event Horizon Telescope results. V.
Physical origin of the asymmetric ring,
Astrophys. J. Lett. 875 (2019) L5.


\bibitem{EHT-6}
The Event Horizon Telescope Collaboration,
First M87 Event Horizon Telescope results. VI.
The shadow and mass of the central black hole,
Astrophys. J. Lett. 875 (2019) L6.


\bibitem{Falcke-2000}
H. Falcke, F. Melia, E. Agol,
Viewing the shadow of the black hole at the Galactic center,
Astrophys. J. Lett. 528 (2000) L13



\bibitem{Melia-Falcke-2001}
F. Melia and H. Falcke,
The Supermassive Black Hole at the Galactic Center,
Annual Review of Astronomy and Astrophysics 39 (2001) 309




\bibitem{Broderick-Loeb-2005}
A. E. Broderick, A. Loeb,
Imaging bright-spots in the accretion flow near the black hole horizon of Sgr A*,
Mon. Not. Roy. Astron. Soc. 363 (2005) 353


\bibitem{Moscibrodzka-2009}
M. Mo\'{s}cibrodzka, Ch.F. Gammie, J. C. Dolence, H. Shiokawa, Po Kin Leung,
Radiative models of Sgr A* from GRMHD simulations,
Astrophys. J. 706 (2009) 497



\bibitem{Dexter-2009}
J. Dexter, E. Agol, P. Chris Fragile,
Millimeter flares and VLBI visibilities from relativistic simulations of magnetized accretion onto the Galactic center black hole,
Astrophys. J. Lett. 703  (2009) L142



\bibitem{Broderick-Fish-2011}
A.E. Broderick, V.L. Fish, Sh.S. Doeleman,  A. Loeb,
Evidence for low black hole spin and physically motivated accretion models from millimeter-VLBI observations of Sagittarius A*,
Astrophys. J. 735 (2011) 110



\bibitem{Broderick-Johannsen-2014}
A.E. Broderick, T. Johannsen, A. Loeb, D. Psaltis,
Testing the no-hair theorem with Event Horizon Telescope observations of Sagittarius A*,
Astrophys. J. 784 (2014) 7


\bibitem{Gralla-2019}
S.E. Gralla, D. E. Holz, R.M. Wald, 
Black hole shadows, photon rings, and lensing rings,
Phys. Rev. D 100 (2019)  024018


\bibitem{Narayan-2019}
R. Narayan, M.D. Johnson, C.F. Gammie, 
The shadow of a spherically accreting black hole
Astrophys. J. Lett. 885 (2019) L33


\bibitem{Johnson-2020}
M.D. Johnson, A. Lupsasca, A. Strominger, et al, 
Universal interferometric signatures of a black hole's photon ring,
Science Advances 6 (2020) eaaz1310




\bibitem{Bronzwaer-2021}
Th. Bronzwaer, J. Davelaar, Z. Younsi et al, 
Visibility of black hole shadows in low-luminosity AGN,
Mon. Not. Roy. Astron. Soc. 501 (2021) 4722



\bibitem{Kocherlakota-2021}
P. Kocherlakota et al. (EHT Collaboration),
Constraints on black-hole charges with the 2017 EHT observations of M87$^*$,
Phys. Rev. D 103 (2021) 104047




\bibitem{Broderick-2021}
A.E. Broderick, P. Tiede, D.W. Pesce, R. Gold,
Measuring Spin from Relative Photon Ring Sizes,
arXiv:2105.09962 (2021)



\bibitem{Bronzwaer-Falcke-2021}
Th. Bronzwaer and H. Falcke,
The Nature of Black Hole Shadows,
arXiv:2108.03966 (2021)





\bibitem{Zakharov-Paolis-2005-New-Astronomy}
A.F. Zakharov, A.A. Nucita, F. DePaolis, G. Ingrosso, 
Measuring the black hole parameters in the galactic center with RADIOASTRON,
New Astronomy 10 (2005) 479



\bibitem{Kardashev-Millimetron-2014}
N. S. Kardashev, I. D. Novikov, V. N. Lukash, et al,
Review of scientific topics for the Millimetron space observatory,
Physics-Uspekhi 57 (2014) 1199







\bibitem{Andrianov-2021}
A. S. Andrianov, A. M. Baryshev, H. Falcke, et al,
Simulations of M87 and Sgr A* imaging with the Millimetron Space Observatory on near-Earth orbits,
Monthly Notices of the Royal Astronomical Society 500 (2021) 4866



\bibitem{Likhachev-2021}
S. F. Likhachev, A. G. Rudnitskiy, M. A. Shchurov, et al,
High Resolution Imaging of a Black Hole Shadow with Millimetron Orbit around Lagrange Point L2,
arXiv:2108.03077 (2021)







\bibitem{Novikov-Millimetron-2021}
I. D. Novikov, S. F. Likhachev, Yu. A. Shchekinov, et al,
Objectives of the Millimetron Space Observatory science program and technical capabilities of its realization,
Physics-Uspekhi 64 (2021) 386








\bibitem{Luminet-1979}
J.-P. Luminet, 
Image of a spherical black hole with thin accretion disk,
Astron. Astrophys.75 (1979) 228


\bibitem{Viergutz-1993}
S. U. Viergutz, 
Image generation in Kerr geometry. I. Analytical investigations 
on the stationary emitter-observer problem,
Astron. Astrophys. 272 (1993) 355


\bibitem{JamesEtAl2015}
O. James, E. von Tunzelmann, P. Franklin, K. S. Thorne, 
Gravitational lensing by spinning black holes in astrophysics, 
and in the movie \emph{Interstellar},
Class. Quantum Grav. 32 (2015) 065001


\bibitem{Synge-1966}
J.~L. Synge,
The escape of photons from gravitationally intense stars,
Mon. Not. Roy. Astron. Soc. 131 (1966) 463


\bibitem{Zeld-Novikov-1965}
Ya.B. Zeldovich, I.D. Novikov, 
Relativistic astrophysics. II., 
Sov. Phys. Usp. 8  (1966) 522,
Russian original: Usp. Fiz. Nauk 86 (1965) 447 


\bibitem{Bardeen-1973}
J.~M. Bardeen, 
Timelike and null geodesics in the Kerr metric, 
in {\em Black Holes}, eds. C. DeWitt and B. DeWitt,
Gordon and Breach, New York (1973) p. 215



\bibitem{Chandra-1983}
S. Chandrasekhar, 
The Mathematical Theory of Black Holes,
Oxford Univ. Press, Oxford (1983)



\bibitem{Young-1976}
P. J. Young,
Capture of particles from plunge orbits by a black hole,
Phys. Rev. D 14 (1976) 3281



\bibitem{Luminet-2018}
J.-P. Luminet,
Seeing black holes: From the computer to the telescope,
Universe 4 (2018)  86



\bibitem{Dymnikova-1986}
I. G. Dymnikova, 
Motion of particles and photons in the gravitational field of a rotating body,
Sov. Phys. Usp. 29 (1986) 215



\bibitem{Fukue-2003}
J. Fukue,
Silhouette of a dressed black hole, 
Publ. Astron. Soc. Japan 55 (2003) 155



\bibitem{Broderick-Loeb-2009}
A.E. Broderick, A. Loeb, 
Imaging the black hole silhouette of M87: Implications for jet formation and black hole spin,
Astrophy. J. 697 (2009) 1164




\bibitem{Johannsen-2010}
T. Johannsen, D. Psaltis, 
Testing the no-hair theorem with observations in the electromagnetic spectrum. II. Black hole images,
Astrophys. J. 718 (2010) 446






\bibitem{Bambi-2017}
C. Bambi, 
Black Holes: A Laboratory for Testing Strong Gravity,
Springer, Singapore (2017)



\bibitem{Gren-Perlick-2014}
A. Grenzebach, V. Perlick, C. L{\"a}mmerzahl,
Photon regions and shadows of Kerr-Newman-NUT black holes with a cosmological constant,
Phys. Rev. D 89 (2014) 124004


\bibitem{Gren-Perlick-2015}
A. Grenzebach, V. Perlick and C. L{\"a}mmerzahl,
Photon regions and shadows of accelerated black holes,
Int. J. Mod. Phys. D 24 (2015) 1542024




\bibitem{BK-2019}
G.S. Bisnovatyi-Kogan, O.Yu. Tsupko, V. Perlick,
Shadow of black holes at local and cosmological distances,
Proceedings of Science, Multifrequency Behaviour of High Energy Cosmic Sources - XIII. 3-8 June 2019. Palermo, Italy. Online at https://pos.sissa.it/cgi-bin/reader/conf.cgi?confid=362, id.9; arXiv: 1910.10514 (2019)






\bibitem{EHT-press}
https://eventhorizontelescope.org/press-release-april-10-2019-astronomers-capture-first-image-black-hole



\bibitem{BK-Tsupko-Universe-2017}
G. S. Bisnovatyi-Kogan, O. Yu. Tsupko,
Gravitational lensing in presence of plasma: Strong lens systems, black hole lensing and shadow,
Universe 3 (2017) 57




\bibitem{Hilbert-1917}
D. Hilbert, 
Die Grundlagen der Physik. (Zweite Mitteilung),
Nachr. Gesellsch. Wissensch. G\"{o}ttingen, Math.-Phys. Kl. (1917) 53




\bibitem{MTW-1973}
C.W. Misner, K.S. Thorne, J.A. Wheeler, 
Gravitation,
Freeman, San Francisco (1973)


\bibitem{Cunha-Herdeiro-2018}
P.V.P. Cunha, C.A.R. Herdeiro,
Shadows and strong gravitational lensing: a brief review,
Gen. Rel. Grav. 50 (2018)  42 








\bibitem{Johannsen-2013}
T. Johannsen,
Photon rings around Kerr and Kerr-like black holes,
Astrophy. J. 777 (2013) 170


\bibitem{Virbhadra-2000}
K.S. Virbhadra, G.F.R. Ellis,
Schwarzschild black hole lensing,
Phys. Rev. D 62 (2000) 084003




\bibitem{Bozza-2001}
V. Bozza, S. Capozziello, G. Iovane, G. Scarpetta,
Strong field limit of black hole gravitational lensing,
Gen. Rel. Grav. 33 (2001) 1535



\bibitem{BK-Tsupko-2008}
G.S. Bisnovatyi-Kogan, O.Yu. Tsupko,
Strong gravitational lensing by Schwarzschild black holes,
Astrophysics 51 (2008) 99



\bibitem{Psaltis-2020}
D. Psaltis et al. (EHT Collaboration),
Gravitational Test beyond the First Post-Newtonian Order with the Shadow of the M87 Black Hole,
Phys. Rev. Lett. 125 (2020) 141104




\bibitem{Cunningham-1973}
C.T. Cunningham, J.M. Bardeen,
The optical appearance of a star orbiting an extreme Kerr black hole
Astrophys. J. 183 (1973) 237 



\bibitem{Dokuchaev-2020}
V.I. Dokuchaev, N.O. Nazarova,
Silhouettes of invisible black holes,
Physics-Uspekhi 63 (2020)  583



\bibitem{Chael-2021}
A. Chael, M. D. Johnson, A. Lupsasca,
Observing the Inner Shadow of a Black Hole: A Direct View of the Event Horizon,
arXiv: 2106.00683 (2021)






\bibitem{HassePerlick2002}
W. Hasse and V. Perlick,
Gravitational lensing in spherically symmetric static spacetimes with centrifugal force reversal,
Gen. Rel. Grav. 34 (2002) 415


\bibitem{Pande-1986}
A.K. Pande, M.C. Durgapal,
Trapping of photons in spherical static configurations,
Class. Quantum Grav. 3 (1986) 547


\bibitem{Cvetic-2016}
M. Cveti\v{c}, G.W. Gibbons, C.N. Pope,
Photon spheres and sonic horizons in black holes from supergravity and other theories,
Phys. Rev. D 94 (2016)  106005


\bibitem{Atkinson-1965}
R. d'E. Atkinson,
On light tracks near a very massive star,
Astronomical Journal 70 (1965) 517





\bibitem{Virbhadra-Ellis-2001}
K. S. Virbhadra and G. F. R. Ellis,
Gravitational lensing by naked singularities,
Phys. Rev. D 65 (2002) 103004



\bibitem{Claudel-Virbhadra-2001}
Clarissa-Marie Claudel, K.S. Virbhadra, G.F.R. Ellis,
The geometry of photon surfaces,
J. Math. Phys. 42 (2001) 818

\bibitem{Bozza-2002}
V. Bozza,
Gravitational lensing in the strong field limit,
Phys. Rev. D 66 (2002)  103001


\bibitem{Perlick-Tsupko-BK-2015}
V. Perlick, O. Yu. Tsupko, G. S. Bisnovatyi-Kogan,
Influence of a plasma on the shadow of a spherically symmetric black hole,
Phys. Rev. D 92 (2015) 104031

\bibitem{Konoplya-2019}
R.A. Konoplya,
Shadow of a black hole surrounded by dark matter,
Phys. Lett. B 795 (2019) 1


\bibitem{Konoplya-et-al-2020}
R.A. Konoplya, Th. Pappas, A. Zhidenko,
Einstein-scalar-Gauss-Bonnet black holes: Analytical approximation for the metric and applications to calculations of shadows,
Phys. Rev. D 101 (2020)  044054



\bibitem{Konoplya-2020}
R.A. Konoplya,
Quantum corrected black holes: Quasinormal modes, scattering, shadows,
Phys. Lett. B 804 (2020) 135363


\bibitem{Grenzebach-2015}
A. Grenzebach,
Aberrational effects for shadows of black holes, 
in D. Puetzfeld, C. Laemmerzahl, B. Schutz (eds.), 
Equations of Motion in Relativistic Gravity, Springer, Heidelberg (2015)



\bibitem{Grenzebach-2016-book}
A. Grenzebach,
The Shadow of Black Holes: An Analytic Description,
Springer, Heidelberg (2016)




\bibitem{Eiroa-2002}
E.F. Eiroa, G.E. Romero, D.F. Torres,
Reissner-Nordstr\"{o}m black hole lensing,
Phys. Rev. D 66 (2002) 024010

\bibitem{Zakharov-2014}
A. F. Zakharov,
Constraints on a charge in the Reissner-Nordstr\"{o}m metric for the black hole at the Galactic center,
Phys. Rev. D 90 (2014) 062007

\bibitem{Zakharov-1994}
A. F. Zakharov,
Particle capture cross sections for a Reissner--Nordstr\"{o}m black hole,
Class. Quantum Grav. 11  (1994) 1027



\bibitem{Zakharov-2005-AA}
A. F. Zakharov, F. De Paolis, G. Ingrosso and A. A. Nucita,
Direct measurements of black hole charge with future astrometrical missions,
Astronomy and Astrophysics 442 (2005) 795



\bibitem{Zakharov-2012-review}
A. F. Zakharov, F. De Paolis, G. Ingrosso and A. A. Nucita,
Shadows as a tool to evaluate black hole parameters and a dimension of spacetime,
New Astronomy Reviews 56 (2012) 64




\bibitem{Zakharov-2021}
A. F. Zakharov,
Constraints on a tidal charge of the supermassive black hole in M87* with the EHT observations in April 2017,
arXiv:2108.01533 (2021)






\bibitem{Alexeyev-2019}
S.O. Alexeyev, B.N. Latosh, V.A. Prokopov, E.D. Emtsova,
Phenomenological extension for tidal charge black hole,
J. Exp. Theor. Phys. 128 (2019) 720



\bibitem{Kottler-1918}	
F. Kottler,	
\"{U}ber die physikalischen Grundlagen der Einsteinschen Gravitationstheorie,
Ann. Phys. (Berlin) 361 (1918) 401



\bibitem{Lake-Roeder-1977}
K. Lake and R.C. Roeder,
Effects of a nonvanishing cosmological constant on the spherically symmetric vacuum manifold,
Phys. Rev. D 15 (1977)  3513



\bibitem{Stuchlik-1983}
Z. Stuchl{\'\i}k,
The motion of test particles in black-hole backgrounds with non-zero cosmological constant,
Bull. Astron. Inst. Czechoslovakia 34 (1983) 129



\bibitem{Stuchlik-1999}	
Z. Stuchl\'{i}k, S. Hled\'{i}k, 
Some properties of the Schwarzschild–de Sitter and Schwarzschild–anti-de Sitter space-times,
Phys. Rev. D 60 (1999) 044006




\bibitem{Perlick-Tsupko-BK-2018}
V. Perlick, O. Y. Tsupko, G. S. Bisnovatyi-Kogan,
Black hole shadow in an expanding universe with a cosmological constant,
Phys. Rev. D 97 (2018) 104062

\bibitem{Carter1968}
B. Carter,
Global structure of the Kerr family of gravitational fields,
Phys. Rev. 174 (1968) 1559




\bibitem{Teo-2021}
E. Teo,
Spherical orbits around a Kerr black hole,
Gen. Rel. Grav. 53 (2021) 10



\bibitem{Teo-2003}
E. Teo, 
Spherical photon orbits around a Kerr black hole,
Gen. Rel. Grav. 35 (2003) 1909 



\bibitem{Hod-2013}
Sh. Hod,
Spherical null geodesics of rotating Kerr black holes,
Phys.  Lett. B 718 (2013) 1552



\bibitem{Igata-2019}
T. Igata, H. Ishihara, Yu. Yasunishi,
Observability of spherical photon orbits in near-extremal Kerr black holes,
Phys. Rev. D 100 (2019) 044058


\bibitem{Cunha-Thesis-2015}
P. V. P. da Cunha,
Black Hole Shadows, Master thesis, University of Coimbra, Portugal (2015)

\bibitem{stein-orbits}
L.C. Stein,
https://duetosymmetry.com/tool/kerr-circular-photon-orbits/

\bibitem{HassePerlick2006}
W. Hasse V. Perlick,
A Morse-theoretical analysis of gravitational lensing in the Kerr-Newman spacetime,
J. Math. Phys. 47 (2006) 042503

\bibitem{Perlick2005}
V. Perlick,
On totally umbilic submanifolds of semi-Riemannian manifolds,
Nonlinear Anal. 63 (2005) e511

\bibitem{Cederbaum2015}
C. Cederbaum, 
Uniqueness of photon spheres in static vacuum asymptotically flat spacetimes,
Contemp. Math. 667 (2015) 86

\bibitem{YazadjievLazov2015}
S. Yazadjiev, B. Lazov, 
Uniqueness of the static Einstein-Maxwell spacetimes with a photon sphere,
Class. Quantum Grav. 32 (2015) 165021

\bibitem{CederbaumGalloway2016}
C. Cederbaum, G.J. Galloway, 
Uniqueness of photon spheres in electro-vacuum spacetimes,
Class. Quantum Grav. 33 (2016) 075006

\bibitem{Yazadjiev2015}
S. Yazadjiev, 
Uniqueness of the static spacetimes with a photon sphere in Einstein-scalar field theory,
Phys. Rev. D 91 (2015) 123013

\bibitem{YazadjievLazov2016}
S. Yazadjiev, B. Lazov, 
Classification of the static and asymptotically flat Einstein-Maxwell-dilaton spacetimes with a photon sphere,
Phys. Rev. D 93 (2016) 083002

\bibitem{Rogatko2016}
M. Rogatko, 
Uniqueness of photon sphere for Einstein-Maxwell-dilaton black holes with arbitrary coupling constant,
Phys. Rev. D 93 (2016) 064003


\bibitem{Yoshino2017}
H. Yoshino,  
Uniqueness of static photon surfaces: Perturbative approach,
Phys. Rev. D 95 (2017) 044047

\bibitem{Semerak1996}
O. Semer{\'a}k,
Photon escape cones in the Kerr field,
Helv. Phys. Acta 69 (1996) 69

\bibitem{Cunningham-1972}
C.T. Cunningham, J.M. Bardeen,
The optical appearance of a star orbiting an extreme Kerr black hole,
Astrophy. J. Lett. 173 (1972) L137



\bibitem{Frolov-Zelnikov-2011}
V. P. Frolov, A. Zelnikov, 
Introduction to Black Hole Physics,
Oxford Univ. Press, New York (2011)

\bibitem{Bozza-2005}
V. Bozza, F. De Luca, G. Scarpetta, M. Sereno,
Analytic Kerr black hole lensing for equatorial observers in the strong deflection limit,
Phys. Rev. D, 72 (2005) 083003


\bibitem{Bozza-2006}	
V. Bozza, F. De Luca, G. Scarpetta,
Kerr black hole lensing for generic observers in the strong deflection limit,
Phys. Rev. D 74 (2006) 063001


\bibitem{Bozza-2008}
V. Bozza,
Optical caustics of Kerr space-time: The full structure
Phys. Rev. D 78 (2008) 063014


\bibitem{Bozza-2010}
V. Bozza,
Gravitational lensing by black holes,
Gen. Rel. Grav. 42 (2010)  2269


\bibitem{Tsupko-2017}
O. Yu. Tsupko,
Analytical calculation of black hole spin using deformation of the shadow,
Phys. Rev. D 95 (2017) 104058



\bibitem{Zakharov-1986}
A. F. Zakharov,
Types of unbound geodesics in the Kerr metric,
Sov. Phys. J. Exp. Theor. Phys. 64 (1986) 1.







\bibitem{Gralla-Lups-2018}
S.E. Gralla, A. Lupsasca and A. Strominger,
Observational signature of high spin at the Event Horizon Telescope,
Mon. Not. Roy. Astron. Soc. 475 (2018) 3829

\bibitem{Gralla-Lups-2020c}
S.E. Gralla, A. Lupsasca,
Observable shape of black hole photon rings,
Phys. Rev. D 102 (2020)  124003



\bibitem{Rauch-Blandford-1994}
K.P. Rauch, R.D. Blandford,
Optical Caustics in a Kerr space-time and the origin of rapid X-ray variability in Active Galactic Nuclei,
Astrophys. J. 421 (1994) 46







\bibitem{Chan-Psaltis-2013}
C. Chan, D. Psaltis, F. \"{O}zel,
GRay: A Massively parallel GPU-based code for ray tracing in relativistic space-times,
Astrophys. J. 777 (2013)  13

\bibitem{Takahashi-2004}
R. Takahashi, 
Shapes and positions of black hole shadows in accretion disks and spin parameters of black holes,
Astrophys. J. 611 (2004) 996



\bibitem{Psaltis-2019-review}
D. Psaltis,
Testing general relativity with the Event Horizon Telescope,
Gen. Rel. Grav. 51 (2019) 137

\bibitem{Hioki-Maeda-2009}
K. Hioki and K. Maeda,
Measurement of the Kerr spin parameter by observation of a compact object's shadow,
Phys. Rev. D 80 (2009) 024042 


\bibitem{Li-Bambi-2014}
Z. Li, C. Bambi,
Measuring the Kerr spin parameter of regular black holes from their shadow,
J. Cosm. Astropart. Phys. 01 (2014) 041


\bibitem{Tsukamoto-Li-Bambi-2014}
N. Tsukamoto, Z. Li, C. Bambi, 
Constraining the spin and the deformation parameters from the black hole shadow,
J. Cosm. Astropart. Phys.  06 (2014) 043


\bibitem{Abdu-Rezzolla-Ahmedov-2015}
A. A. Abdujabbarov, L. Rezzolla and B. J. Ahmedov,
A coordinate-independent characterization of a black hole shadow, 
Mon. Not. Roy. Astron. Soc. 454 (2015) 2423

\bibitem{Yang-Li-2016}
L. Yang, Z. Li, 
Shadow of a dressed black hole and determination of spin and viewing angle,
Int. J. Mod. Phys. D 25 (2016) 1650026


\bibitem{Wei-2019-Rapid}
S. Wei, Y. Liu, R.B. Mann,
Intrinsic curvature and topology of shadows in Kerr space-time,
Phys. Rev. D  99 (2019) 041303


\bibitem{Kumar-Ghosh-2020}
R. Kumar, S.G. Ghosh,
Black hole parameter estimation from its shadow,
Astrophys. J. 892 (2020) 78

\bibitem{Farah-2020}
J. R. Farah, D. W. Pesce, M. D. Johnson, L. Blackburn,
On the approximation of the black hole shadow with a simple polar curve
Astrophys. J. 900 (2020) 77


\bibitem{Vries-2000}
A. de Vries,
The apparent shape of a rotating charged black hole, closed photon orbits and the bifurcation set $A_4$,
Class. Quantum Grav. 17 (2000)  123

\bibitem{KonoplyaRezzollaZhidenko2016}
R. Konoplya, L. Rezzolla, A. Zhidenko,
General parametrization of axisymmetric black holes in metric theories of gravity,
Phys. Rev. D 93 (2016) 064015

\bibitem{YounsiEtAl2016}
Z. Younsi, A. Zhidenko, L. Rezzolla, R. Konoplya, Y. Mizuno, 
New method for shadow calculations: Application to parametrized axisymmetric black holes,
Phys. Rev. D 94 (2016) 084025


\bibitem{KonoplyaStuchlikZhidenko2018}
R.A. Konoplya, Z. Stuchl{\'\i}k, A. Zhidenko, 
Axisymmetric black holes allowing for separation of variables in the Klein-Gordon and Hamilton-Jacobi equations,
Phys. Rev. D 97 (2018) 084044


\bibitem{KonoplyaZhidenko2021}
R.A. Konoplya, A. Zhidenko,
Shadows of parametrized axially symmetric black holes allowing for separation of variables,
arXiv:2103.03855 (2021)

\bibitem{Tsukamoto-2018}
N. Tsukamoto,
Black hole shadow in an asymptotically flat, stationary, and axisymmetric spacetime: The Kerr-Newman and rotating regular black holes,
Phys. Rev. D 97 (2018) 064021 



\bibitem{GlampedakisPappas2019}
K. Glampedakis and G. Pappas,
 Modification of photon trapping orbits as a diagnostic of non-Kerr spacetimes,
Phys. Rev D 99 (2019) 124041




\bibitem{Cunha-Herdeiro-PRL-2020}
P. V. P. Cunha and C. A. R. Herdeiro,
Stationary Black Holes and Light Rings,
Phys. Rev. Lett. 124, 181101 (2020)











\bibitem{Mars-2018}
M. Mars, C.F. Paganini, M.A. Oancea, 
The fingerprints of black holes - Shadows and their degeneracies,
Class. Quantum Grav. 35 (2018) 025005


\bibitem{Sen-1992}
A. Sen,
Rotating charged black hole solution in heterotic string theory,
Phys. Rev. Lett. 69 (1992)  1006



\bibitem{XavierEtAl-2020}
S.V.M.C.B. Xavier, P.V.P. Cunha, L.C.B. Crispino, C.A.R. Herdeiro, 
Shadows of charged rotating black holes: Kerr-Newman versus Kerr-Sen,
Int. J. Mod. Phys. D 29 (2020) 2041005



\bibitem{Haroon-2019}
S. Haroon, M. Jamil, K. Jusufi, K. Lin, R.B. Mann,
Shadow and deflection angle of rotating black holes in perfect fluid dark matter with a cosmological constant,
Phys. Rev. D 99 (2019) 044015



\bibitem{Neves-2020a}
J.C.S. Neves,
Upper bound on the GUP parameter using the black hole shadow,
Eur. Phys. J. C 80 (2020a) 343


\bibitem{AmarillaEiroa2012}
L. Amarilla, E.F. Eiroa,
Shadow of a rotating braneworld black hole,
Phys.Rev.D 85 (2012) 064019

\bibitem{Eiroa-2018}
E.F. Eiroa, C.M. Sendra,
Shadow cast by rotating braneworld black holes with a cosmological constant,
Eur. Phys. J. C 78 (2018) 91



\bibitem{Neves-2020b}
J.C.S. Neves,
Constraining the tidal charge of brane black holes using their shadows,
Eur. Phys. J. C 80 (2020b)  717



\bibitem{NewmanJanis-1965}
E.T. Newman, A.I. Janis, 
Note on the Kerr spinning-particle metric,
J. Math. Phys. 6 (1965) 915



\bibitem{Azreg-2014}
M. Azreg-A{\"i}nou, 
Generating rotating regular black hole solutions without complexification,
Phys. Rev. D 90 (2014) 064041



\bibitem{LimaEtAl-2020}
H.C.D. Lima Junior, L.C.B. Crispino, P.V.P. Cunha, C.A.R. Herdeiro,
Spinning black holes with a separable Hamilton-Jacobi equation from a modified Newman-Janis algorithm,
Eur. Phys. J. C 80 (2020) 1036


\bibitem{AbdolrahimiEtAl2015}
S. Abdolrahimi, R.B. Mann, C. Tzounis,
Distorted local shadows,
Phys. Rev. D 91 (2015) 084052



\bibitem{Grover-2018}
J. Grover, J. Kunz, P. Nedkova, A. Wittig, S. Yazadjiev,
Multiple shadows from distorted static black holes,
Phys. Rev. D 97 (2018) 084024


\bibitem{CunhaEtAl2015}
P.V.P. Cunha, C.A.R. Herdeiro, E. Radu, H.F. R{\'u}narsson,
Shadows of Kerr black holes with scalar hair,
Phys. Rev. Lett. 115 (2015) 211102


\bibitem{VincentEtAl2016a}
F. H. Vincent, E. Gourgoulhon, C. Herdeiro, E. Radu,
Astrophysical imaging of Kerr black holes with scalar hair,
Phys. Rev. D 94 (2016) 084085



\bibitem{BohnEtAl2015}
A. Bohn, W. Throwe, F. H{\'e}bert, K. Henriksson, 
D. Bunandar, M. A. Scheel, N. W. Taylor,
What does a binary black hole merger look like?
Class. Quantum Grav. 32 (2015) 065002.

\bibitem{NittaEtAl2011}
D. Nitta, T. Chiba, N. Sugiyama,
Shadows of colliding black holes,
Phys. Rev. D 84 (2011) 063008



\bibitem{YumotoEtAl2012}
A. Yumoto, D. Nitta, T. Chiba, N. Sugiyama,
Shadows of multi-black holes: Analytic exploration,
Phys. Rev. D 86 (2012) 103001


\bibitem{CunhaEtAl2018}
P.V.P. Cunha, C.A.R. Herdeiro, M.J. Rodriguez,
Shadows of exact binary black holes,
Phys. Rev. D 98 (2018) 044053



\bibitem{Shipley-Dolan-2016}
J. O. Shipley, S. R. Dolan,
Binary black hole shadows, chaotic scattering and the Cantor set,
Class. Quantum Grav. 33 (2016) 175001



\bibitem{Yunes-2019}
H. Gott, D. Ayzenberg, N. Yunes, A. Lohfink,
Observing the shadows of stellar-mass black holes with binary companions,
Class and Quantum Grav. 36 (2019) 055007



\bibitem{AbramowiczEtAl2002}
M.A. Abramowicz, W. Klu{\'z}niak, J.-P. Lasota,
No observational proof of the black-hole event-horizon,
Astr. Astrophys.396 (2002) L31


\bibitem{CardosoEtAl2014}
V. Cardoso, L.C.B. Crispino, C.F.B. Macedo, H. Okawa, P. Pani, 
Light rings as observational evidence for event horizons: Long-lived modes, ergoregions and nonlinear instabilities of ultracompact objects,
Phys. Rev. D 90  (2014) 044069


\bibitem{EinsteinRosen1935}
A. Einstein, N. Rosen,
The particle problem in the general theory of relativity,
Phys. Rev. 48 (1935) 73



\bibitem{Ellis1973}
H. Ellis, 
Ether flow through a drainhole: A particle model in general relativity,
J.Math.Phys. 14 (1973) 104



\bibitem{MorrisThorne1988}
M. S. Morris, K. S. Thorne,
Wormholes in space-time and their use for interstellar travel: A tool for teaching general relativity,
Amer. J. Phys. 56 (1988) 395





\bibitem{Kanti-2011}
P. Kanti, B. Kleihaus, and J. Kunz,
Wormholes in Dilatonic Einstein-Gauss-Bonnet Theory,
Phys. Rev. Lett. 107 (2011) 271101




\bibitem{Bronnikov-2017}
K. A. Bronnikov, K. A. Baleevskikh, and M. V. Skvortsova,
Wormholes with fluid sources: A no-go theorem and new examples,
Phys. Rev. D 96 (2017) 124039



\bibitem{Antoniou-2020}
G. Antoniou, A. Bakopoulos, P. Kanti, et al,
Novel Einstein-scalar-Gauss-Bonnet wormholes without exotic matter,
Phys. Rev. D 101 (2020) 024033











\bibitem{ChetouaniClement1984}
L. Chetouani, G. Cl{\'e}ment,
Geometrical optics in the Ellis geometry,
Gen. Rel. Grav. 16 (1984) 111


\bibitem{Perlick-2004b}
V. Perlick,
Exact gravitational lens equation in spherically symmetric and static space-times,
Phys. Rev. D69 (2004) 064017



\bibitem{NandiZhangZakharov2006}
K.K. Nandi, Y.-Zh. Zhang, A.V. Zakharov,
Gravitational lensing by wormholes,
Phys. Rev. D 74 (2006) 024020


\bibitem{Mueller2008}
T. M{\"u}ller,
Exact geometric optics in a Morris-Thorne wormhole space-time,
Phys. Rev. D 77 (2008) 044043


\bibitem{NakajimaAsada2012}
K. Nakajima, H. Asada,
Deflection angle of light in an Ellis wormhole geometry,
Phys. Rev. D 85 (2012) 107501



\bibitem{Ohgami-2015}
T. Ohgami, N. Sakai,
Wormhole shadows,
Phys. Rev. D 91  (2015) 124020


\bibitem{WielgusEtAl2020}
M. Wielgus, J. Hor{\'a}k, F. Vincent, M. Abramowicz,
Reflection-asymmetric wormholes and their double shadows,
Phys. Rev. D 102 (2020) 084044 

\bibitem{Tsukamoto-2021}
N. Tsukamoto, 
Linearization stability of reflection-asymmetric thin-shell wormholes with double shadows,
Phys. Rev. D 103 (2021) 064031





\bibitem{Teo1998}
E. Teo,
Rotating traversable wormholes,
Phys. Rev. D 58 (1998) 024014


\bibitem{NedkovaEtAl2013}
P.G. Nedkova, V.K. Tinchev, S.S. Yazadjiev,  
Shadow of a rotating traversable wormhole,
Phys. Rev. D 88 (2013) 124019



\bibitem{Shaikh2018}
R. Shaikh,
Shadows of rotating wormholes,
Phys. Rev. D 98 (2018) 024044


\bibitem{GyulchevEtAl2018}	
G. Gyulchev, P. Nedkova, V. Tinchev,  S. Yazadjiev, 
On the shadow of rotating traversable wormholes,
Eur. Phys. J. C 78 (2018) 544



\bibitem{Galtsov-2019}
D.V. Gal'tsov and K.V. Kobialko,
Photon trapping in static axially symmetric space-time,
Phys. Rev. D 100 (2019) 104005




\bibitem{VincentEtAl2016b}
F. H. Vincent, Z. Meliani, P. Grandcl{\'e}ment, 
E. Gourgoulhon, O. Straub, 
Imaging a boson star at the Galactic center,
Class. Quantum Grav. 33 (2016) 105015


\bibitem{HerdeiroEtAl2021}
C.A.R. Herdeiro, A.M. Pombo, E. Radu, P.V.P. Cunha, N. Sanchis-Gual,
The imitation game: Proca stars that can mimic the Schwarzschild shadow,
Journal of Cosmology and Astroparticle Physics, 04 (2021) 051


\bibitem{GomezEtAl2016}
L.G. G{\'o}mez, C.R. Arg{\"u}elles, V. Perlick, 
J.A. Rueda, R. Ruffini, 
Strong lensing by fermionic dark matter in galaxies
Phys. Rev. D 94 (2016) 123004


\bibitem{AmesThorne1968}
W.L. Ames, K.S. Thorne, 
The optical appearance of a star that is collapsing through its gravitational radius,
Astrophys. J. 151 (1968)  659



\bibitem{Jaffe1969}
J. Jaffe,  
Collapsing objects and the backward emission of light,
Ann. Phys. (NY) 55 (1969) 374



\bibitem{LakeRoeder1979}
K.~Lake, R.C.~Roeder, 
Note on the optical appearance of a star collapsing through its gravitational radius,
Astrophys. J.  232 (1979)  277

	
\bibitem{FrolovKimLee2007}
V.~P.~Frolov, K.~Kim, H.~K.~Lee, 
Spectral broadening of radiation from relativistic collapsing objects,
Phys. Rev. D 75 (2007)  087501


\bibitem{KongMalafarinaBambi2014}
L.~Kong, D.~Malafarina, C.~Bambi,
Can we observationally test the weak cosmic censorship conjecture?
Eur. Phys. J. C 74 (2014)  2983


\bibitem{KongMalafarinaBambi2015}
L.~Kong, D.~Malafarina, C.~Bambi,
Gravitational blueshift from a collapsing object,
Phys. Lett. B 741 82 (2015)


\bibitem{OrtizSarbachZannias2015a}
N.~Ortiz, O.~Sarbach, T.~Zannias, 
Shadow of a naked singularity,
Phys. Rev.D 92 (2015) 044035


\bibitem{OrtizSarbachZannias2015b}
N.~Ortiz, O.~Sarbach, T.~Zannias, 
Observational distinction between black holes and naked singularities: the role of the redshift function,
Class. Quantum Grav. 32 (2015)  247001


\bibitem{SchneiderPerlick2018}
S. Schneider, V. Perlick,
The shadow of a collapsing dark star,
Gen. Rel. Grav. 50  (2018) 58


\bibitem{OppenheimerSnyder1939}
J. R. Oppenheimer, H.~Snyder,
On continued gravitational contraction,
Phys. Rev. 56 (1939)  455
  
	
\bibitem{Mattig-1958}	
W. Mattig,
\"{U}ber den Zusammenhang zwischen Rotverschiebung und scheinbarer Helligkeit,
Astron. Nachr. 284 (1958) 109

\bibitem{Zeldovich-1964}
Ya. B. Zeldovich,
Observations in a universe homogeneous in the mean,
Sov. Astron. 8 (1964) 13

\bibitem{Dashevsk-Zeldovich-1965}
V. M. Dashevskii, Ya. B. Zeldovich,
Sov. Astron. 8 (1965)  854



\bibitem{Zeldovich-Novikov-book-2}	
Ya. B. Zeldovich, I.D. Novikov, 
Relativistic Astrophysics. Vol.2: The structure and Evolution of the Universe 
Univ. Chicago Press, Chicago (1983)



\bibitem{Hobson}
M.P. Hobson, G.P. Efstathiou, A.N. Lasenby, 
General Relativity: An Introduction for Physicists.
Cambridge Univ. Press, Cambridge (2006)


\bibitem{Mukhanov-book}
V. Mukhanov, 
Physical Foundations of Cosmology, 
Cambridge Univ. Press, Cambridge (2005).



\bibitem{Einstein-Straus-1945}
A. Einstein, and E. G. Straus,
The influence of the expansion of space on the gravitation fields surrounding the individual stars,
Rev. Mod. Phys. 17 (1945) 120


\bibitem{Einstein-Straus-1946}
A. Einstein,  E. G. Straus,
Corrections and additional remarks to our paper: The influence of the expansion of space on the gravitation fields surrounding the individual stars,
Rev. Mod. Phys. 18 (1946) 148


\bibitem{Schucking-1954}
E. Sch\"{u}cking,
Das Schwarzschildsche Linienelement und die Expansion des Weltalls,
Z. Phys. 137 (1954) 595


\bibitem{Stuchlik-1984}
Z. Stuchl\'{i}k,
An Einstein-Strauss-de Sitter model of the universe,
Bull. Astron. Inst. Czechoslovakia 35 (1984) 205


\bibitem{Balbinot-1988}
R. Balbinot, R. Bergamini, A. Comastri,
Solution of the Einstein-Strauss problem with a $\Lambda$ term,
Phys. Rev. D 38 (1988) 2415

\bibitem{Schucker-2009}
T. Sch\"{u}cker,
Strong lensing in the Einstein-Straus solution,
Gen. Rel. Grav. 41 (2009) 1595


\bibitem{Schucker-2010}
T. Sch\"{u}cker,
Lensing in an interior Kottler solution,
Gen. Rel. Grav. 42 (2010) 1991




\bibitem{McVittie1933}
G. C. McVittie,
The mass-particle in an expanding universe,
Mon. Not. Roy. Astron. Soc. 93 (1933) 325


\bibitem{Nolan-1}
B. C. Nolan,
A point mass in an isotropic universe: Existence, uniqueness, and basic properties,
Phys. Rev. D  58 (1998) 064006


\bibitem{Nolan-2}
B. C. Nolan,
A point mass in an isotropic universe: II. Global properties,
Class. Quantum Grav. 16 (1999a) 1227


\bibitem{Nolan-3}
B. C. Nolan,
A point mass in an isotropic universe: III. The region $R \le 2m$,
Class. Quantum Grav., \textbf{16} (1999b) 3183.


\bibitem{Kaloper-2010}
N. Kaloper, M. Kleban, D. Martin,
McVittie's legacy: Black holes in an expanding universe,
Phys. Rev. D 81 (2010) 104044


\bibitem{Carrera-Giulini-2010a}
M. Carrera, D. Giulini,
Generalization of McVittie's model for an inhomogeneity in a cosmological space-time,
Phys. Rev. D 81 (2010) 043521



\bibitem{Carrera-Giulini-2010-review}
M. Carrera, D. Giulini,
Influence of global cosmological expansion on local dynamics and kinematics,
Rev. Mod. Phys. 82 (2010) 169



\bibitem{Anderson-2011}
M. Anderson,
Horizons, singularities and causal structure of the generalized McVittie space-times,
J. Phys.: Conf. Ser. 283 (2011) 012001



\bibitem{Lake-Abdelqader-2011}
K. Lake, M. Abdelqader,
More on McVittie's legacy: A Schwarzschild-de Sitter black and white hole embedded in an asymptotically $\Lambda$CDM cosmology
Phys. Rev. D 84 (2011) 044045



\bibitem{Hobson-2012a}
R. Nandra, A. N. Lasenby, M. P. Hobson,
The effect of a massive object on an expanding universe,
Mon. Not. Roy. Astron. Soc. 422 (2012) 2931


\bibitem{Hobson-2012b}
R. Nandra, A. N. Lasenby and M. P. Hobson,
The effect of an expanding universe on massive objects,
Mon. Not. Roy. Astron. Soc. 422 (2012) 2945


\bibitem{Silva-2013}
A.M. da Silva, M. Fontanini, D.C. Guariento,
How the expansion of the Universe determines the causal structure of McVittie space-times,
Phys. Rev. D 87 (2013) 064030




\bibitem{Nolan-2014}
B.C. Nolan,
Particle and photon orbits in McVittie space-times,
Class. Quantum Grav. 31 (2014) 235008


\bibitem{Piattella-PRD-2016}
O. F. Piattella,
Lensing in the McVittie metric,
Phys. Rev. D 93 (2016) 024020.


\bibitem{Piattella-Universe-2016}
O. F. Piattella,
On the effect of the cosmological expansion on the gravitational lensing by a point mass,
Universe 2 (2016) 25


\bibitem{Nolan-2017}
B. C. Nolan,
Local properties and global structure of McVittie space-times with non-flat Friedmann-Lema{\^\i}tre-Robertson-Walker backgrounds,
Class. Quantum Grav. 34 (2017) 225002



\bibitem{Faraoni-2017}
V. Faraoni, M. Lapierre-L\'{e}onard,
Beyond lensing by the cosmological constant,
Phys. Rev. D 95 (2017) 023509



\bibitem{Aghili-2017}
M.E. Aghili, B. Bolen, L. Bombelli,
Effect of accelerated global expansion on the bending of light,
Gen. Rel. Grav. 49 (2017) 10





\bibitem{Stuchlik-2006}
P. Bakala, P. {\v C}erm{\` a}k, S. Hled{\' i}k, Z. Stuchl{\'\i}k, K. Truparov{\`a},
A virtual trip to the Schwarzschild-de Sitter black hole,
arXiv:gr-qc/0612124 (2006).


\bibitem{Stuchlik-2007}
P. Bakala, P. \v{C}erm\'{a}k, S. Hled\'{i}k, Z. Stuchl\'{i}k, and K. Truparov\'{a},
Extreme gravitational lensing in vicinity of Schwarzschild-de Sitter black holes,
Central Europ. J. Phys. 5 (2007) 599



\bibitem{Stuchlik-2018}
Z. Stuchl\'{i}k, D. Charbul\'{a}k, J. Schee,
Light escape cones in local reference frames of Kerr-de Sitter black hole space-times and related black hole shadows,
Eur. Phys. J. C 78 (2018) 180




\bibitem{Tsupko-BK-2020b}
O.Yu. Tsupko, G.S. Bisnovatyi-Kogan,
First analytical calculation of black hole shadow in McVittie metric,
Int. J. Mod. Phys. D 29 (2020) 2050062



\bibitem{Roy-2020}
R. Roy, S. Chakrabarti,
Study on black hole shadows in asymptotically de Sitter space-times,
Phys. Rev. D 102 (2020) 024059



\bibitem{Chang-Zhu-2020}
Zhe Chang and Qing-Hua Zhu,
Black hole shadow in the view of freely falling observers,
J. Cosmol. Astropart. Phys. 06
(2020) 055




\bibitem{Islam-1983}
J.N. Islam,
The cosmological constant and classical tests of general relativity,
Phys. Lett. A 97 (1983) 239


\bibitem{Rindler-Ishak-2007}	
W. Rindler, M. Ishak,
Contribution of the cosmological constant to the relativistic bending of light revisited,
Phys. Rev. D 76 (2007) 043006


\bibitem{Hackmann-2008a}
E. Hackmann, C. L{\"a}mmerzahl,
Complete analytic solution of the geodesic equation in Schwarzschild-(anti-)de Sitter space-times,
Phys. Rev. Lett. 100 (2008) 171101



\bibitem{Hackmann-2008b}
E. Hackmann, C. L{\"a}mmerzahl,
Geodesic equation in Schwarzschild-(anti-)de Sitter space-times: Analytical solutions and applications,
Phys. Rev. D 78 (2008) 024035



\bibitem{Lebedev-2013}
D. Lebedev, K. Lake,
On the influence of the cosmological constant on trajectories of light and associated measurements in Schwarzschild de Sitter space,
arXiv: 1308.4931 (2013)


\bibitem{Lebedev-2016}
D. Lebedev, K. Lake,
Relativistic aberration and the cosmological constant in gravitational lensing I: Introduction,
arXiv: 1609.05183 (2016)



\bibitem{Jones-book}
B.J.T. Jones,
Precision Cosmology: The First Half Million Years,
Cambridge Univ. Press, Cambridge (2017)





\bibitem{BK-Tsupko-2018}
G.S. Bisnovatyi-Kogan, O.Yu. Tsupko,
Shadow of a black hole at cosmological distances,
Phys. Rev. D 98 (2018) 084020



\bibitem{Li-Guo-2020}
P.-C. Li, M. Guo, B. Chen,
Shadow of a spinning black hole in an expanding universe,
Phys. Rev. D 101 (2020) 084041



\bibitem{SN-1998}
A.G. Riess, A.V. Filippenko, P. Challis et al,
Observational evidence from supernovae for an accelerating universe and a cosmological constant,
Astron. J. 116 (1998) 1009



\bibitem{SN-1999}
S. Perlmutter, G. Aldering, G. Goldhaber et al,
Measurements of $\Omega$ and $\Lambda$ from 42 high-redshift supernovae,
Astrophys. J. 517 (1999) 565


\bibitem{sirens-1986}
B.F. Schutz,
Determining the Hubble constant from gravitational wave observations,
Nature 323 (1986) 310


\bibitem{sirens-2005}
D.E. Holz, S.A. Hughes,
Using gravitational-wave standard sirens,
Astrophys. J. 629 (2005) 15



\bibitem{sirens-2017}
B.P. Abbott, R. Abbott, T.D. Abbott et al,
A gravitational-wave standard siren measurement of the Hubble constant,
Nature 551 (2017) 85


\bibitem{Tsupko-Fan-BK-2020}
O.Yu. Tsupko, Z. Fan,  G.S. Bisnovatyi-Kogan,
Black hole shadow as a \textit{standard ruler} in cosmology,
Class. Quantum Grav. 37 (2020) 065016




\bibitem{Qi-Zhang-2019}
J.-Zh. Qi, X. Zhang,
A new cosmological probe using super-massive black hole shadows,
Chinese Phys. C 44 (2020) 055101





\bibitem{Vagnozzi-2020}
S. Vagnozzi, C. Bambi, L. Visinelli,
Concerns regarding the use of black hole shadows as standard rulers,
Class. Quantum Grav. 37 (2020) 087001




\bibitem{Eubanks-2020}
T.M. Eubanks,
Anchored in shadows: Tying the celestial reference frame directly to black hole event horizons,
arXiv:2005.09122 (2020)

\bibitem{Breuer-Ehlers-1980}
R. Breuer, J. Ehlers,
Propagation of high-frequency electromagnetic waves through a magnetized plasma in curved space-time. I,
Proc. Roy. Soc. London.  A 370 (1980) 389


\bibitem{Synge-1960}
J.L. Synge,
Relativity: The General Theory,
North-Holland, Amsterdam (1960)

\bibitem{Bicak-1975}
J. Bi\v{c}\'{a}k, P. Hadrava,
General-relativistic radiative transfer theory in refractive and dispersive media,
Astron. Astrophys. 44 (1975) 389

\bibitem{Kulsrud-Loeb-1992}
R. Kulsrud, A. Loeb,
Dynamics and gravitational interaction of waves in nonuniform media,
Phys. Rev. D 45 (1992) 525




\bibitem{MuhlemanJohnston1966}
D.O. Muhleman, I.D. Johnston,
Radio propagation in the solar gravitational field,
Phys. Rev. Lett. 17 (1966) 455


\bibitem{Bliokh-Minakov-1989}
P. V. Bliokh, A. A. Minakov, 
Gravitational Lenses (in Russian),
Naukova Dumka, Kiev, (1989) 


\bibitem{Perlick-2000b}
V.~Perlick, 
Ray Optics, Fermat's Principle, and Applications to General Relativity,
Springer, Berlin (2000)



\bibitem{BK-Tsupko-2009}
G. S. Bisnovatyi-Kogan, O. Yu. Tsupko,
Gravitational radiospectrometer,
Gravitation and Cosmolog 15 (2009) 20


\bibitem{BK-Tsupko-2010}
G. S. Bisnovatyi-Kogan, O. Yu. Tsupko,
Gravitational lensing in a non-uniform plasma,
Mon. Not. Roy. Astron. Soc. 404 (2010) 1790



\bibitem{Tsupko-BK-2013}
O. Yu. Tsupko, G. S. Bisnovatyi-Kogan,
Gravitational lensing in plasma: Relativistic images at homogeneous plasma,
Phys. Rev. D 87 (2013) 124009


\bibitem{Perlick-Tsupko-2017}
V. Perlick, O. Yu. Tsupko,
Light propagation in a plasma on Kerr space-time: Separation of the Hamilton-Jacobi equation and calculation of the shadow,
Phys. Rev. D 95 (2017) 104003


\bibitem{Yan-2018}
H. Yan,
Influence of a plasma on the observational signature of a high-spin Kerr black hole,
Phys. Rev. D 99 (2019) 084050


\bibitem{Huang-2018}
Yang Huang, Yi-Ping Dong, Dao-Jun Liu,
Revisiting the shadow of a black hole in the presence of a plasma,
Int. J. Mod. Phys. D, 27 (2018) 1850114





\bibitem{Morozova-2013}
V.S. Morozova, B.J. Ahmedov, A.A. Tursunov,
Gravitational lensing by a rotating massive object in a plasma,
Astrophys. Space Sci. 346 (2013) 513.


\bibitem{Er-Mao-2014}
X. Er, S. Mao,
Effects of plasma on gravitational lensing,
Mon. Not. Roy. Astron. Soc. 437 (2014) 2180



\bibitem{Rogers-2015}
A. Rogers,
Frequency-dependent effects of gravitational lensing within plasma,
Mon. Not. Roy. Astron. Soc. 451 (2015) 17


\bibitem{Rogers-2017a}
A. Rogers,
Escape and trapping of low-frequency gravitationally lensed rays by compact objects within plasma,
Mon. Not. Roy. Astron. Soc. 465 (2017) 2151


\bibitem{Rogers-2017b}
A. Rogers,
Gravitational lensing of rays through the levitating atmospheres of compact objects,
Universe 3 (2017) 3




\bibitem{BK-Tsupko-2015}
G. S. Bisnovatyi-Kogan, O. Yu. Tsupko,
Gravitational lensing in plasmic medium,
Plasma Phys. Rep. 41 (2015) 562



\bibitem{Er-Rogers-2018}
X. Er, A. Rogers,
Two families of astrophysical diverging lens models,
Mon. Not. Roy. Astron. Soc. 475 (2018) 867



\bibitem{Perlick-2017}
K. Schulze-Koops, V. Perlick, D.J. Schwarz, 
Sachs equations for light bundles in a cold plasma,
Class. Quantum Grav. 34 (2017) 215006




\bibitem{Crisnejo-Gallo-2018}
G. Crisnejo, E. Gallo,
Weak lensing in a plasma medium and gravitational deflection of massive particles using the Gauss-Bonnet theorem. A unified treatment,
Phys. Rev. D 97 (2018) 124016



\bibitem{Crisnejo-Gallo-Rogers-2019}
G. Crisnejo, E. Gallo, A. Rogers,
Finite distance corrections to the light deflection in a gravitational field with a plasma medium,
Phys. Rev. D 99 (2019) 124001


\bibitem{Crisnejo-Gallo-Villanueva-2019}
G. Crisnejo, E. Gallo, J.R. Villanueva,
Gravitational lensing in dispersive media and deflection angle of charged massive particles in terms of curvature scalars and energy-momentum tensor,
Phys. Rev. D 100 (2019) 044006

\bibitem{Crisnejo-Gallo-Jusufi-2019}
G. Crisnejo, E. Gallo, K. Jusufi,
Higher order corrections to deflection angle of massive particles and light rays in plasma media for stationary space-times using the Gauss-Bonnet theorem,
Phys. Rev. D, 100 (2019) 104045




\bibitem{Kimpson-2019a}
T. Kimpson, K. Wu, S. Zane,
Spatial dispersion of light rays propagating through a plasma in Kerr space-time,
MNRAS, 484 (2019a)  2411



\bibitem{Kimpson-2019b}
T. Kimpson, K. Wu, S. Zane,
Pulsar timing in extreme mass ratio binaries: a general relativistic approach,
Mon. Not. Roy. Astron. Soc. 486, 360 (2019b)



\bibitem{Tsupko-BK-2020a}
O.Yu. Tsupko, G.S. Bisnovatyi-Kogan,
Hills and holes in the microlensing light curve due to plasma environment around gravitational lens,
Mon. Not. Roy. Astron. Soc. 491 (2020) 5636





\bibitem{Wagner-Er-2020}
J. Wagner, X. Er,
Plasma lensing in comparison to gravitational lensing -- Formalism and degeneracies,
arXiv:2006.16263 (2020)


\bibitem{Sareny-2019}
M. S\'{a}ren\'{y}, V. Balek,
Effect of black hole--plasma system on light beams,
Gen. Rel. Grav. 51 (2019) 141



\bibitem{Matsuno-2020}
K. Matsuno,
Light deflection by squashed Kaluza-Klein black holes in a plasma medium,
Phys. Rev. D 103 (2021) 044008


\bibitem{Jin-2020}
Xing-Hua Jin, Yuan-Xing Gao, Dao-Jun Liu,
Strong gravitational lensing of a 4-dimensional Einstein-Gauss-Bonnet black hole in homogeneous plasma,
Int. J. Mod. Phys. D 29 (2020) 2050065



\bibitem{Chowdhuri-2020}
A. Chowdhuri, A. Bhattacharyya,
Shadow analysis for rotating black holes in the presence of plasma for an expanding universe,
arXiv:2012.12914 (2020)



\bibitem{Li-Wang-2020}
Q. Li, T. Wang,
Gravitational effect of a plasma on the shadow of Schwarzschild black holes,
arXiv:2102.00957 (2020)




\bibitem{Broderick-Blandford-2003}
A. Broderick, R. Blandford,
Covariant magnetoionic theory -- I. Ray propagation,
Mon. Not. Roy. Astron. Soc. 342 (2003)  1280


\bibitem{Broderick-Blandford-2004}
A. Broderick, R. Blandford,
Covariant magnetoionic theory -- II. Radiative transfer,
Mon. Not. Roy. Astron. Soc. 349 (2004)  994


\bibitem{Tsupko-2021}
O.Yu. Tsupko,
Deflection of light rays by a spherically symmetric black hole in a dispersive medium,
Phys. Rev. D 103 (2021) 104019




\bibitem{EHT-2021}
The Event Horizon Telescope Collaboration,
First M87 Event Horizon Telescope results. VIII. Magnetic field structure near the event horizon,
Astrophys. J. Lett. 910 (2021) L13




\bibitem{BK-1974}
G. S. Bisnovatyi-Kogan, A. A. Ruzmaikin,
The accretion of matter by a collapsing star in the presence of a magnetic field,
Astrophys. Space Sci. 28 (1974) 45 

\bibitem{BK-1976}
G. S. Bisnovatyi-Kogan, A. A. Ruzmaikin,
The accretion of matter by a collapsing star in the presence of a magnetic field. II. Selfconsistent stationary picture,
Astrophys. Space Sci. 42 (1976) 401


\bibitem{Igumenshchev-2003}
I.V. Igumenshchev, R. Narayan, M.A. Abramowicz,
Three-dimensional magnetohydrodynamic simulations of radiatively inefficient accretion flows,
Astrophys. J. 592 (2003) 1042


\bibitem{Narayan-2003}
R. Narayan, I.V. Igumenshchev, M.A. Abramowicz,
Magnetically Arrested Disk: an energetically efficient accretion flow,
Publ. Astron. Soc. Japan 55 (2003) L69



\bibitem{BK-2019-review}
G.S. Bisnovatyi-Kogan,
Accretion into black hole, and formation of magnetically arrested accretion disks,
Universe 5 (2019) 146








\end{thebibliography}
\end{document}